\documentclass[12pt]{aastex631}

\newcommand{\DELETED}[1]{\relax}%
{\relax}%

\definecolor{violet}  {rgb}{1.0,0.0,1.0}

\definecolor{dviolet} {rgb}{0.75,0.0,1.0}

\definecolor{blue}    {rgb}{0.0,0.7,1.0}

\definecolor{lblue}   {rgb}{0.5,1,1}

\definecolor{dblue}   {rgb}{0.0,0.0,1.0}

\definecolor{blgr}    {rgb}{0.70,0.80,1.00}

\definecolor{navy}    {rgb}{0.00,0.00,0.48}

\definecolor{green}   {rgb}{0.7,1.0,0.0}
\newcommand{\green}   {\color{green}}
\definecolor{dgreen}  {rgb}{0.0,0.5,0.0}

\definecolor{lgreen}  {rgb}{0.0,0.8,0.0}

\definecolor{dg}      {rgb}{0.0,0.6,0.0}

\definecolor{orange}  {rgb}{1.0,0.5,0.0}

\definecolor{dorange} {rgb}{1.0,0.6,0.0}

\definecolor{brown}   {rgb}{0.1,0.1,0.0}

\definecolor{lbrown}  {rgb}{0.7,0.5,0.0}

\definecolor{red}     {rgb}{1,0.0,0.0}

\definecolor{dred}    {rgb}{0.6,0.0,0.0}

\definecolor{grey}    {rgb}{0.1,0.1,0.1}

\definecolor{lgrey}   {rgb}{0.5,0.5,0.5}

\definecolor{black}   {rgb}{0.0,0.0,0.0}

\newcommand\n            {\noindent}

\newcommand\m            {\medskip}
\newcommand\s            {\smallskip}

\newcommand\si           {\smallskip\indent}
\newcommand\bn           {\bigskip\noindent}
\newcommand\mn           {\medskip\noindent}
\newcommand\sn           {\smallskip\noindent}
\newcommand\cl           {\centerline}
\newcommand\ve           {\vfill\eject}

\newcommand\arcspt       {{$\buildrel{\prime\prime}\over .$}}
\newcommand\degree       {{\ifmmode^\circ\else$^\circ$\fi}} 
\newcommand\arcm         {{\ifmmode {'\ }\else$'     $\fi}} 
\newcommand\arcs         {{\ifmmode{''\ }\else$''    $\fi}} 
\newcommand{\ch}         {} 

\newcommand{\bul}        {$\bullet$\ }
\newcommand{\gbul}       {\ {\green $\bullet$}\ }
\newcommand\cge          {{$\gtrsim$}}
\newcommand\cle          {{$\lesssim$}}
\newcommand\degsq        {{deg$^{2}$}}

\newcommand\sqdeghalfmag {{deg$^{-2}$/0.5 mag}}

\newcommand\eg           {{\it e.g.},}
\newcommand\ie           {{\it i.e.},}
\newcommand\emin         {{$\rm e^{-}$}}

\newcommand\etal         {{et\thinspace al.}} 
\newcommand\fAGN         {{$f_{\rm AGN}$}}

\newcommand\Inu          {{$I_{\nu}$}}

\newcommand\kmsMpc       {{km~s$^{-1}$\ Mpc$^{-1}$}}

\newcommand\magarc       {{mag\ arcsec$^{-2}$}}

\newcommand\mum          {{\micron}}

\newcommand\Mo           {{M$_{\odot}$}}

\newcommand\nWsqmsr      {{nW\ m$^{-2}$\ sr$^{-1}$}}

\newcommand\ProFit       {\texttt{ProFit}}
\newcommand\ProFound     {\texttt{ProFound}}
\newcommand\ProSpect     {\texttt{ProSpect}}
\newcommand\SExtractor   {\texttt{SourceExtractor}}

\newcommand\texp         {{$t_{\rm exp}$}}

\newcommand\zmed         {{$z_{\rm med}$}}

\defcitealias{Driver2016b}{D16}
\defcitealias{Windhorst2011}{W11}
\defcitealias{Carleton2022}{SKYSURF-2}
\defcitealias{Windhorst2022}{SKYSURF-1}

\def\ltsima{$\; \buildrel < \over \sim \;$}
\def\lsim{\lower.5ex\hbox{\ltsima}}
\def\gtsima{$\; \buildrel > \over \sim \;$}
\def\gsim{\lower.5ex\hbox{\gtsima}}

\newlength{\txw}
\newlength{\txh}

\begin{document}

\vspace*{-0.50cm}
\title{JWST's PEARLS: Prime Extragalactic Areas for Reionization and Lensing
Science:\\ Project Overview and First Results}

\author[0000-0001-8156-6281]{Rogier A. Windhorst}
\affiliation{School of Earth and Space Exploration, Arizona State University,
Tempe, AZ 85287-1404, USA}

\author[0000-0003-3329-1337]{Seth H. Cohen} 
\affiliation{School of Earth and Space Exploration, Arizona State University,
Tempe, AZ 85287-1404, USA}

\author[0000-0003-1268-5230]{Rolf A. Jansen} 
\affiliation{School of Earth and Space Exploration, Arizona State University,
Tempe, AZ 85287-1404, USA}

\author[0000-0002-7265-7920]{Jake Summers} 
\affiliation{School of Earth and Space Exploration, Arizona State University,
Tempe, AZ 85287-1404, USA}

\author[0000-0001-9052-9837]{Scott Tompkins} 
\affiliation{School of Earth and Space Exploration, Arizona State University,
Tempe, AZ 85287-1404, USA}

\author[0000-0003-1949-7638]{Christopher J. Conselice} 
\affiliation{Jodrell Bank Centre for Astrophysics, Alan Turing Building, 
University of Manchester, Oxford Road, Manchester M13 9PL, UK}

\author[0000-0001-9491-7327]{Simon P. Driver} 
\affiliation{International Centre for Radio Astronomy Research (ICRAR) and the
International Space Centre (ISC), The University of Western Australia, M468,
35 Stirling Highway, Crawley, WA 6009, Australia}

\author[0000-0001-7592-7714]{Haojing Yan} 
\affiliation{Department of Physics and Astronomy, University of Missouri,
Columbia, MO 65211, USA}

\author[0000-0001-7410-7669]{Dan Coe} 
\affiliation{AURA for the European Space Agency (ESA), Space Telescope Science
Institute, 3700 San Martin Drive, Baltimore, MD 21210, USA}

\author[0000-0003-1625-8009]{Brenda Frye} 
\affiliation{Steward Observatory, University of Arizona, 933 N Cherry Ave,
Tucson, AZ, 85721-0009, USA}

\author[0000-0001-9440-8872]{Norman Grogin} 
\affiliation{Space Telescope Science Institute, 3700 San Martin Drive, 
Baltimore, MD 21210, USA}

\author[0000-0002-6610-2048]{Anton Koekemoer} 
\affiliation{Space Telescope Science Institute, 3700 San Martin Drive,
Baltimore, MD 21210, USA}

\author[0000-0001-6434-7845]{Madeline A. Marshall} 
\affiliation{National Research Council of Canada, Herzberg Astronomy \&
Astrophysics Research Centre, 5071 West Saanich Road, Victoria, BC V9E 2E7, 
Canada}
\affiliation{ARC Centre of Excellence for All Sky Astrophysics in 3 Dimensions
(ASTRO 3D), Australia}

\author[0000-0003-3351-0878]{Rosalia O'Brien} 
\affiliation{School of Earth and Space Exploration, Arizona State University,
Tempe, AZ 85287-1404, USA}

\author[0000-0003-3382-5941]{Nor Pirzkal} 
\affiliation{Space Telescope Science Institute, 3700 San Martin Drive,
Baltimore, MD 21210, USA}

\author[0000-0003-0429-3579]{Aaron Robotham} 
\affiliation{International Centre for Radio Astronomy Research (ICRAR) and the
International Space Centre (ISC), The University of Western Australia, M468,
35 Stirling Highway, Crawley, WA 6009, Australia}

\author[0000-0003-0894-1588]{Russell E. Ryan, Jr.} 
\affiliation{Space Telescope Science Institute, 3700 San Martin Drive, 
Baltimore, MD 21210, USA}

\author[0000-0001-9262-9997]{Christopher N. A. Willmer} 
\affiliation{Steward Observatory, University of Arizona, 933 N Cherry Ave,
Tucson, AZ, 85721-0009, USA}

\author[0000-0001-6650-2853]{Timothy Carleton} 
\affiliation{School of Earth and Space Exploration, Arizona State University,
Tempe, AZ 85287-1404, USA}

\author[0000-0001-9065-3926]{Jose M. Diego} 
\affiliation{Instituto de F\'isica de Cantabria (CSIC-UC). Avenida. Los Castros
s/n. 39005 Santander, Spain}

\author[0000-0002-6131-9539]{William C. Keel} 
\affiliation{Dept. of Physics and Astronomy, University of Alabama, Box 870324,
Tuscaloosa, AL 35404, USA}

\author[0000-0002-6078-0841]{Paolo Porto} 
\affiliation{School of Earth and Space Exploration, Arizona State University,
Tempe, AZ 85287-1404, USA}

\author[0000-0002-9961-2984]{Caleb Redshaw} 
\affiliation{School of Earth and Space Exploration, Arizona State University,
Tempe, AZ 85287-1404, USA}

\author[0000-0001-9497-7338]{Sydney Scheller} 
\affiliation{Yale University, New Haven, CT 06511, USA}

\author[0000-0003-3903-6935]{Stephen M. Wilkins} 
\affiliation{Astronomy Centre, Department of Physics and Astronomy, University
of Sussex, Brighton, BN1 9QH, UK}

\author[0000-0002-9895-5758]{S. P. Willner} 
\affiliation{Center for Astrophysics \textbar\ Harvard \& Smithsonian, 60 
Garden St., Cambridge, MA 02138, USA}

\author[0000-0002-0350-4488]{Adi Zitrin} 
\affiliation{Physics Department, Ben-Gurion University of the Negev, P.O. Box
653, Beer-Sheva 8410501, Israel}

\author[0000-0003-4875-6272]{Nathan J. Adams} 
\affiliation{Jodrell Bank Centre for Astrophysics, Alan Turing Building, 
University of Manchester, Oxford Road, Manchester M13 9PL, UK}

\author[0000-0003-0519-9445]{Duncan Austin} 
\affiliation{Jodrell Bank Centre for Astrophysics, Alan Turing Building, 
University of Manchester, Oxford Road, Manchester M13 9PL, UK}

\author[0000-0001-8403-8548]{Richard G. Arendt} 
\affiliation{UMBC/CRESST2, NASA Goddard Space Flight Center, Greenbelt, MD
21771, USA}

\author[0000-0002-0005-2631]{John F. Beacom} 
\affiliation{Center for Cosmology and AstroParticle Physics (CCAPP), Department
of Physics, Ohio State University, 191 W. Woodruff Ave., Columbus, OH 43210,
USA}

\author[0000-0003-0883-2226]{Rachana A. Bhatawdekar} 
\affiliation{European Space Agency, ESA/ESTEC, Keplerlaan 1, 2201 AZ Noordwijk,
The Netherlands}

\author[0000-0002-7908-9284]{Larry D. Bradley} 
\affiliation{Space Telescope Science Institute, 3700 San Martin Drive,
Baltimore, MD 21210, USA}

\author[0000-0002-5807-4411]{Tom~Broadhurst} 
\affiliation{Department of Theoretical Physics, University of the Basque
Country UPV-EHU, 48040 Bilbao, Spain}
\affiliation{Donostia International Physics Center (DIPC), 20018 Donostia, The
Basque Country}
\affiliation{IKERBASQUE, Basque Foundation for Science, Alameda Urquijo, 36-5
48008 Bilbao, Spain}

\author[0000-0003-0202-0534]{Cheng Cheng} 
\affiliation{Chinese Academy of Sciences, National Astronomical Observatories,
CAS, Beijing 100101, China}

\author[0000-0002-2115-1137]{Francesca Civano} 
\affiliation{Center for Astrophysics \textbar\ Harvard \& Smithsonian, 60 
Garden St., Cambridge, MA 02138, USA}

\author[0000-0003-2091-8946]{Liang Dai} 
\affiliation{Department of Physics, 366 Physics North MC 7300, University of
California, Berkeley, CA 94720, USA}

\author[0000-0002-9767-3839]{Herv\'e Dole} 
\affiliation{Universit\'e Paris-Saclay, CNRS, Institut d'Astrophysique
Spatiale, 91405, Orsay, France}

\author[0000-0002-9816-1931]{Jordan C. J. D'Silva} 
\affiliation{International Centre for Radio Astronomy Research (ICRAR) and the
International Space Centre (ISC), The University of Western Australia, M468,
35 Stirling Highway, Crawley, WA 6009, Australia}

\author[0000-0001-6889-8388]{Kenneth J. Duncan} 
\affiliation{Institute for Astronomy, Royal Observatory, Blackford Hill,
Edinburgh, EH9 3HJ, UK}

\author[0000-0002-0670-0708]{Giovanni G. Fazio} 
\affiliation{Center for Astrophysics \textbar\ Harvard \& Smithsonian, 60 
Garden St., Cambridge, MA 02138, USA}

\author[0000-0002-2012-4612]{Giovanni Ferrami} 
\affiliation{School of Physics, University of Melbourne, Parkville, VIC 3010, 
Australia}
\affiliation{ARC Centre of Excellence for All Sky Astrophysics in 3 Dimensions
(ASTRO 3D), Australia}

\author[0000-0002-8919-079X]{Leonardo Ferreira} 
\affiliation{University of Nottingham, School of Physics \& Astronomy,
Nottingham, NG7 2RD, UK}

\author[0000-0001-8519-1130]{Steven L. Finkelstein} 
\affiliation{Department of Astronomy, The University of Texas at Austin,
Austin, TX 78712, USA}

\author[0000-0001-6278-032X]{Lukas J. Furtak} 
\affiliation{Physics Department, Ben-Gurion University of the Negev, P.O. Box
653, Beer-Sheva 84105, Israel} 

\author[0000-0003-1436-7658]{Hansung B. Gim} 
\affiliation{Department of Physics, Montana State University, P. O. Box 173840, 
Bozeman, MT 59717, USA}

\author[0000-0003-1880-3509]{Alex Griffiths} 
\affiliation{University of Nottingham, School of Physics \& Astronomy,
Nottingham, NG7 2RD, UK}

\author[0000-0001-8751-3463]{Heidi B. Hammel} 
\affiliation{Associated Universities for Research in Astronomy, Inc., 1331 
Pennsylvania Avenue NW, Suite 1475, Washington, DC 20004, USA}

\author[0000-0001-5429-5762]{Kevin C. Harrington} 
\affiliation{European Southern Observatory, Alonso de C{\'o}rdova 3107,
Vitacura, Casilla 19001, Santiago de Chile, Chile}

\author[0000-0001-6145-5090]{Nimish P. Hathi} 
\affiliation{Space Telescope Science Institute, 3700 San Martin Drive,
Baltimore, MD 21210, USA}

\author[0000-0002-4884-6756]{Benne W. Holwerda} 
\affiliation{Department of Physics and Astronomy, University of Louisville,
Louisville KY 40292, USA} 

\author[0000-0002-9984-4937]{Rachel Honor} 
\affiliation{School of Earth and Space Exploration, Arizona State University,
Tempe, AZ 85287-1404, USA}

\author[0000-0001-6511-8745]{Jia-Sheng Huang} 
\affiliation{National Astronomical Observatories of China, A20 Datun Road, 
Beijing, China} 

\author[0000-0003-4738-4251]{Minhee Hyun} 
\affiliation{SNU Astronomy Research Center, Astronomy program, Dept. of Physics
\& Astronomy, Seoul National University, 1 Gwanak-ro, Gwanak-gu, Seoul 08826,
Republic of Korea}
\affiliation{Korea Astronomy and Space Science Institute, 776 Daedeok-daero,
Yuseong-gu, Daejeon 34055, Republic of Korea}

\author[0000-0002-8537-6714]{Myungshin Im} 
\affiliation{SNU Astronomy Research Center, Astronomy program, Dept. of Physics
\& Astronomy, Seoul National University, 1 Gwanak-ro, Gwanak-gu, Seoul 08826,
Republic of Korea}

\author[0000-0002-7593-8584]{Bhavin A. Joshi} 
\affiliation{Center for Astrophysical Sciences, Department of Physics and
Astronomy, The Johns Hopkins University, 3400 N Charles St., Baltimore, MD
21218, USA}

\author[0000-0001-9394-6732]{Patrick S. Kamieneski} 
\affiliation{Department of Astronomy, University of Massachusetts, Amherst, MA
01003, USA}

\author[0000-0003-3142-997X]{Patrick Kelly} 
\affiliation{School of Physics and Astronomy, University of Minnesota, 116
Church Street SE, Minneapolis, MN 55455, USA} 

\author[0000-0003-2366-8858]{Rebecca L. Larson} 
\affiliation{Department of Astronomy, The University of Texas at Austin,
Austin, TX 78712, USA}

\author[0000-0002-8184-5229]{Juno Li} 
\affiliation{International Centre for Radio Astronomy Research (ICRAR) and the
International Space Centre (ISC), The University of Western Australia, M468,
35 Stirling Highway, Crawley, WA 6009, Australia}

\author[0000-0003-4220-2404]{Jeremy Lim} 
\affiliation{Department of Physics, The University of Hong Kong, Pokfulam Road, 
Hong Kong}

\author[0000-0003-3270-6844]{Zhiyuan Ma} 
\affiliation{Department of Astronomy, University of Massachusetts, Amherst, MA
01003, USA}

\author[0000-0002-2203-7889]{Peter Maksym} 
\affiliation{Center for Astrophysics \textbar\ Harvard \& Smithsonian, 60 
Garden St., Cambridge, MA 02138, USA}

\author[0000-0001-8220-2324]{Giorgio Manzoni} 
\affiliation{Jockey Club Institute for Advanced Study, The Hong Kong University 
of Science and Technology, Hong Kong S.A.R., China}

\author[0000-0002-7876-4321]{Ashish Kumar Meena} 
\affiliation{Physics Department, Ben-Gurion University of the Negev, P.O. Box
653, Beer-Sheva 8410501, Israel}

\author[0000-0001-7694-4129]{Stefanie N. Milam} 
\affiliation{Astrochemistry Laboratory, NASA Goddard Space Flight Center, 
Code 691, Greenbelt, MD 20771, USA}

\author[0000-0001-6342-9662]{Mario Nonino} 
\affiliation{INAF-Osservatorio Astronomico di Trieste, Via Bazzoni 2, 34124
Trieste, Italy}

\author[0000-0002-2282-8795]{Massimo Pascale} 
\affiliation{Department of Astronomy, University of California, 501 Campbell
Hall \#3411, Berkeley, CA 94720, USA}

\author[0000-0003-4030-3455]{Andreea Petric} 
\affiliation{Space Telescope Science Institute, 3700 San Martin Drive,
Baltimore, MD 21210, USA}

\author[0000-0002-2361-7201]{Justin D. R. Pierel} 
\affiliation{Space Telescope Science Institute, 3700 San Martin Drive,
Baltimore, MD 21210, USA}

\author[0000-0001-7411-5386]{Maria del Carmen Polletta} 
\affiliation{INAF, Istituto di Astrofisica Spaziale e Fisica Cosmica (IASF)
Milano, Via A. Corti 12, 20133 Milan, Italy}

\author[0000-0001-8887-2257]{Huub J. A. R\"ottgering} 
\affiliation{Leiden Observatory, PO Box 9513, 2300 RA Leiden, The Netherlands}

\author[0000-0001-7016-5520]{Michael J. Rutkowski} 
\affiliation{Minnesota State University-Mankato, Trafton North Science Center, 
Mankato, MN, 56001, USA}

\author[0000-0003-3037-257X]{Ian Smail} 
\affiliation{Centre for Extragalactic Astronomy, Department of Physics, Durham
University, South Road, Durham DH1 3LE, UK}

\author[0000-0002-4772-7878]{Amber N. Straughn} 
\affiliation{Astrophysics Science Division, NASA Goddard Space Flight Center,
8800 Greenbelt Rd, Greenbelt, MD 20771, USA} 

\author[0000-0002-7756-4440]{Louis-Gregory Strolger} 
\affiliation{Space Telescope Science Institute, 3700 San Martin Drive, 
Baltimore, MD 21210, USA}

\author[0000-0003-1778-7711]{Andi Swirbul} 
\affiliation{School of Earth and Space Exploration, Arizona State University,
Tempe, AZ 85287-1404, USA}

\author[0000-0002-9081-2111]{James A. A. Trussler} 
\affiliation{Jodrell Bank Centre for Astrophysics, Alan Turing Building, 
University of Manchester, Oxford Road, Manchester M13 9PL, UK}

\author[0000-0001-7092-9374]{Lifan Wang} 
\affiliation{Texas A\&M University/Physics and Astronomy, College Station,
TX 77842-4242, USA}

\author[0000-0003-1815-0114]{Brian Welch} 
\affiliation{Center for Astrophysical Sciences, Department of Physics and
Astronomy, The Johns Hopkins University, 3400 N Charles St., Baltimore, MD
21218, USA}

\author[0000-0001-7956-9758]{J. Stuart B. Wyithe} 
\affiliation{School of Physics, University of Melbourne, Parkville, VIC 3010, 
Australia}
\affiliation{ARC Centre of Excellence for All Sky Astrophysics in 3 Dimensions
(ASTRO 3D), Australia}

\author[0000-0001-7095-7543]{Min Yun} 
\affiliation{Department of Astronomy, University of Massachusetts, Amherst, MA
01003, USA}

\author[0000-0003-1096-2636]{Erik Zackrisson} 
\affiliation{Observational Astrophysics, Department of Physics and Astronomy,
Uppsala University, Box 516, SE-751 20 Uppsala, Sweden}

\author[0000-0002-3783-4629]{Jiashuo Zhang} 
\affiliation{Department of Physics, The Hong Kong University of Science and
Technology, Clear Water Bay, Kowloon, Hong Kong}

\author[0000-0002-7791-3671]{Xiurui Zhao} 
\affiliation{Center for Astrophysics \textbar\ Harvard \& Smithsonian, 60 
Garden St., Cambridge, MA 02138, USA}

\email{Rogier.Windhorst@asu.edu}


\begin{abstract} 
We give an overview and describe the rationale, methods, and first results from
NIRCam images of the JWST ``Prime Extragalactic Areas for Reionization and
Lensing Science'' (``PEARLS'') project. PEARLS uses up to eight NIRCam filters
to survey several prime extragalactic survey areas: two fields at the North
Ecliptic Pole (NEP); seven gravitationally lensing clusters; two high redshift
proto-clusters; and the iconic backlit VV~191 galaxy system to map its dust
attenuation. PEARLS also includes NIRISS spectra for one of the NEP fields and
NIRSpec spectra of two high-redshift quasars. The main goal of PEARLS is to
study the epoch of galaxy assembly, AGN growth, and First Light. Five fields,
the JWST NEP Time-Domain Field (TDF), IRAC Dark Field (IDF), and three lensing
clusters, will be observed in up to four epochs over a year. The cadence and
sensitivity of the imaging data are ideally suited to find faint variable
objects such as weak AGN, high-redshift supernovae, and cluster caustic
transits. Both NEP fields have sightlines through our Galaxy, providing
significant numbers of very faint brown dwarfs whose proper motions can be
studied. Observations from the first spoke in the NEP TDF are public. This
paper presents our first PEARLS observations, their NIRCam data reduction and
analysis, our first object catalogs, the 0.9--4.5~\mum\ galaxy counts and
Integrated Galaxy Light. We assess the JWST sky brightness in 13 NIRCam
filters, yielding our first constraints to diffuse light at 0.9--4.5~\mum. 
PEARLS is designed to be of lasting benefit to the community.
\end{abstract}

\bn \keywords{Instruments: James Webb Space Telescope---Solar System: 
Zodiacal Light---Stars: Galactic Star Counts---Galaxies: Galaxy Counts---
Cosmology: Extragalactic Background Light }



\n \section{Introduction} \label{sec1}

\sn The James Webb Space Telescope (``JWST'') was designed in the 1990s and
2000s to observe very faint objects at near- and mid-infrared wavelengths from
the Sun--Earth L2 Lagrange point \citep[\eg][]{Rieke2005, Gardner2006,
Beichman2012, Windhorst2008}. With its 6.5-meter aperture and state-of-the-art
scientific instruments,\footnote{\url{https://www.stsci.edu/jwst/}\ \ and\ \ 
\url{https://www.stsci.edu/jwst/instrumentation/}} JWST builds on the
scientific results from two of NASA's previous flagship missions: the Hubble
Space Telescope \citep[HST; for a review of 27 years of HST imaging data, see,
\eg][]{Windhorst2022} and the Spitzer Space Telescope \citep[\eg][]{Werner2004,
Soifer2008,Werner2022}. The NASA/ESA/CSA JWST was successfully launched on 2021
December 25 on an Ariane~V launch vehicle into a direct-insertion trajectory to
L2. JWST was subsequently deployed, cooled to its intended cryogenic
temperatures behind its giant sunshield,\footnote{\eg\
\url{https://webb.nasa.gov}} and its instruments were successfully commissioned
and calibrated \citep[\eg][]{Rigby2022}.\footnote{
\url{https://www.stsci.edu/contents/news/jwst/2022/learn-about-jwsts-known-scientific-performance.html}}
In its 96-minute low earth orbit (LEO), the Hubble Space Telescope (HST) has
experienced over 175,000 sunrises and sunsets since its launch on 1990 April
24. This, for instance, leads to HST's ``orbital breathing'' and time-dependent
point-spread functions \citep[PSFs; \eg][]{Mechtley2012, Mechtley2016,
Marshall2020, Marshall2021}, as well as its significant orbital-phase-dependent
sky surface-brightness (sky-SB) levels \citep[\eg][]{Carleton2022,
Windhorst2022}. In contrast, JWST was designed to have exactly one sunrise and
one sunset during its planned 10$^+$-year mission: its one and only sunrise
occurred when the Ariane launch fairing opened on Christmas Day 2021, and its
one-and-only sunset came when its sunshield fully deployed in early January
2022. Compared to HST, JWST will have more stable PSFs and foreground sky-SB
levels, which depend primarily on its component temperatures and its pointing
direction (pitch angle), respectively. The resulting very dark and stable L2
environment makes JWST particularly suited for faint-object detection in the
observatory's 0.6--29~\mum\ wavelength range, as well as assessing its sky-SB
which PEARLS will pursue at 0.9--4.5~\mum. 

From the start of observatory design in the early 2000s, JWST had four main
science themes that drove its performance requirements: First Light \&
Reionization, Assembly of Galaxies, Birth of Stars and Protoplanetary Systems,
and Planetary Systems and Origins of Life \citep[\eg][]{Gardner2006}. Now
almost twenty years later, these remain key research areas with major
unknowns, and these themes are reflected in the Cycle~1 proposals from the
astronomical community and in the observing time granted.\footnote{\eg\ 
\url{https://www.stsci.edu/jwst/science-execution/approved-programs/}} As part
of the science planning of JWST, R.\ Windhorst was chosen as a JWST
Interdisciplinary Scientist in 2002 June. His 20$^+$-year effort and commitment
comes with 110 hours of Guaranteed Time Observations (GTO)\null. This paper 
gives an overview and describes the rationale, methods, and first scientific
results of our project ``Prime Extragalactic Areas for Reionization and Lensing
Science'', or ``PEARLS.'' 

PEARLS' main science goals address JWST's first two themes: First Light \&
Reionization and Assembly of Galaxies, including supermassive black-hole
(SMBH) growth. Specifically, PEARLS will observe three ``blank'' fields, seven
galaxy clusters that show strong gravitational lensing, two high-redshift
proto-clusters, two high-redshift quasars, and one nearby spiral galaxy backlit
by a neighboring elliptical galaxy. Two of the blank fields are especially
suited for time-domain science \citep[\eg][]{Jansen2018, Yan2018}. These 
reside in program PID 2738 (PIs R.\ Windhorst \& H.\ Hammel). All other PEARLS
observations reside in PID 1176 (PI Windhorst). In collaboration with GTO
programs by C.\ Willott (PID 1208) and M.\ Stiavelli (PID 1199), two of the
lensing clusters (MACS0416 and MACS1149) will have a significant additional 
time baseline to search for caustic transits of stars at redshifts $z\ga 1$
\citep[\eg][]{Kelly2018, Diego2018, Chen2022} or even individual
highly-magnified stars or stellar-mass black-hole accretion disks at $z\ga 6$
\citep[\eg][]{Windhorst2018, Meena2022, Welch2022a, Welch2022c}. 

Section~\ref{sec2} of this paper describes the PEARLS rationale and target
selection along with the planning and scheduling of the JWST observations.
Section~\ref{sec3} describes the first PEARLS JWST/NIRCam data and their
initial calibration. Section~\ref{sec4} presents the NIRCam catalogs of the
first PEARLS blank-field survey including their completeness, the star--galaxy
classification procedure, and the object counts in broad-band filters covering
0.9--4.5~\mum. Section~\ref{sec5} describes the detected and extrapolated
integrated galaxy light (IGL) as derived from the 0.9--4.5~\mum\ galaxy counts,
and analyses the JWST sky-SB in 13 NIRCam filters to assess what is required to
set limits to diffuse light, including any diffuse Extragalactic Background
Light (EBL)\null. Section~\ref{sec6} discusses the significance of our early
PEARLS results, and Section~\ref{sec7} summarizes our results and future
prospects. PEARLS is designed to be of lasting benefit to the community, and we
hope that it will catalyze a variety of multi-wavelength studies during the
lifetime of JWST. 

This paper uses Planck cosmology \citep{Aghanim2020}: $H_0=67.4\pm0.5$~
\kmsMpc, matter density parameter $\Omega_{M}=0.315\pm0.007$ and vacuum energy
density $\Omega_{\Lambda}=0.685$. These give the Universe an age of 13.8~Gyr.
To compare our NIRCam results to decades of previous work, our object fluxes are
in AB units.\footnote{Defined as AB-mag = --2.5 log($F_{\nu}$) + 8.90, where the
flux density $F_{\nu}$ is in Jy units.} \citep{OkeGunn1983}. Surface brightness
(SB) values are in units of AB \magarc\ or in MJy/sr.\footnote{All JWST pixel
values are in units of MJy/sr, which can be converted to units of \nWsqmsr\ by
multiplying by 10$^{-11}$(c/$\lambda_{c})$, where $\lambda_{c}$ is the filter
pivot wavelength in microns \citep[\eg\ Equation~A15 of][]{Bessell2012}.} 


\n \section{PEARLS Rationale} \label{sec2} 

\n \subsection{PEARLS Science Objectives } \label{sec21} 

\sn PEARLS targets for First Light and Reionization studies include
high-redshift Lyman-$\alpha$ galaxies and protoclusters. In light of
observations with HST WFC3 over the past 13 years, PEARLS will also image
several rich galaxy clusters that boost the signal of faint, high-redshift
objects via strong gravitational lensing. Blank-field surveys will contribute
to the First Light theme via number counts. To study the Assembly of Galaxies,
we will observe galaxies up to the highest redshifts, and lowest masses and
luminosities, in different environments. We will also investigate SMBH growth
by observing high-redshift galaxies having an active nucleus: quasars and radio
galaxies. The blank fields at high Ecliptic latitude will contribute
time-domain information. PEARLS will also study VV~191, a nearby, overlapping
galaxy pair, to provide a benchmark dust-attenuation profile for studying
high-redshift, dusty environments. Table~1 summarizes the PEARLS fields
observed thus far (as of 2022 July 31), and Table~2 the PEARLS fields to be
observed subsequently. The 112.3 calendar hours allocated to PEARLS include 2.3
hours from H.\ Hammel. In all, PEARLS will image 16 NIRCam and four NIRISS
fields in up to eight filters to $\rm AB \la28.5$--29 mag and will cover
$\sim$165.66~arcmin$^2$ or 0.046~deg$^2$, equivalent to $\sim$34 HUDF/XDF
fields \citep[\eg][]{Beckwith2006, Koekemoer2013}. 

\begin{figure*}[!hptb]
\n\cl{
\includegraphics[width=1.00\txw]{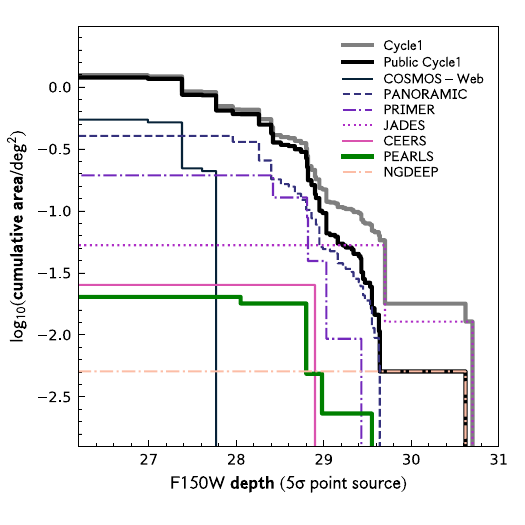}
}

\vspace*{-0.50cm}
\n \caption{
Summary of the area and depth covered by JWST Cycle~1 surveys. Data came from
G.\ Brammer's website 
{\url{https://erda.ku.dk/vgrid/Gabriel\_Brammer/JWST-Cycle1/full\_timeline.html}}. 
Colors and line types identify seven GO and GTO surveys as shown in the legend,
and the order in the legend matches the maximum survey area as shown along the
left ordinate. The thick black line shows the total area--depth of all public
Cycle 1 surveys, and the thick grey line shows the same including surveys with
proprietary data. The thick green line indicates PEARLS, which combines a
smaller area of significant depth with a much larger area of 16 shallower 
fields to significantly average over Cosmic Variance (Section~\ref{sec2} \&
\ref{sec45}), and include 7 lensing clusters. 
} 
\label{fig:fig1}
\end{figure*}

PEARLS will obtain data over at least 13 independent lines of sight more than
three degrees apart from each other and is therefore more robust against Cosmic
Variance (CV) at $\rm AB\la 28.5$~mag than programs that image only a few areas
\citep[\eg][]{Somerville2004, Driver2010, Windhorst2022}.
Figure~\ref{fig:fig1} compares the area and depth covered by PEARLS to other
JWST Cycle~1 surveys. While not as deep or wide as other {\it contiguous} JWST
Cycle 1 surveys, PEARLS covers more fields across the sky to decrease the
effects of CV\null. The expected CV for PEARLS fields can be found with the
calculator\footnote{\url{https://cosmocalc.icrar.org/}} of \citet{Driver2010}
based on the areas covered and sensitivity limits Tables~1 and ~2. To $\rm
AB\la 28.5$~mag, the PEARLS fields sample a typical redshift range of
$z\simeq0.3$--8 with a median redshift of $z\simeq1$--2 (see
Section~\ref{sec45} and Appendix~\ref{secAppB2}). The NIRCam FOV covers
$\sim$0.0026 \degsq\ (Section~\ref{sec31} or $\sim$1.1$\times$2.2~Mpc), over
which its CV is then predicted to be $\sim$30\%. For the two PEARLS fields with
galaxy counts presented here in Section~\ref{sec45}, CV is expected to be \cle
9\%. At the end of JWST Cycle 1, large JWST NIRCam parallel programs like
PANORAMIC (PID 2514; C. Williams PI) may push CV of the sampled objects to
$\sim$1--2\%. 

In four of our NIRCam pointings, coordinated NIRISS grism and imaging parallels
will cover a significant portion of our NIRCam images (Table~2), while
UV-optical images are available from HST WFC3+ACS. The coordinated NIRISS
parallels will be used for both object characterization and redshifts, and to
expand the area and time-baseline of time-domain studies. The coordinated
parallel observations are critical to obtain imaging and grism data that is as
homogeneous as possible, over as large an area as possible, and in the least
amount of time feasible with JWST. 

Two of the PEARLS blank fields and two galaxy cluster fields will be observed
more than once. This time-domain component will allow us to find and study
Galactic brown dwarfs via high proper motion or atmospheric variability,
variable AGN, high-redshift supernovae, and any time-varying objects seen
behind lensing clusters, including possible cluster caustic transits. The
PEARLS time-domain data at the North Ecliptic Pole (NEP) may also reveal some
faint moving objects at high Ecliptic latitude in our outer solar system.

To encourage immediate use of JWST data by the community and follow-up
proposals by JWST Cycle~2 GO proposers, we will make the first epoch of our
JWST NEP Time-Domain Field (TDF) public immediately (\#112.* in Table~2). The
other three JWST NEP TDF epochs will be released together with the v1 data
products as soon as we have these. Also public right away are the Cycle 25
(R. Jansen \etal\ 2022, in preparation), 28, and 29 HST WFC3/UVIS F275W and
ACS/WFC F435W+F606W observations of the NEP TDF and other ancillary data across
the electromagnetic spectrum, as these become available and their data
reduction is completed. These include 600~ks NuSTAR 3--24~keV images
\citep[][F. Civano \etal\ in preparation]{Zhao2021}, 900~ks of Chandra ACIS
0.2--10~keV images (P. Maksym \etal\ 2022, in preparation), 31~hours of
JCMT/SCUBA-2 plus 66~hours of SMA data at 0.85~mm, as well as 70~hours of VLA
3~GHz A+B-array images (M. Hyun \etal\ 2022, in preparation), 147~hrs of VLBA
4.5~GHz data at milliarcsec (mas) resolution to sub-microJy levels, and
75~hours of LOFAR 150~MHz images including LOFAR VLBI. The presence of a
$S_{3GHz}\simeq$239~mJy quasar at $z=1.4429$ in the JWST NEP TDF that is
unresolved at VLBI m.a.s. resolution is used as phase calibrator to provide
high resolution VLA/VLBA and LOFAR/VLBI images of very high dynamic range in the
NEP TDF. The NEP TDF database also includes multi-epoch Large Binocular
Telescope/LBC + Subaru/HSC $Ugiz$ images to AB\cle 26.0 mag, Gran Telescopio 
Canarias/HiPERCAM $ugriz$ images to $\rm AB\la 27$ mag, Multiple Mirror
Telescope/MMIRS images to $YJHK\la 24$--23 mag, and MMT/Binospec and MMIRS
spectra to 22--24~mag (C. Willmer \etal\ 2022, in preparation), plus JPAS
56-narrow-band spectrophotometry to provide confirmation of the astrometric,
photometric, and spectroscopic calibration of our JWST NIRCam+NIRISS
observations. 


\n \subsection{PEARLS Target Selection} \label{sec22} 

\sn PEARLS target selection began in the early 2010's, when it became clear
that JWST had a viable path towards launch and that it could perform as
designed. The largest blank field is in the JWST continuous viewing zone (CVZ)
near the North Ecliptic Pole (NEP). This NEP Time Domain Field has the best
combination of low foreground extinction and absence of AB\cle 16 mag stars
\citep{Jansen2018}. A second blank field is within the IRAC Dark Field, which
is a Spitzer/IRAC calibration field near the NEP observed repeatedly for over
15 years. These historical light curves offer several examples of what might be
high-redshift, dusty supernovae (SNe) in ultra luminous infrared galaxies
selected by Herschel \citep{Yan2018}. Figure~1a \& 2a of \citet{Jansen2018}
give a layout of the JWST CVZ in the NEP, where the IRAC Dark Field (IDF) is
$\sim$3.56\degree\ NE of the TDF. Our Figure~\ref{fig:fig2} shows the
first-epoch NIRCam observation of the JWST IDF (hereafter the JWIDF)\null. The
final blank field is in the WFC3 ERS area \citep{Windhorst2011}, which is in the
northern part of the GOODS-South area.

PEARLS gravitational-lensing clusters were selected to have high mass and
central compactness or to have apparent double-cluster nature. The latter
could result in higher transverse motions and therefore makes caustic transits
more likely. Possible transiting sources include distant, luminous single
stars, double stars, and possibly stellar-mass black-hole accretion disks 
\citep[\eg][]{MiraldaEscude1991}. All of our selected clusters show 
gravitationally lensed arcs, and all have lensing magnification maps produced
with multiple independent lensing models, which will be refined with the JWST
data. Other lensing clusters were similarly selected because of their high mass
and high central compactness, and their lower IntraCluster Light (ICL)
content, which could make it easier to detect caustic transits with less
microlensing by foreground stars in the cluster ICL
\citep[\eg][]{Windhorst2018}. The PEARLS lensing clusters are:

\begin{itemize}

\item The HFF \citep{Lotz2017} cluster MACS J0416.1$-$2403 at $z\simeq0.397$.
This field will be covered by three JWST epochs about six months apart to
maximize the chance of seeing caustic transits at $z\ge 6$
\citep[\eg][]{Windhorst2018, Welch2022a, Welch2022c}. A number of plausible
caustic transits at $z\simeq0.9$--1.5 have already been observed for this
cluster by HST \citep{Dai2018, Kelly2018, Dai2021}. 

\item Abell 2744 at $z\simeq0.31$ and MACS J1149.5+2223 at $z\simeq0.54$. These
are likewise HFF \citep{Lotz2017} clusters. They will have additional GTO
observations by C.\ Willott and the NIRISS GTO team, and by M.\ Stiavelli and
his team to look for variable objects in or behind these clusters, and by the
GLASS team (PID 1324; PI: Treu). This allows us to monitor potential
high-redshift caustic transits on timescales longer than a year.

\item The cluster known as El Gordo at $z\simeq0.87$ \citep{Menanteau2012,
Zitrin2013, Cerny2018, Diego2020, Caputi2021}. This cluster was selected
because of its enormous mass
\citep[][$\sim$2$\times$10$^{15}$\Mo]{Menanteau2012}, its elongation to
maximize the probability of caustic transits \citep{Windhorst2018}, and its
rich collection of distant lensed source candidates. The field includes a
background galaxy grouping at z$\simeq$4.3 \citep{Caputi2021} that is lensed by
the cluster. Figure~\ref{fig:fig3} shows the module around El Gordo that did not
cover the central part of the cluster (hereafter referred to as the
``non-cluster'' module; see Section~\ref{sec45}). 

\item PLCK G165.7+67.0 (G165) is a double cluster at $z\simeq0.35$ selected by
its FIR colors, and not by the Sunyaev-Zeldovich effect
\citep[\eg][]{Canameras2015, Harrington2016, PlanckColl2016}. Many
caustic-crossing arcs are detected, which are well-suited to transient
science. One example is a strongly-lensed red and dusty Sub-Millimeter Galaxy
(SMG) detected in the HST imaging, whose counter image appears in the
LBT/LUCI+ARGOS laser-guided AO K-band images \citep[\eg][]{Frye2019,
Rabien2019}. Spectral Energy Distributions (SEDs) fit to the optical--near-IR
images yield photometric redshift estimates for some image families, which
constrain the lens model \citep{Pascale2022}. The model confirms the bi-modal
mass distribution of this ongoing merger that is only a low-luminosity X-ray
source. The JWST observations aim to constrain the dynamical state of this
cluster and detect a significant number of lensed background sources.

\item The Clio cluster at $z\simeq0.42$ from the GAMA survey \citep{Driver2010, 
Alpaslan2017} is a massive, compact cluster selected to have significant 
potential for lensing background sources. A ground-based VLT image
\citep{Griffiths2018} already showed strongly lensed arcs and a
lower-than-average amount of IntraCluster Light. This is attractive because
low-mass IntraCluster Medium (ICM) stars can significantly lengthen
caustic-transit times \citep[\eg][]{Diego2018, Windhorst2018} and complicate
their lensing analysis.

\item The cluster RXC J1212+27=A1489 at $z\simeq0.35$. This cluster was chosen
because of strong gravitational lensing, using the automated implementation
\citep{Zitrin2012} of the Light Traces Mass (LTM) method
\citep[\eg][]{Zitrin2016, Zitrin2017}. HST images showed a significant number
of lensed sources that resulted in a good lensing model \citep{Zitrin2020}. 

\end{itemize}

The first public JWST images \citep{Pontoppidan2022} released starting 2022 
July 12 have already inspired a number of further studies. Relevant for PEARLS,
a possible caustic transit candidate has been suggested at z$\simeq$3 in some of
the public JWST images of the cluster SMACS0723 \citep[\eg][]{Chen2022}, for
which mass models were made by \citet[\eg][]{Pascale2022}, and in which also a
significant number of red spirals were identified \citep[\eg][]{Ferreira2022,
Fudamoto2022}. Indeed, some very high redshift candidates were already
suggested in some of the very first JWST ERS images \citep[][]{Adams2022,
Finkelstein2022}. The PEARLS high redshift protoclusters are:

\begin{itemize}

\item PHz G191.24+62.04 (G191) is a protocluster candidate at $z=2.55$ with one
of the highest star formation rates (SFR$\simeq$23000 \Mo/yr) in the parent
sample of Planck high-z sources \citep[PHz;][]{PlanckColl2016}. G191 hosts an
overdensity of red Spitzer sources \citep{PlanckColl2015, Martinache2018},
containing $\sim$14 objects/arcmin$^2$ with IRAC 3.6--4.5~\mum\ colors \cge
--0.1 mag. Two of the Herschel sources have spectroscopic redshifts and a large
estimated SFRs$\simeq$1000--1500 \Mo/yr \citep{Polletta2022}, \ie\ high enough
that they present challenges for theoretical models \citep{Granato2015,
Lim2021, Gouin2022}. The JWST observations will constrain the stellar mass
assembly and fueling mechanism (\eg\ major mergers, cold accretion) occurring
in this highly star-forming high-z structure.

\item TNJ1338$-$1942 is a protocluster at $z=4.1$ that was discovered with the
VLT as 60 Ly$\alpha$-emitters near a luminous, steep-spectrum radio source
\citep[\eg][]{DeBreuck1999, DeBreuck2000, Venemans2002, Miley2004, Intema2006, 
Saito2015}. The radio source's AGN activity and outflow will be studied by a
JWST Cycle 1 GO program (PID 1964, PI R.\ Overzier). PEARLS has imaged the
field in the five NIRCam medium-band filters that best straddle the
Balmer/4000\AA\ break at $z=4.1$ to help delineate the ages of $\sim$30 of the
Ly$\alpha$-emitters. Figure~\ref{fig:fig4} shows part of the NIRCam image
around TNJ1338$-$1942. To maximize the scientific return on TNJ1338$-$1942, 
the analysis of the PEARLS and GO data will be coordinated. 

\end{itemize}

In addition to our above two protocluster targets, the $z\simeq4.3$ group of 
galaxies behind El Gordo \citep{Caputi2021} may also turn out to be a
protocluster candidate. Additional PEARLS targets are:

\begin{itemize}

\item Two QSOs, QSO 1425+3254 \citep[or NDWFS J142516.3+325409 at $z=5.85$;
\eg][]{Mechtley2012}, and QSO J0005$-$0006 (or SDSS J000552.35$-$000655.6 at
$z=5.86$). The first has a number of possible $z\simeq6$ companions
\citep[\eg][]{Marshall2020, Marshall2021}. PEARLS IFU observations will address
whether these form a group around the QSO\null. QSO J0005$-$0006 was selected
because it lacks both hot and cold dust \citep{Wang2008, Jiang2010}. It
therefore represents a rare sub-population of dust-free high-$z$ quasars. 

\item The VV 191 system (Figure~\ref{fig:fig5}) consists of a foreground spiral
galaxy with an unassociated elliptical galaxy behind it
\citep[\eg][]{Keel2013}. Light from the elliptical suffers extinction from dust
in the spiral. PEARLS NIRCam imaging maps the extinction and determine its
wavelength dependence \citep{Keel2022}. 

\end{itemize}


\n \subsection{PEARLS' Observation Planning} \label{sec23} 

\n \subsubsection{JWST Observation Planning of PEARLS Targets} \label{sec231} 

\sn Most PEARLS targets will be imaged with NIRCam in a set of eight broad-band
filters, as shown in Table~1. In a few fields, fewer filters are needed to
accomplish the intended science. The NIRISS Grism mode and NIRSpec Prism mode
are used in a few fields. One field (TNJ1338$-$1942) will be observed in five
NIRCam medium-band filters and one broad-band filter, as summarized in Table~2. 

Four PEARLS fields have a time-domain component on time-scales of hours to a
year. This could reveal objects with high parallax in our solar system,
Galactic brown dwarfs with high proper motion and/or atmospheric variability,
variable AGN, high redshift supernovae, and caustic transits behind galaxy
clusters. The two PEARLS fields at high ecliptic latitude and with multiple
visits also enable searches for solar-system objects in high-inclination
orbits. To increase the search effectiveness, H.\ Hammel allocated a portion of
her GTO time to the NEP observations. The combined observations make up PID
2738, which is an efficient combination of three epochs of observation in the
JWIDF and four epochs in the TDF\null. Where possible, visits with similar
orientations were combined to save JWST overhead time. The PEARLS programs
1176+2738 require $62.0+53.7$ spacecraft hours and give 68.9 hours of net
exposure time (Tables~1 and~2). The observing efficiency is therefore
$\sim$59.5\%. This is less than the maximum JWST spacecraft efficiency of
$\sim$70\% achievable for very long integrations on deep fields, but it is in
line with the efficiency of JWST observations of average duration. Accepting
somewhat lower efficiencies was a deliberate choice to address CV. 

The PEARLS time-domain fields are:

\begin{enumerate}

\item The TDF\null. The field layout is four ``spokes'' with orientations 
differing by 90\degree. This is accomplished by observing at three-month
intervals. Each spoke is a 2$\times$1 mosaic of pointings with 57\% overlap to
fill the NIRCam inter-chip gaps of each module. At each pointing in the
mosaic, four dithers fill in the gaps in the NIRCam SW detector module. All
eight broad-band filters are used. The TDF observations include coordinated
parallel observations with NIRISS/WFSS in the orthogonal low-resolution grisms
GR150C and GR150R\null. A broad-band filter must be used simultaneously to 
define the sampled spectral wavelength range and so limit spectral overlap. The
F200W.\footnote{Note all JWST NIR filter and grism names are numbered in units
of 10 nm, \ie\ GR150C or F200W indicate an effective wavelength of 1.5 or
2.0~\mum, respectively.} broad-band filter was used for this purpose to explore
a new wavelength range not sampled by the HST WFC/IR G102 or G140 grisms. The
field dimensions were chosen to make the NIRISS footprints maximally overlap
each NIRCam mosaic that was taken $\sim$183 days earlier or later, \ie\
180\degree\ different position angle \citep[Figure~7b of][]{Jansen2018}. The
grism spectra will allow object characterization and yield redshifts, and the
direct NIRISS F200W images --- needed to identify which grism spectrum is which
object --- will give an additional 2.0~\mum\ epoch image for time-domain
studies. In order to match the number of primary and parallel exposures, the
exposure time in several of the NIRCam filters is split over two observing
sequences. 

\item The JWIDF. The field covers a single rectangle of
$\sim$5\farcm9$\times$2\farcm4. It will be observed in three epochs six months
apart, giving position angles that differ by 180\degree, \ie\ covering the same
area at each epoch. A 6-point, full-box dither pattern fills both intra-module
and intra-chip gaps. Four broad-band filters (SW: F150W and F200W; LW: F356W
and F444W) are observed. Our hope is that many future epochs will be observed
in GO time to provide long-duration monitoring, including dusty high redshift 
supernovae in Herschel selected galaxies.

\item MACS 0416$-$24, MACS J1149.5+2223, and Abell 2744 (see also
Section~\ref{sec22}). We will observe MACS0416 in three different epochs to
search for caustic transits. Time intervals between epochs on the JWST Long
Range Plan (LRP) are scheduled $\sim$3 and $\sim$12 months after the first
epoch, as listed in Table~1. 

\end{enumerate}

JWST scheduling is a complex, ongoing and constantly changing process.
Information about when PEARLS observations have been, or will be, carried out
is available on
\url{https://www.stsci.edu/cgi-bin/get-visit-status?id=1176&markupFormat=html&observatory=JWST}
for most PEARLS targets, and on
\url{https://www.stsci.edu/cgi-bin/get-visit-status?id=2738&markupFormat=html&observatory=JWST}
for the JWIDF and TDF. Full details of observations are in the JWST ``APT
files'' also available at these websites.\footnote{For the APT tool, see 
\url{https://www.stsci.edu/scientific-community/software/astronomers-proposal-tool-apt}}


\n \subsubsection{PEARLS' Primary NIRCam Observations and Areas}
\label{sec232} 

\sn NIRCam consists of two modules A and B, each imaging two sky regions
2\farcm15$\times$2\farcm15 in size separated by $\sim$0\farcm7.\footnote{
\url{https://jwst-docs.stsci.edu/jwst-near-infrared-camera/nircam-instrumentation/nircam-filters}}
In each module, dichroics direct short-wavelength (SW, 0.6--2.3~\mum) and
long-wavelength (LW, 2.3--5.2~\mum) light from each sky region onto
corresponding detectors. The LW modules have a single detector with
2040$\times$2040 illuminated pixels at a scale of 0\farcs0629 per pixel. Each
SW module has four detectors to cover the same sky area at 0\farcs0312 per
pixel. There are 4\farcs5 gaps between the four SW detectors in each module.
With 4-step dithering across the 4\farcs5 gaps, the total area covered by each
exposure is about 2\farcm2$\times$2\farcm2 in each of the two modules or about
9.6~arcmin$^2$ for a full all-detector exposure.

Our PEARLS NIRCam imaging uses both modules (A and B) and both detector units
(SW and LW) for a total of 10 detector readouts per integration. A 4-point
INTRAMODULEBOX dither pattern is used to filter out the cosmic ray (CR) flux in
JWST's L2 orbit. The on-the-ramp readout patterns are typically either MEDIUM8
with seven groups per integration or SHALLOW4 with 8--10 groups per integration,
whichever produced the required sensitivity according to the NIRCam exposure
time calculator (ETC).\footnote{\url{https://jwst.etc.stsci.edu}} For some
shallower targets, a FULLBOX dither pattern was used with six primary dithers
(6TIGHT) and STANDARD dither-type to cover the SW inter-chip gaps in order to
make them schedulable. The resulting total net exposure times in each NIRCam
filter and their ETC sensitivities are listed in Tables~1 or~2. These
sensitivities will be verified with the object counts of
Section~\ref{sec44}--\ref{sec45}. Dither steps were made large enough to cover
the small SW intra-module gaps. Most PEARLS targets are small enough to fit in
the field of view (FOV) of a single NIRCam module. The exceptions are the NEP 
TDF and the JWIDF\null. For those targets, dithers need to cover the 43\arcsec\
inter-module gap and for the TDF also the FOV covered by the NIRISS parallels
\citep{Jansen2018}. This results in four NIRCam spokes
2\farcm15$\times$6\farcm36 with a total area of 13.67~arcmin$^2$ per spoke, and
a total area for the four TDF epochs of 54.79~arcmin$^2$ at the nominal
4-dither point depth. 

\n \subsubsection{PEARLS' Coordinated NIRISS Parallel Observations and Areas} 
\label{sec233} 

NIRISS covers a single 133\arcsec$\times$133\arcsec\ FOV with 
2040$\times$2040 light sensitive pixels\footnote{
\url{https://jwst-docs.stsci.edu/jwst-near-infrared-imager-and-slitless-spectrograph/niriss-instrumentation/niriss-detector-overview}}
at a scale of 0\farcs065 per pixel to cover an area of 4.9~arcmin$^2$. Its
wavelength coverage is 0.8--5.2~\mum. The NIRISS parallels in the TDF will
consist of 2$\times$1 mosaics with orthogonal grisms GR150C and GR150R. Each
position will include finder images in the F200W filter that are expected to
reach AB\cle 29.5 mag. Each of the dispersed NIRISS images must be bracketed by
F200W images to enable source identification and wavelength calibration, so
there are a total of four such direct images per pointing. The NIRISS F200W
images thus 6456~s total integration time, more than the non-overlapping
outskirts of the NIRCam F200W images, and therefore may reach $\sim$0.4~mag
deeper (Table~2).

The NIRISS grism exposures will use the readout pattern ``NIS'' and have
typically 13 groups per integration and two integrations per exposure. By
necessity, the NIRISS coordinated parallels have the same dither pattern as the
NIRCam primary images, and thus NIRISS covers about 2\farcm22$\times$4\farcm2
or 9.32~arcmin$^2$ in each spoke. This will give 37.35~arcmin$^2$ at the
nominal 4-dither depth for the four TDF epochs combined. Because of the larger
NIRISS pixels, the optimal NIRCam dither pattern is not optimal for NIRISS, so
the NIRCam F200W primary images will provide a better-sampled F200W image, but
the very faintest objects in F200W will be detected only by NIRISS. Net
exposure times in F200W and the two grisms and their ETC sensitivities are
listed in Table~2.

When all four NEP TDF epochs are taken as planned, the NIRISS area of each
spoke will nearly perfectly overlap with the NIRCam spoke observed $\sim$183
days earlier or later. Details are given in \citet{Jansen2018}. The resulting
NIRCam spokes with total area 54.79~arcmin$^2$ will have a depth of $\rm AB\la
28.5$ in most filters (5$\sigma$ for point sources, Table~2). NIRISS will
provide $R\sim150$ spectra ranging from 1.75 to 2.22~reaching $\rm AB\la 25.9$
mag for objects in the coverage area, which is a total of 37.35~arcmin$^2$ when
combining all four NIRISS spokes. This is the most efficient way of getting
both JWST NIRCam images and NIRISS spectra of the same area.

According to the JWST ETC, typical 5-sigma sensitivities obtained for point
sources from our shallowest ($\sim$2~hours) to our deepest ($\sim$6~hours)
mosaics are $\sim$28--28.5~mag to $\sim$28.5--29~mag per target. According to
the ETC the reddest (3--5~\mum) filters may be less sensitive than the bluer
(0.9--3~\mum) ones in both NIRCam and NIRISS\null. However, as we will see in
Section~\ref{sec4}, the wider PSF of the redder filters more than compensates
for any lower sensitivity in detection of very faint and slightly extended
galaxies. Modest variations in sensitivity occur from field-to-field,
depending on exactly how much time could be fit into the scheduled APTs for
each field within our total GTO allocations, and on the actual Zodiacal-light
brightness and the straylight contributions in each field (Section~\ref{sec5}). 


\n \section{PEARLS Calibration, Mosaicing and Data Quality } \label{sec3} 

\n \subsection{Initial Calibration} \label{sec31} 

\sn Calibration of PEARLS data obtained as of 2022 July used the calibration
files on the STSCI JWST website as of 2022 July~12. All data were processed
with the standard STScI pipeline
\texttt{CALWEBB},\footnote{\url{https://jwst-docs.stsci.edu/getting-started-with-jwst-data}}
which comes in three stages: 1) detector-level corrections to the raw
individual exposures to produce count-rate images from the non-destructive
readouts (``ramps''); 2) photometric and astrometric calibration of the
individual exposures; and 3) drizzling the calibrated and 
distortion-corrected images into mosaics. The NIRCam pipeline \texttt{CALWEBB}
was used to process all our images. The Calibration Reference Data System
(CRDS) provides the latest reference files that we used to calibrate our
data.\footnote{
\url{https://jwst-crds.stsci.edu/static/users\_guide/web\_site\_use.html}}
CRDS version 11.13.1 was used for all images. 

Our NIRCam TNJ1338, VV191, and JWIDF images taken in early July 2022 were {\ch
initially} reduced with Pipeline version v1.6.1.dev2+g408c711 and context file
jwst\_0916.pmap\_filters, which contained the pre-launch ZP values {\ch
available then}. Our more recent El Gordo images of 2022 July 29 were reduced
with Pipeline version 1.6.2 in early August 2022 using context file
jwst\_0942.pmap\_filters, which implemented \citet{Rigby2022} in-flight ZPs
affecting all NIRCam filters. We refer to these calibrations and their
resulting mosaics as version v0.5. {\ch When more accurate on-orbit NIRCam
flat-field and ZP-calibrations became available in early October 2022, we
reprocessed our PEARLS images into a version v1 with context file
jwst\_0995.pmap\_filters and Pipeline version 1.7.2. We will make v1 of the NEP
TDF available to the community as soon as its catalogs are completed and
verified (R. Jansen \etal\ 2022, in preparation). Further details of the October
2022 calibration improvements are given in Sections~\ref{sec33} \&
\ref{secAppB1}. }

Performance of the NIRCam detectors is relevant to depth, calibration quality,
and accuracy of the sky-SB values discussed in Section~\ref{sec4}--\ref{sec5}.
SW and LW module characteristics relevant for PEARLS are:\footnote{
\url{https://jwst-docs.stsci.edu/jwst-near-infrared-camera/nircam-instrumentation/nircam-detector-overview/nircam-detector-performance}\ 
see also\ \url{http://ircamera.as.arizona.edu/nircam}}

\begin{itemize}

\item Average NIRCam read-noise values are $\sim$16.2 and
$\sim$13.5~\emin~pixel$^{-1}$ using correlated double-sampling. 

\item Average dark current values in typical exposures are very low: 
$\sim$0.0019$\pm$0.002 and $\sim$0.027$\pm$0.005~\emin/s. 

\item Average detector gains are $\sim$2.05 and $\sim$1.82 \emin~ADU$^{-1}$, 
respectively. 

\item Persistence of charge from a previous equal-length exposure is
$\la$0.01\% of the original charge detected in the previous image. 

\end{itemize}

\n \subsection{Mosaicing of the PEARLS Images } \label{sec32} 

The first step in mosaicing was to anchor all individual frames into the Gaia
DR3 reference frame \citep{Gaia2022}. This step used catalogs based on recent,
deep ground-based or HST images already referenced to Gaia DR3\null. Both stars
and galaxies were used for the correction with star positions corrected for
proper motion from the epoch of the reference image to the JWST observing
epoch (Table~1). Coordinate differences between the catalog and JWST frame were
measured, and each frame's center position and position angle were adjusted to
minimize the differences. Typical adjustments were \cle 10 m.a.s., and the
final uncertainty in each frame's position is about 2--3 m.a.s. rms. We used
AstroDrizzle\footnote{
\url{https://hst-docs.stsci.edu/drizzpac/chapter-5-drizzlepac-software-package/5-2-astrodrizzle-the-new-drizzle-workhorse}}\ 
\citep{Koekemoer2013, Avila2015} to drizzle the NIRCam images as calibrated in
the \texttt{CALWEBB} pipeline Stages~1 and~2 into two mosaics for each field. 
Pixel sizes used were 0\farcs0300 and 0\farcs0600 for SW and LW, respectively.
For the LW module, we also provide 0\farcs0300/pixel mosaics to facilitate
aligned analysis. The higher resolution of the former samples the NIRCam PSFs
better in the SW channel, while the bigger pixels of the latter have better
SB-sensitivity for the generally short PEARLS exposures. All NIRCam images
were drizzled after we removed wisps as well as possible, applied a $1/f$
correction, flagged the snowballs before object detection, and subtracted a
surface-fit to the sky-SB between all the detected objects. Details on these
aspects are given below. 


\n \subsection{First Assessment of Calibration Quality } \label{sec33} 

\sn {\bf \bul NIRCam $1/f$ Noise Pattern Removal:}\ NIRCam images are read out
simultaneously in four vertical 510$\times$2040-pixel strips. Differing gains
or zero points in the four amplifiers causes banding or striping across the
images as they are read out. \citet{Rest2014} presented a mathematical method
to separate this $1/f$ noise from the random read noise\footnote{
\url{http://www.stsci.edu/files/live/sites/www/files/home/jwst/documentation/technical-documents/_documents/JWST-STScI-004118.pdf}}
and derived the time dependence of the $1/f$ noise. \citet{Rest2014} found that
the $1/f$ noise strongly correlates between the amplifiers of a given detector
because it is caused by a common reference voltage. \citet{Rest2014} also found
that the $1/f$ noise has reproducible spatial structure at the 10--20\% level
down to spatial scales of tens of pixels, and this structure does not seem to
change on timescales of months. \citet{Schlawin2020} described methods to
reduce $1/f$-noise patterns in the highly demanding NIRCam grism time-series
observations of exoplanets. As a bonus, their spatial background subtraction
also efficiently removes many random detector defects, including preamplifier
offsets, amplifier discontinuities, and even--odd column offsets. The NIRCam
$1/f$ noise can also be reduced by subtracting the values from background
pixels or reference pixels that are read closely in time. E.g.,
\citet{Schlawin2020} removed the $1/f$ noise as a step in the pipeline {\it
after} the superbias correction.\footnote{
\url{https://tshirt.readthedocs.io/en/latest/specific\_modules/ROEBA.html}}
\citet{Hilbert2016} presented a method to subtract $1/f$ noise from NIRCam
integrations {\it before} averaging the data to produce superbias maps. This
method produced superbias images with significantly lower noise levels than
images produced using the more traditional approach. C.~Willott provided a more
recent code to remove $1/f$ noise that runs on the calibrated files.\footnote{
\url{https://github.com/chriswillott/jwst.git}} M.~Bagley presented a code to
remove both the detector-level offsets in the SW modules and the $1/f$ noise
patterns. This code was produced for the CEERS project and is part of their
SDR1 release.\footnote{ \url{https://ceers.github.io/releases.html\#sdr1}}

We used both the Willott code and the \ProFound\ code \citep{Robotham2017,
Robotham2018} to remove the $1/f$ noise patterns. Together with the low-level
pedestal removal between the SW detectors below, the \ProFound-package resulted
in images that are mostly visually flat without major row-based artifacts. We
visually verified that the $1/f$ noise-removal parameter settings did not
introduce new artifacts in the final images. Further details are given in
Appendix~\ref{secAppA}, which compares the results from both $1/f$-noise
removal methods and verifies that these do not noticeably affect object 
photometry at S/N-levels \cge 5$\sigma$. 

\sn {\bf \bul SW Detector-Level Offsets:}\ The eight detectors in the SW
modules A and B have detector-level offsets that are a combination of additive
and multiplicative corrections, although most of the effect seems to be
additive. Except in very crowded fields, these detector-level offsets are
relatively easy to remove early in our data reduction workflow with the
\ProFound-based code \citep{Robotham2018}, as implemented in our previous HST
WFC3/IR work \cite[\eg][]{Windhorst2022}. In short, with \ProFound\ we created
a new {\sc fits} extension {\it SKY} that removed bad pixels and real objects
detected to $AB\la 28.5$~mag (Section~\ref{sec4}) and that interpolated the
local sky-SB plus its local rms noise underneath each object. With a number of
images in the NIRCam broad-band filters now available, we created a
low-frequency super-sky image {\it ``SKY\_SUPER''} as a clipped mean over these
{\it SKY} images. The {\it SKY} images were then subtracted from each science
{\it SCI} image using: 
\begin{equation} 
{\mathrm
(\hbox{\it SCI} - \hbox{\it SKY}) = \hbox{\it SCI} - (M \times \hbox{\it SKY\_SUPER\/} + P)\quad,
}
\label{eq:eq1} 
\end{equation}
\n where $M$ and $P$ describe the linear model and pedestal, respectively, 
of the {\it SKY} image pixels made for each detector and filter combination from
these {\it SKY\_SUPER} frames. In this process, we also determined the lowest 
object-free sky-SB in each detector following Section~4.2 of
\cite{Windhorst2022}, which is used for sky-SB estimates in our 13 PEARLS
filters in Section~\ref{sec5}. By design, our medium-deep PEARLS images come
from relatively short integrations (Tables~1--2), and we adjusted our
signal-to-noise-ratio criteria for object detection (Section~\ref{sec4}) to
achieve uniform detection above any residual $1/f$-noise patterns. Further
details of the $1/f$ and pedestal removal procedures are given in
Appendix~\ref{secAppA}.

\sn {\bf \bul NIRCam SW Detector Wisps:}\ The so-called ``wisps'' are caused by
straylight hitting a secondary mirror support bar and then being reflected
into the main light path. Only four NIRCam detectors are affected (SW A3, A4,
B3 and B4) and mainly in the F090W, F115W, F150W, F182M, 200W, and F210M
filters. Wisp positions are fixed on the telescope and each detector, and
therefore a sky-flat made in a given filter from available images is able to
subtract much of the wisp pattern locally. We used our NIRCam images to improve
available wisp templates to be subtracted from the images that show visible
wisps. We then subtracted the wisp template from each image that {\ch best}
matches the wisp amplitude. This amplitude needs to be high enough that no
significant positive wisp signal is left but not so high that a negative hole
is created in the local sky-SB. Some wisp residuals are left in some of the
images, and we masked these areas in the faint-object-detection phase
(Section~\ref{sec41}) and in the sky-SB analysis (Section~\ref{sec5}). Steady
collection of NIRCam images over time is expected to improve the wisp templates
to allow more accurate subtraction in future image reductions. 

\sn {\bf \bul NIRCam Snowballs:}\ NIRCam detectors can show large artifacts
that resemble ``snowballs'' when an energetic cosmic ray impact 
occurs.\footnote{
\url{https://jwst-docs.stsci.edu/data-artifacts-and-features/snowballs-artifact}}
The vast majority of CR hits impact only a few detector pixels, but snowballs
can affect several hundred pixels. They occur approximately at a rate of 50
snowballs per 1000~s of exposure time in each NIRCam detector. Our four-point
dithers enable the pipeline to remove many of the snowball artifacts, but this
process is not perfect. E.g., Figure~\ref{fig:fig2} shows {\ch a few} dim green
rings left over from snowballs that were not fully removed in the drizzling
process of the JWIDF F200W images. Because the drizzle weight-maps give these
pixels a lower weight, we can still derive catalogs from this image.
Nevertheless, we vet these catalogs carefully and mask out areas that are
visibly affected by residual snowballs when doing faint object counts as in
Section~\ref{sec4}. That is, any large remaining defects that generate a local excess
in the object catalogs are masked before running our final catalogs. 

\sn {\bf \bul NIRCam PSFs:}\ JWST NIRCam Point Spread Functions are detailed
on the STSCI website.\footnote{
\url{https://jwst-docs.stsci.edu/jwst-near-infrared-camera/nircam-performance/nircam-point-spread-functions}} 
We generated JWST PSFs with the \texttt{WebbPSF} tool.\footnote{
\url{https://www.stsci.edu/jwst/science-planning/proposal-planning-toolbox/psf-simulation-tool}}
Brighter star images (Section~\ref{sec41}) have PSF FWHM values consistent with
a diffraction limited telescope at 1.1~\mum\ wavelength \citep{Rigby2022}, \ie\
much better than the JWST Observatory requirement of a diffraction limit that
was designed and kept during development at 2.0~\mum\ (Section~\ref{sec44}).

\sn {\bf \bul NIRCam Flat Fields:}\ The accuracy of the NIRCam flat-fields is
better than 7\% rms (B. Sunnquist; private communication) and has improved to
$\sim$2\% with the release of the flat fields captured in
jwst\_0952.pmap\_filters {\ch and most recently jwst\_0995.pmap\_filters. We
processed our images with this latest} context file to estimate the sky-SB
values in each exposure and assess their quality in Section~{sec5}. Table~3
summarizes the parameters that characterize the 0.9--4.5~\mum\ galaxy counts
and integrated galaxy light, which are needed in Tables~4--5. (See
Section~\ref{sec5}.) Table~4 gives the predicted and observed sky-SB for the
three PEARLS targets observed in 4--8 broad-band NIRCam filters, and Table~5
gives the values for TNJ1338 and its five medium-band filters. 

\sn {\bf \bul NIRCam Zeropoints:}\ Measured zeropoints (ZPs) for all
JWST+NIRCam+Filters are on the STScI website.\footnote{
\url{https://jwst-pipeline.readthedocs.io/en/stable/jwst/photom/main.html\#imaging-and-non-ifu-spectroscopy}}
The JWST Mission Requirement is that the absolute ZPs of the imaging filters are
known to better than 5\%,\footnote{
\url{https://jwst-docs.stsci.edu/jwst-data-calibration-considerations/}} and
they are stated to be good to $\sim$4\% or better for Cycle~1 \citep{Rigby2022}.
The ``Throughput'' column in their Table~4 lists the in-flight zeropoints in
units of DN/sec/nJy. 

{\ch Results based on the initial v0.5 calibrations are recorded in our first
submission of this paper on \url{https://arxiv.org/abs/2209.04119}.
\citet{Boyer2022}\footnote{
\url{https://www.stsci.edu/contents/news/jwst/2022/an-improved-nircam-flux-calibration-is-now-available.html}}
and their cited URLs analyzed new standard star observations and updated the
NIRCam ZPs. Our v1 results below use this latest calibration, which more
accurately corrects for ZP variations between each of the 10 NIRCam detectors.
Typical ZP changes for individual detectors were \cle 10--20\%. The new NIRCam
F356W and F444W zeropoints produce photometry in the JWIDF field consistent
with the deepest Spitzer images available to within 2.6--2.9\% \citep{Yan2018}.
The v1 calibration also tightened the dispersion between the bright end of the
NIRCam galaxy counts and the faint end of the ground-based, HST, and Spitzer
galaxy counts. The uncertainty in our estimates of the Integrated Galaxy Light
in Section~\ref{sec4} went down, the rms variation between the sky-SB
measurements decreased, and our limits on diffuse light in Section~\ref{sec5}
improved. Further details of the calibration improvements are given in
Appendix~\ref{secAppB1}. }

For context,all HST ZPs were defined in units of AB-mag for a count rate of 1.000~\emin/pixel/s. This definition permitted monitoring of the ZPs' 
wavelength and time dependence over many decades \citep[\eg][]{Calamida2022,
Windhorst2011, Windhorst2022}. (For a summary, see Section~4 and references
therein of \citealt{Windhorst2022}.) However, all JWST ZPs instead have been
defined in units of MJy/sr. Conversion between the two sets of units can be
made using the footnotes of Section~\ref{sec1} but also requires knowledge of
the drizzled pixel scale in the case of JWST\null. For completeness, we
therefore list both the drizzled pixel scale and the resulting equivalent ZP in
AB-mag for our PEARLS NIRCam images in the footnotes of Table~1 and in 
Appendix~\ref{secAppB1}. 

Note that given this different JWST ZP definition, the equivalent JWST ZPs in
AB-mag are {\it no longer wavelength dependent}---unlike the case of HST---but
{\it only} depend on the image pixel scale, which therefore should always be
stated. To leave no further ambiguity, for our {\it basic drizzled pixel scale}
of 0\farcs0300/pixel, the JWST ZP for 1.000 MJy/sr converted to AB-mag would
thus be the following {\it same value for every wavelength}:
\begin{equation} 
{\mathrm{ 
ZP\ =\ 8.900-2.5 \log [10^6/((360\times 3600)/(2\times\pi\times0.0300))^2]\ =\ 
28.0865~ AB\hbox{-}mag~per~pixel.
}}
\label{eq:eq2} 
\end{equation}
\n The constant 28.0865 in Equation~\ref{eq:eq2} will be valid at all
wavelengths for all our images at 0\farcs0300/pixel {\it if} the flux
calibration is correct. {\ch With the v1 calibration, this appears to be the
case to within 3--4\% (see Appendix~\ref{secAppB1}.)}

\mn {\bf \bul NIRCam Straylight Levels:}\ JWST has an open-architecture
Optical Telescope Element (OTE), and it will have more straylight (SL) than a
closed-tube design such as HST or Spitzer. The JWST Project designed the JWST
sunshield and baffles to minimize SL with expected levels $\la$20--40\% (worst
case) of the Zodiacal SB in a given direction. {\ch Bright near-IR sources like
the Zodiacal cloud, Galactic Center and Galactic plane---the brightest NIR
sources in the sky other than the Sun, Earth, and Moon---can add SL that
scatters off dust accumulated on the primary mirror into the telescope FOV}.
Estimates of the SL levels have been made by \citet[\eg][]{Lightsey2016} and
are incorporated in extensions of the JWST ETC predictions. Tables~4--5 show
the predicted SL levels for our PEARLS targets. The predicted SL levels are
modest at 0.9--3.5~\mum, but they increase at wavelengths longer than 4.5~\mum\
due to the increased thermal foreground from the telescope and Zodiacal belt.
Further details on the adopted SL levels are given in Section~\ref{sec5} and
Appendix~\ref{secAppC}.

\si Once we have verified that the astrometry and zeropoints of the images are
robust, our v1 mosaics and catalogs will be made available via our PEARLS
websites.\footnote{\url{https://sites.google.com/view/jwstpearls}}


\n \section{NIRCam Catalogs } \label{sec4} 

\n \subsection{PEARLS NIRCam Catalog Construction} \label{sec41}

We used both the \SExtractor\ \citep{Bertin1996} and \ProFound\ 
\citep{Robotham2018} packages to generate object catalogs from our processed 
NIRCam images. Both packages were designed to deblend close objects and find the
object total fluxes. Details of these procedures are given by
\citet{Windhorst2011} and \citet{Windhorst2022}, and we applied similar
procedures to the PEARLS NIRCam images. The current paper focuses on
single-filter PEARLS object catalogs, and so we use the {\it single-filter
mode} of object detection with \SExtractor. That is, we defer the \SExtractor\
steps necessary to produce accurate object colors, such as dual image-mode
extraction, the production of band-merged catalogs, and the application of
PSF-matching and aperture corrections to future papers, which will study the
colors of faint stars and galaxies (R. Ryan \etal\ 2022, in preparation) as well 
as high-redshift dropout candidates \citep[\eg][]{Yan2022}. 

The \SExtractor\ input parameters for the NIRCam images used a minimum detection
threshold above sky of $1.5\sigma$ and {\it nine} connected pixels above this
threshold for inclusion in the catalog. We found that using fewer connected
pixels resulted in too many small spurious sources, particularly in the LW
images, where the original pixel size is larger. While the more stringent
nine-pixel requirement may result in missing a few real sources at the faint
end, we found this to be a good compromise between reliability and
completeness at all wavelengths. To detect sources, we used a 5$\times$5 pixel
convolution filter with Gaussian FWHM of 3.0 pixels (0\farcs090). This value
is close to the median size of the faintest galaxies in
Figures~\ref{fig:fig6}--\ref{fig:fig8} (\ie\ with object $\rm FWHM\sim0\farcs1$
or half-light radii $r_e\simeq0\farcs05$), which enables us to better detect
very faint, low-SB, or clumpy galaxies. The \SExtractor\ parameter
DEBLEND\_MINCONT was set to 0.06 to assure that real objects were not
over-deblended. These parameters were chosen as a balance between extracting
objects deep enough to achieve sample completeness to approximately the
5$\sigma$ detection level (Section~\ref{sec42}) but not so deep that a visible
number of bogus objects were detected around remaining low-level image
artifacts (Section~\ref{sec33}).

For all images and filters, the corresponding weight maps were used to account
for image borders and properly characterize the photometric uncertainties and
the effective areas for each drizzled mosaic. We compared our catalogs to the
actual images and weight maps to look for any visible excess objects near
residual image structures and snowballs and where needed applied additional
masking to the images and their weight maps. Most masked regions are due to
edge effects, low-exposure regions, or strips due to the dither pattern
adapted for our shallow PEARLS exposures. Some bright stellar diffraction
spikes (\eg\ Figure~\ref{fig:fig3}) and residual wisp patterns also required
masking. All of these masked areas were excluded when calculating galaxy number
counts (Section~\ref{sec45}). 

The first line of Table~1 lists the J2000 tangent point to which the images in
all filters of each target were drizzled, the observing date, the APT visit
number, the area covered, the net exposure time per filter and total net hours,
and the spacecraft efficiency of that visit. The second line for each filter
lists the PSF FWHM, and the third line lists the 5$\sigma$ point source
sensitivity in AB-mag predicted by the {\it pre-launch} ETC for the net
integration time on the first line of each target. The fourth line in each
filter indicates the achieved $\sim$5$\sigma$ detection limits. The values were
derived from Figures~\ref{fig:fig6}--\ref{fig:fig8} as the median AB magnitude
where \SExtractor\ reports flux error bars of 0.20~mag. The fifth line
indicates the AB level in Figures~\ref{fig:fig6}--\ref{fig:fig8} where the
galaxy counts are $\sim$80\% complete compared to a power law extrapolation, as
derived from the figures in Section~\ref{sec45}. 

\n \subsection{PEARLS Star--Galaxy Classification Procedure } \label{sec42}

Separating stars from galaxies is important, especially for the PEARLS NEP
fields, which are located at Galactic latitudes +31\fdg6 (JWIDF) and +33\fdg6
(TDF)\null. Fields at these latitudes are expected to have more faint brown
dwarfs than fields at higher Galactic latitudes \citep[\eg][]{Ryan2011,
Ryan2017, Jansen2018, Ryan2022}. The PEARLS star--galaxy classification
procedure is based on the method described by \citet{Windhorst2011} for the
10-band WFC3 ERS images and by \citet{Windhorst2022} for the HST Archival
Legacy project SKYSURF sample of $\sim$249,000 HST images.
Figures~\ref{fig:fig6}--\ref{fig:fig8} show the object detection,
classification, and count diagnostics used for the JWIDF, as well as for the El
Gordo non-cluster module, respectively. 

For each detected object, the left panels show the \SExtractor\ magnitude error
vs.\ MAG\_AUTO AB-mag. The horizontal dashed line indicates the adopted
detection limit where the MAG\_AUTO error is $\ge$0.20 mag, which approximately
corresponds to a $\sim$5$\sigma$ detection for point sources. Table~1 lists the
average AB-mag values where the MAG\_AUTO error reaches $\ge$0.20 mag, as
derived from the left panels of Figures~\ref{fig:fig6}--\ref{fig:fig8}. 

The middle panels shows the star-galaxy classification diagram using 
\SExtractor\ MAG\_AUTO AB-magnitudes versus image FWHM\null. The NIRCam
diffraction limit is indicated in by the full-drawn left-most vertical line,
and the FWHM of the PSF is listed in the legend of each middle panel and in
Table~1. Objects with FWHM $<$ FWHM(PSF) have been flagged and removed from
this plot as spurious detections or border imperfections. In short, objects
detected by \SExtractor\ with sizes straddling the NIRCam diffraction limit to
a certain magnitude limit are classified as stars (red dots), following
\citet{Windhorst2011}. Stars are generally located in a thin nearly vertical
column bordered by the right-most vertical line. The remaining objects are
classified as galaxies (blue dots). The green dot-dashed lines indicate the
$5\sigma$ sensitivity limits of each image, with the horizontal part showing
the $\sim$5$\sigma$ point-source limit and the slanted part showing the SB
limit. For further details, see \citet{Windhorst2022}. 

The right panels of Figures~\ref{fig:fig6}--\ref{fig:fig8} show the resulting 
star counts (red filled circles) and galaxy counts (blue filled circles). At
most wavelengths \cge 0.9~\mum\ and at intermediate to high Galactic latitudes,
the star counts generally have a very flat slope ($\gamma\simeq0.04$~dex/mag),
while the galaxy counts have relatively steep slopes
$\gamma\simeq0.21$--0.25~dex/mag, continuing the trend seen at 0.2--1.6~\mum\
wavelengths by \citet{Windhorst2011}. As a consequence, galaxies generally
dominate the object counts for $\rm AB\ga 18$~mag and far outnumber Galactic
stars at fainter magnitudes. Hence, reliable identification of faint objects as
stars becomes difficult for $\rm AB\ga 26$--27~mag, and we have treated all
objects fainter than this as galaxies. {\ch (The specific limiting values for
each filter are shown in the right panels of
Figures~\ref{fig:fig6}--\ref{fig:fig8}.)}

Future work may be able to expand the identification of somewhat fainter stars
through color--color diagrams and comparison with theoretical stellar loci. In a
prior HST study \citep{Windhorst2011}, comparison of the 10-band WFC3 ERS star
counts to Galactic-structure model predictions verified the star--galaxy
classification procedure {\it a posteriori}. A similar check for the TDF data
is described by R. Ryan \etal\ (2022, in preparation). \citet{Windhorst2011} also
checked their star counts against spectroscopic ones from the HST ACS
$R\sim100$ grism survey ``PEARS'' \citep[Probing Evolution And Reionization
Spectroscopically;][]{Pirzkal2009}. Such {\it a posteriori} verification of our
star--galaxy classification procedure will be possible for the PEARLS TDF data
when we receive the NIRISS grism spectra in all four NEP TDF epochs later in
Cycle~1. 

Star--galaxy classification in JWST NIRcam images is---ironically---hardest at
the {\it bright} magnitude levels of $\rm AB\simeq18$--20 mag, as can be seen
from Figures~\ref{fig:fig6}--\ref{fig:fig8}. These bright objects are
generally unsaturated in the NIRCam images, but their significant diffraction
spikes make FWHM an unreliable indicator. The spikes also make it difficult to
derive accurate magnitudes and colors for identifying Galactic stars, as was
done for the WFC3/IR images of \citet{Windhorst2011}. We therefore used the
Gaia DR3 \citep[\eg][]{Gaia2022} catalog to identify stars with AB\cle 19~mag
through their non-zero proper motions and the SDSS DR7 spectroscopic catalog to
identify stars with available spectra at AB\cle 17.5 mag. 

\n \subsection{Reliability of the PEARLS Star--Galaxy Classification Procedure
}\label{sec43}

{\ch As indicated above, the complex shape of the JWST PSF makes star--galaxy
separation more difficult for objects with 18\cle AB\cle 21~mag. Another
complication is that an AGN can appear as a point source strong enough to hide
the host galaxy and make the object appear stellar even in the JWST images. To
verify whether the automated star--galaxy separation of
Figures~\ref{fig:fig6}--\ref{fig:fig8} was done correctly, we therefore
proceeded as follows: 

\sn (a) Two independent visual observers inspected all objects (both those
classified as stars and as galaxies) with 18\cle AB\cle 21 mag to arrive at a
consensus on which bright objects are stars and which are compact galaxies with
or without weak AGN. This was done in all 4--8 filters in the JWIDF and El
Gordo non-cluster fields. In the JWIDF, we found no objects classified amongst
the 20 brightest stars (18\cle AB\cle 24.4 mag) at the shorter wavelengths that
were classified as galaxies at the long wavelengths. In El Gordo, we found 8
possible galaxies amongst the 20 brightest objects classified as stellar. So in
total the fraction of brighter stellar objects that are misclassified galaxies
is about 20\%. 

\sn (b) We required that stars needed to be classified as such in at least 2
out of 4--8 NIRCam filters using using the method in the middle panels of 
Figures~\ref{fig:fig6}--\ref{fig:fig8}. \cite{Windhorst2011} required a
stellar object to be classified as such in at least 3 out 10 their HST ACS or 
WFC3 filters. This method found 69 objects classified as stellar in 2--4 JWIDF
filters and 23 objects classified as stellar in 2--8 filters in the El Gordo
non-cluster module. The surface density of stellar objects in 
Figures~\ref{fig:fig6}--\ref{fig:fig8} is thus about 1000--2000 \sqdeghalfmag\
in the El-Gordo non-cluster and JWIDF fields, respectively. The JWIDF is at
lower Galactic latitude, and so has a higher surface density of stars. The
results of (a) and (b) were largely consistent, and together reduced the number
of bright stars that were misclassified as galaxies. This is reflected in the
galaxy counts of Figures~\ref{fig:fig9}--\ref{fig:fig10}.

Despite the above procedures, our objects classified as ``stellar'' could still
be contaminated by faint, compact galaxies or weak AGN\null. To estimate the
contamination level by (weak) AGN, we must consider their expected surface
density at near-IR wavelengths. In the optical, the QSO surface density is
known to be \cle 15 \sqdeghalfmag\ to $B\la 21$~mag \citep{Boyle2000} and \cle
125 \sqdeghalfmag\ to $B\la 23$~mag \citep{Koo1982}. To predict surface
densities at the JWST NIRCam wavelengths, we used the UV--far-IR data sets from
the GAMA \citep{Bellstedt2020a} and DEVILS \citep{Davies2021} surveys with
multiwavelength SED fits by \citet{Bellstedt2020b}, \citet{Thorne2021}, and
\citet{Thorne2022}. These codes used \ProSpect\ \citep{Robotham2020}, which fits
stellar and AGN SEDs with an AGN fraction \fAGN\ at 1--2~\mum\ wavelengths as a
free parameter. We used the GAMA and DEVILS databases to estimate the surface
densities of objects with 20\cle AB\cle 25 mag and with SEDs in the VISTA
$ZYJH$ filters yielding \fAGN\cge 0.5--0.9. Similar to the behavior at the
optical wavelengths above, the surface density of such weak AGN converges to
\cle 100--50 \sqdeghalfmag\ for AB(1--2 \mum)\cle 25 mag and \fAGN\cge
0.5--0.9, respectively. This amounts to $\sim$10\% of our observed surface
density of stellar objects of 1000--2000 \sqdeghalfmag\ in the El-Gordo
non-cluster and JWIDF fields, comparable to our estimated $\sim$20\%
contamination rate of stellar samples by galaxies with weak AGN above. In
conclusion, we expect that $\sim$10--20\% of our stellar samples may be
contaminated by extragalactic objects. Future work that includes
matched-aperture SED fits and PSF subtraction will be able to make a more
accurate assessment of the fraction of compact galaxies or weak AGN remaining
in our announced stellar samples. }

\n \subsection{The Wavelength-Dependent Completeness of the PEARLS 
0.9--4.5~\mum\ Object Counts} \label{sec44}

Figures~\ref{fig:fig6}--\ref{fig:fig8} show that the stellar locus moves
steadily towards smaller FWHM values as wavelength decreases from 4.5 to
0.9~\mum. This is a consequence of the JWST diffraction limit being much better
than its requirement at 2.0~\mum. That is, the stellar locus keeps moving to
smaller object sizes from 2.0 to 1.5~\mum\ and continues to do so down to 
1.15~\mum, although the stellar region does become somewhat wider in the F090W
filter and in some fields in the F115W filter. Achieving a diffraction limit
well below 2.0~\mum\ wavelength is a major accomplishment for the JWST Project
team, as this had to be planned between 2003 and 2005 without further driving
up the Project cost. This was achieved by keeping the diffraction limit
requirement at 2.0~\mum\ but polishing the JWST mirrors well enough that the
high-frequency wavefront error would have a low enough rms to make a
diffraction limit below 2.0~\mum\ possible as long as the actuators below the
mirrors could remove the main low- and mid-frequency errors well. The telescope
in L2 is indeed able to do this \citep{Rigby2022}. 

The consequences of this achievement by the JWST Project are far-reaching, as
seen in the current paper: the median size of the faintest galaxies in the
PEARLS images is about FWHM$\simeq$0\farcs1 as shown in {\it all} the SW
panels of Figures~\ref{fig:fig6}--\ref{fig:fig8}. Hence, at JWST's diffraction
limited SW resolution at $\sim$1.1--2.0~\mum, essentially all faint galaxies in
Figures~\ref{fig:fig6}--\ref{fig:fig8} are {\it resolved} by NIRCam. In the LW
2.7--4.5~\mum\ panels of Figures~\ref{fig:fig6}--\ref{fig:fig8}, a significant
fraction of faint galaxies remain {\it unresolved} and therefore bunch up
against the diffraction limits. {\it Therefore, faint galaxies with flat SEDs
are more easily detectable with the wider LW PSF of $\sim$0\farcs17 FWHM
compared to the SW PSF, which has 0\farcs06--0\farcs08 FWHM\null.} This is
especially visible in the filters F277W and longwards. 

Table~1 quantifies the detection limits, using the 80\% galaxy-count
completeness limits as a fiducial. These limits were determined from power-law
fits to the counts and their extrapolations (Section~\ref{sec45}). In all
NIRCam SW filters, the 80\% galaxy count completeness limits typically appear
to be about {\ch $-$0.3 to $-$0.9 mag} {\it brighter} than the 5$\sigma$ point
source detection limits predicted by the pre-launch ETC\null. This is indicated
by the $\rm \Delta AB_{lim}(80\%-ETC)$ values on the sixth line for each target
in Table~1. A small part of this difference is due to the fact that some SW
filters can be $\sim$10\% less sensitive than the pre-launch predictions
\citep{Rigby2022}, but for the most part this ``apparent loss'' in SW point
source sensitivity occurs because the large majority of faint galaxies observed
in NIRCam SW are {\it no longer point sources}. The simple reverse is true for
the faintest galaxies in most NIRCam LW filters: our PEARLS LW galaxy counts
appear to be between {\ch +0.0 to +0.7 mag} {\it more sensitive} than the
pre-launch ETC prediction for point sources. Part of this difference occurs
because most LW filters are 20--30\% more sensitive than the pre-launch
predictions of \citet{Rigby2022}. But in addition, the detection of the
faintest galaxies in the LW images is surely aided by the fact that a
significant fraction of the faintest galaxy sizes no longer exceeds the size of
the PSF FWHM in the LW filters. The tendency of faint galaxies to bunch up
against the HST diffraction limit at brighter flux levels was first suggested
based on the Hubble Deep Field images by \citet[\eg][]{Odewahn1996} and
\citet{Windhorst1998a} and later by \citet[\eg][]{Welch2022b},
\citet{Windhorst2021}, and references therein based on more recent HST images.

In conclusion, at the FWHM of the NIRCam PSFs delivered by the JWST Project,
faint galaxies with flat SEDs are noticeably easier to detect in the LW
filters compared to the SW filters. Late-type stars are bluer than galaxies
and are point sources in all NIRCam filters, so this PSF-advantage at longer
wavelengths for galaxies does not apply to stars. Current and future JWST
surveys that search for high-redshift dropouts and other red objects will need
to keep this rather strongly wavelength-dependent sensitivity to the typical
faint-galaxy sizes into account. 


\n \subsection{The Combined 0.9--4.5~\mum\ Galaxy Counts } \label{sec45}

\sn {\bf \bul The PEARLS 0.9--4.5~\mum\ Galaxy Counts:}\ Our first PEARLS
galaxy counts are based on the data in two of our four fields observed so far
with the best available data in Table~1. We used all 10 NIRCam detectors in the
JWIDF plus the four SW and one LW detector in the El Gordo non-cluster module.
A comparison of the two sets of galaxy counts in the four filters where they
overlap then enables us to quantify whether the El Gordo non-cluster module was
biased in a measurable way by the presence of the $z=0.870$ cluster in the
adjacent NIRCam module. We defer reporting galaxy counts in the VV~191 field to
a later paper, owing to the presence of the two large, targeted galaxies and
bright objects in the same nearby galaxy group that cover some of the other
detectors. We also do not present object counts in the five medium-band filters
of the TNJ-1338$-$1942 protocluster at $z=4.1$ because these medium-band images
are shallower than the broad-band images in Table~1 and because there are no
reference filters available from the ground for comparison to brighter object
counts. Nonetheless, we did carry out their star--galaxy classification and
object counts, as was done for the JWIDF and El Gordo in
Figures~\ref{fig:fig6}--\ref{fig:fig8}, to check on the reliability of our
procedures and to report their 5$\sigma$ point source sensitivities in Table~1
as compared to the ETC predictions. 

{\ch We also inspected all 18\cle AB\cle 21 mag objects in the non-cluster
module of El-Gordo and found 10 galaxies close to the cluster outskirts that
were in the non-cluster module and have colors very similar to the El Gordo
cluster galaxies. These are likely part of the outskirts of El Gordo. We
removed these 10 objects from the galaxy counts. The galaxy counts in the JWIDF
and El Gordo non-cluster module at the JWST bright end (18\cle AB\cle 20 mag)
are consistent to within their (large) error bars and are close to the average
counts from the previous ground-based, HST, and Spitzer surveys, which have
much smaller error bars over this magnitude range. In any case, the error bars
on the bright end of the JWST galaxy counts (18\cle AB\cle 20 mag) are large
enough that they do not weigh significantly into the spline fits of the galaxy
counts (middle and right panels of Figures~\ref{fig:fig9}--\ref{fig:fig10}. }

Our PEARLS galaxy counts in the JWIDF and El Gordo non-cluster module are shown
in Figures~\ref{fig:fig9}--\ref{fig:fig10}. These are based on our object
catalogs of Section~\ref{sec41} with objects identified as stars removed
(Section~\ref{sec42}--\ref{sec43}). Error bars reflect the statistical
uncertainties in the remaining galaxy counts. At the bright end
(AB$\simeq$18-21 mag), larger discrepancies are seen in the counts between the
two fields due to the uncertainties in the star--galaxy classification
procedure (Section~\ref{sec42}--\ref{sec43}) and due to cosmic variance.
Because we used two NIRCam fields far apart in the sky, their cosmic variance is
expected to be \cle 9\% \citep[\eg][and Section~\ref{sec21} above]{Driver2010}.
This estimate uses the JWIDF+El Gordo survey area of Table~1 and the redshift
distribution expected for NIRCam objects with 18\cle AB\cle 28 mag, assuming
most objects are in the redshift range $0.3\la z\la 8$. Further details are
given in Appendix~\ref{secAppB2}. For the brighter fluxes of 18\cle AB\cle 21
mag, CV can be as high as 20\% for redshifts $ 0.1\la z\la 1$, explaining some
of the remaining statistical variations seen at the bright end of the counts in
Figures~\ref{fig:fig9}--\ref{fig:fig10}. 

The bright end of the 0.9--4.5~\mum\ JWIDF and El Gordo counts in
Figures~\ref{fig:fig9}--\ref{fig:fig10} are consistent with each other to
within their error bars. Hence, we see no evidence that the galaxy counts in
the El Gordo non-cluster module are significantly higher than those in the
JWIDF, which is a random survey field. We will thus proceed with the JWIDF and
El Gordo non-cluster module counts as representative for the 0.9--4.5~\mum\
galaxy counts, and will use a CV error of $\sim$9\% in our discrete IGL error
budget over the entire magnitude range of 18\cle AB\cle 28.5 mag in
Figures~\ref{fig:fig9}--\ref{fig:fig10}. 

A few more objects may reside in the large-scale structure associated with the
El Gordo cluster at $z=0.870$. These may be removed from the galaxy counts in
the El Gordo non-cluster module when more spectra of the cluster and its
surroundings become available. Future work will improve the accuracy of galaxy
counts when done over more JWST fields. Given the high surface density of
background galaxies detected by NIRCam and the negative magnification bias
predicted by the shallow NIR count slopes in
Figures~\ref{fig:fig9}--\ref{fig:fig10}, more accurate background-galaxy counts
combined with a weak shear analysis of the same images may improve mass
profile measurements of the galaxy clusters \citep[\eg][]{Umetsu2011}.

\sn {\bf \bul Comparison to Previous Galaxy Counts at 0.9--4.5~\mum:}\ The
PEARLS 0.9--4.5~\mum\ galaxy counts are compared to those at brighter levels
from the combined GAMA \citep{Driver2010, Driver2011, Driver2022} and DEVILS 
\citep{Davies2021} surveys at similar wavelengths in
Figures~\ref{fig:fig9}--\ref{fig:fig10} and at the shorter wavelength
(0.9--1.6~\mum) also with the deepest available HST counts. A possible check of
catalog completeness and reliability is to compare our JWST/NIRCam object
counts to the deepest available HST counts in the same or similar filters and
see whether the agreement is good to within the known ZP, rms counting, and
cosmic-variance errors. This will also help verify the flux level at which
catalog incompleteness sets in. 

The GAMA and COSMOS/DEVILS survey data were compiled over a wide range of
wavelengths by \citet[][and references therein]{Driver2016b}, building on the
GAMA panchromatic data release of \citet{Driver2016a} and the COSMOS data as
reanalyzed by \citet{Andrews2017}. This compendium was later extended by
\citet{Davies2018} and \citet{Davies2021} as part of the DEVILS survey, while 
\citet{Bellstedt2020a} extended the GAMA data to also include the ESO VST KiDS
data. These updates were reported by \citet{Koushan2021}, whose work forms the
basis of the galaxy-counting data used here. These compendia also include the
ESO $K$-bands counts of \citet{Fontana2014}, deep Spitzer 3.6 and 4.5~\mum\
counts \citep[\eg][]{Ashby2009, Mauduit2012, Ashby2015}, and the WISE counts of
\citet{Jarrett2017}. The typical combined ZP uncertainties in the combined
GAMA+DEVILS surveys in the $z$-band to $K$-band are 2--3\% after bringing the
flux scale in every filter onto the flux scale of the VISTA survey filters,
which incorporated the GAMA and DEVILS surveys \citep[Table~4 
of][]{Koushan2021}. 

The deepest panchromatic HST ACS+WFC3 galaxy counts come from the combined HUDF
images, whose database has grown considerably over time since the launch of
WFC3 in May 2009 \citep[see \eg][and references therein]{Windhorst2011,
Koekemoer2013, Rafelski2015}. Following the rich HUDF database summarized in
these papers, the panchromatic galaxy counts were once more repeated on the
deepest available HUDF images in 2015 with the same procedures as in
Sections~\ref{sec41}--\ref{sec42} and were included by \citet{Driver2016b} and
\citet{Koushan2021}. [This deepest 2015 realization of the HUDF counts is
listed in the legend of Figures~\ref{fig:fig9}--\ref{fig:fig10} as
``\citet[][$^+$]{Windhorst2011}'' and follows the same methods.] The ZP 
errors in the HST ACS, WFC/UVIS, and WFC3/IR images over the decades resulted in
flux scales accurate to 1--3\% as summarized in Section~4.1.5 and Table~5 of
\citet{Windhorst2022}, \ie\ comparable to or slightly better than the 2--3\% ZP
accuracy of the VISTA filters of \citet{Koushan2021}. 

To within these ZP errors, our 0.9--1.5~\mum\ PEARLS galaxy counts that come
from shallow NIRCam exposures are consistent with the deeper HUDF counts in the
HST ACS and WFC3/IR filters F850LP, F105W/F125W, and F160W (green open circles
in Figures~\ref{fig:fig9}--\ref{fig:fig10}). It is important to realize that our
PEARLS galaxy counts come from 1890--3157~s NIRCam exposures in the F090W,
F115W, and F150W filters and reach $\sim$28.5 mag (Table~1), while the HUDF
galaxy counts reach $\sim$29.5--28.5 mag in $\sim$156--87 HST orbits
\citep[$\simeq$117--65 hours of net exposure time;][]{Beckwith2006,
Koekemoer2013} in the F850LP, F105W/F125W, and F160W filters, respectively.
Compared to the total HUDF ACS exposure time of 421.6~ks in F850LP, JWST
NIRCam thus reaches approximately $\sim$5$\times$ deeper per unit time in its
F090W filter, while compared to the total HUDF WFC3/IR exposure time of
236.1~ks in F160W, NIRCam reaches about $\sim$9--11$\times$ deeper per unit
time in its F150W filter, respectively. In the light of the discussion in
Section~{sec44}, this is an impressive performance improvement, especially because
NIRCam was optimized for performance longwards of 2.0~\mum. 

Because our PEARLS 0.9--4.5~\mum\ galaxy counts are done in NIRCam filters
whose effective wavelengths and bandpasses can be somewhat different from the
VISTA, GAMA, and DEVILS surveys, Section~\ref{secAppB2} addresses whether
additional corrections to the NIRCam flux scale are needed to compare the
galaxy counts in similar filters. For this, we used the fiducial flux scale of
the VISTA filters into which the GAMA and DEVILS surveys were anchored
\citep{Koushan2021}. Section~\ref{secAppB2} shows that the corrections needed to
transform the NIRCam AB-mag scale onto the fiducial VISTA/IRAC filters are \cle
3--4\% with combined uncertainties of $\la$3--6\%, and therefore no corrections
to the NIRCam AB-mag scale needed to be applied when plotting the results in
Figures~\ref{fig:fig9}--\ref{fig:fig10}. However, we folded this \cle 3--6\%
uncertainty into our error budget of the 0.9--4.5~\mum\ IGL of
Section~\ref{sec46}. We now have all the ingredients in place to compare our
0.9--4.5~\mum\ NIRCam galaxy counts to previous work at brighter levels and
can do so without further wavelength dependent ZP corrections. 

\sn {\bf \bul The Faint-End Slope of the Combined 0.9--4.5~\mum\ Galaxy
Counts:}\ Figures~\ref{fig:fig9}--\ref{fig:fig10} compare our PEARLS
0.9--4.5~\mum\ galaxy counts to previous work summarized above and extend it to
AB\cle 29~mag over this wavelength range. Over the AB$\simeq$16--29 mag range
for which they are available at 0.9--4.5~\mum, the \citet{Yung2022} models are
consistent with the combined GAMA/DEVILS ground-based, Spitzer/WISE, and our
PEARLS NIRcam galaxy counts. The faint-end slopes of our observed galaxy
counts have an average value $\gamma\simeq0.23 \pm 0.04$~dex/mag for 
$22\la\rm AB\la29$~mag, where the counts are a nearly straight power-law
(Figures~\ref{fig:fig9}--\ref{fig:fig10}, Table~3). The middle panels of
Figures~\ref{fig:fig9}--\ref{fig:fig10} showing the 0.9--4.5~\mum\ IGL energy
counts were derived from the left panels by dividing by the 0.4 slope. The
spline extrapolations (grey lines and error fans) in the middle panels show
that energy counts are clearly converging at all these wavelengths. The
resulting IGL integral is shown in the right panels of
Figures~\ref{fig:fig9}--\ref{fig:fig10}. Under the assumption that the faint
galaxy counts at AB\cge 29~mag continue as a power-law with the same slope as
observed for AB$\simeq$22--29 mag, these spline extrapolations will form the
basis of our IGL values used in Section~\ref{sec46} and~\ref{sec5}. Further
details on the faint-end slope of the galaxy counts and its wavelength
dependence are given by S. Tompkins \etal\ (2022, in preparation). 

A magnitude slope $\gamma\simeq0.23$~mag/dex corresponds to a faint end slope 
$\alpha\simeq-1.58\pm0.1$ in flux units, where $\alpha = -1 -2.5\gamma$.
Ground-based spectroscopic surveys with VLT/MUSE and the spectro-photometric
survey 3D-HST with Hubble suggest that faint galaxies with AB$\simeq$23--29 mag
have a median redshift in the range \zmed$\simeq$1--2
\citep[\eg][]{Skelton2014, Inami2017} with the caveat that the completeness of
these surveys becomes more difficult to quantify at fainter magnitudes. Around
this median redshift, the faint end of the galaxy counts samples the power-law
part of the Schechter luminosity function (LF), which also has a faint-end
flux slope $\alpha\simeq-1.4$ to $-$1.5 at $z\simeq1.5$
\citep[\eg][]{Hathi2010, Finkelstein2016}. {\it In conclusion, over the
AB$\simeq$22--29 magnitude range sampled by our 0.9--4.5~\mum\ NIRCam images,
the PEARLS galaxy counts have a slope consistent with the faint-end slope of
the Schechter LF at the median redshift sampled by these objects.} Fainter JWST
galaxy counts would then be expected to continue with the same slope, {\it if}
the LF over this redshift range were to continue with the same slope towards 
fainter luminosities. Upcoming ultradeep JWST NIRCam GTO surveys (M. Rieke PI) 
are designed to cast light on this issue. 

\n \subsection{Characteristics of the Integrated Galaxy Light at 0.9--4.5~\mum} 
\label{sec46}

\sn The parameters that best characterize the 0.9--4.5~\mum\ galaxy counts and
integrated galaxy light are summarized Figure~\ref{fig:fig11}. The errors on
these parameters are summarized in Table~3 for each filter as the quadratic sum
of the NIRCam ZP uncertainties from Section~\ref{secAppB1}, and the uncertainty
in bringing the flux scale of the NIRCam filters onto the effective wavelengths
of the VISTA/IRAC filters used as fiducial wavelengths for the galaxy counts
(Section~\ref{sec45}). Some interesting trends can be seen in the galaxy
counts over the wavelength range 0.9--4.5~\mum\ from
Figures~\ref{fig:fig9}--\ref{fig:fig10}. (These trends are best seen if all PNG
files of Figures~\ref{fig:fig9}--\ref{fig:fig10} are shown on a computer screen
at high magnification in rapid succession --- all panels are plotted at exactly
the same scale for this purpose). The smooth behavior of the data in
Figure~\ref{fig:fig11} suggests that these trends in the IGL are real and
meaningful: 

\begin{enumerate}

\item The left panels of Figures~\ref{fig:fig9}--\ref{fig:fig10} all show a
clear change in slope from a steep non-converging slope ($\rm \ge 0.4$~dex/mag)
to shallow and converging slope ($\rm < 0.4$~dex/mag). This change sets in
around AB$\sim$19.3--20.3 mag at 0.9--4.5~\mum\ (top panel of
Figure~\ref{fig:fig11}). This peak AB-mag is the flux level where most of the
IGL is generated, and is a clear function of wavelength. 

\item In more detail, the normalized differential counts reaches the highest
SB-level around 2--3~\mum\ wavelength, and declines to both longer and shorter
wavelengths (first and second panel of Figure~\ref{fig:fig11}, where the
second panel shows the peak SB-value of the IGL at the AB-magnitude peak of the
top panel). 

\item The magnitude range over which most of the IGL is generated (here called
the ``IGL FWHM'' and measured as the interquartile 25\%--75\% range of the
middle panels in Figures~\ref{fig:fig9}--\ref{fig:fig10}) decreases from
$\sim$4.5~mag at wavelengths \cle 1.25~\mum\ to $\sim$3~mag at 4.5~\mum\ (third
panel of Figure~\ref{fig:fig11}). This reflects the luminosity function and
redshift distribution of the older, earlier-type galaxies that dominate the
3.5--4.5~\mum\ galaxy counts. These filters sample the rest-frame 
$\sim$1.5~\mum\ peak in the stellar emission of early-type galaxies as caused
by their stellar mass distribution at z$\simeq$1--2. The JWST images at these
wavelengths are visibly dominated by earlier-type galaxies. At shorter
wavelengths, we sample a larger fraction of galaxies of later-types, which have
a wider range of ages and extinction, and extend to lower luminosities and
redshifts, causing their larger contribution to the IGL at bluer wavelengths
\citep[\eg][]{Driver1995, Driver2016b, Andrews2018, Koushan2021}. This can be
seen in the larger IGL-width at bluer wavelengths in the third panel of
Figure~\ref{fig:fig11}.

\item The discrete IGL with the highest energy (in units of \nWsqmsr) comes
from wavelengths between 1 and 2~\mum, as shown in the bottom panel of
Figure~\ref{fig:fig11}. These are derived from the converging integrals from
the right panels of Figures~\ref{fig:fig9}--\ref{fig:fig10}, and are the values 
we plot in the last Figure of Section~\ref{sec5}, which provides further
discussion of the total IGL and diffuse light. 

\end{enumerate}

The new JWST results are most noticeable at the faint end of the AB-magnitude
scale plotted in Figures~\ref{fig:fig9}--\ref{fig:fig10}. The bright end of the
galaxy counts can be --- and has been --- done from the ground (\ie\ the $z$,
$Y$-, $J$-, $H$-, and $K$-band filters) or from space with WISE and Spitzer at
$L$- (3.5~\mum) and $M$-band (4.5~\mum). Nevertheless, JWST NIRCam and also HST
WFC3/IR below 1.6~\mum\ have unique filters that are valuable for object
counts. These include WFC3 F140W, and NIRCam F277W and F410M, as well as the 
other medium-band filters in Table~1. These filters have {\it no} ground-based
counterparts because telluric water vapor blocks these wavelengths. Therefore
bright-end galaxy counts to make a full energy integral as in
Figures~\ref{fig:fig9}--\ref{fig:fig10} are absent for the F277W and F410M as
well as the medium-band filters. 

Because of this, \citet{Carleton2022} had to interpolate the IGL integral in the
WFC3/IR F140W filter from the adjacent WFC3/IR F125W and F160W filters, which
have extensive ground-based coverage of the bright end counts. Fortunately,
this is straightforward because the IGL SB is flat between 1.25 and 1.65 \mum\
wavelengths (bottom panel of Figure~\ref{fig:fig11}). We therefore give
low-order spline functions fit to the IGL parameters vs. wavelength in Table~3
and plot these in Figure~\ref{fig:fig11}. We use these splines to interpolate
the IGL SB-values for the JWST NIRCam F277W and F410M filters, which are
tabulated as the PEARLS-IGL values in Tables~4--5. This includes the values
beyond the PEARLS NIRCam detection limits of AB$>$28.5 mag, which were derived
from the integrals in the right-hand panels in
Figures~\ref{fig:fig9}--\ref{fig:fig10}. This allows us to estimate the IGL as
a function of wavelength for the JWST filters at 0.9--4.5~\mum\ wavelength,
including the \cle 2.5\% of the IGL that is not included in our faint object
counts to AB\cle 28.5 mag (see Section~\ref{sec52} here, and Section~3.4.3 of
\citealp{Carleton2022} for its procedure). Wider-area JWST surveys such as \eg\
COSMOS-Webb (PI J. Kartaltepe) will become available during JWST's lifetime to
improve the bright-end of the galaxy counts, as they have for HST ACS and WFC3
during the last two decades \citep[\eg][]{Windhorst2022}.


\n \section{NIRCam 13-band sky-SB Estimates and Limits on Diffuse Light}
\label{sec5} 

In this section we give our estimates of the sky-SB as measured in between the
detected discrete objects in the 13-band NIRCam filters observed with PEARLS,
and assess if we can set meaningful limits to diffuse light in excess of the
Integrated Galaxy Light from Section~\ref{sec46}. In this process, we account
for the NIRCam systematics summarized in Section~\ref{sec3} and
Appendix~\ref{secAppB}--\ref{secAppC}, and include these in our error budget.

\n \subsection{JWST sky-SB in the Context of Previous Diffuse Light Limits} 
\label{sec51}

JWST's ability to work {\it continuously} in a dark-sky environment makes it
especially suitable for measurements of sky-SB\null. This is in contrast with
HST, which at best gets complete dark time for at most $\sim$30 minutes of its
96-minute orbit \citep[\eg][]{Caddy2021, Caddy2022, Windhorst2022}. 
Figures~\ref{fig:fig12}--\ref{fig:fig13} summarize the astrophysical foreground
and background energy relevant to PEARLS compared to recent data \citep[as
summarized by \eg][]{Driver2016b, Koushan2021, Carleton2022} and IGL models
\citep[\eg][]{Andrews2018}. Figure~\ref{fig:fig13} shows the PEARLS discrete 
IGL measurements of Section~\ref{sec46} compared to those of 
\citetalias{Driver2016b} and \citet{Koushan2021}. The IGL is the sum of the
integrated (observed) galaxy counts (iEBL) and extrapolated galaxy counts
(eEBL) derived in Section~\ref{sec45}. The black line shown in
Figure~\ref{fig:fig13} is a modification of the \citet{Andrews2018} IGL model
for accumulated star-formation in spheroids (red dashed), disks (green
dashed), and unobscured AGN (purple dashed lines). Here we have adjusted these
contributing elements from the published \citet{Andrews2018} model, as
described in the caption, to better fit the IGL data including the PEARLS
points. 

JWST was meticulously designed and built to have the darkest possible sky as
seen from L2. Here we explore its capability to constrain potential levels of
diffuse light. \citet{Windhorst2022} stated that over 95\% of the
0.6--1.25~\mum\ photons in the HST archive come from the Zodiacal light in the
interplanetary dust (IPD) cloud, \ie\ from distances $<$5~AU. This can also be
seen by comparing the typical Zodiacal light levels (green line) to the IGL
counts in Figure~\ref{fig:fig13}. The Zodiacal/IGL ratio {\it decreases
significantly} towards longer wavelengths in the 1.5--3.5~\mum\ wavelength
range. This is because the Sun is a zero redshift 5770~K G-star, and the IGL is
the summation over multiple stellar populations, including hotter and cool
stars, spanning a wide redshift range \citep[\eg][]{Madau2014}. Longwards of
3.5~\mum, thermal radiation from the Zodiacal belt and also from the JWST
telescope and instruments make increasing contributions to the SB levels. The
wavelength dependence of the diffuse light is precisely what JWST can explore
from its first images, bearing in mind the significant JWST and NIRCam
calibration uncertainties that we expect (Section~\ref{sec33} and
Appendix~\ref{secAppB1}--\ref{secAppB3}). 

Obtaining Diffuse Light (DL) estimates requires accurate modeling of the
Zodiacal Light (ZL) and Diffuse Galactic Light (DGL)\null. ZL can be 
$\sim$10--70$\times$ higher than the discrete iEBL+eEBL, dependent on the 
direction and time of observation
(Figures~\ref{fig:fig12}--Figure~\ref{fig:fig13}). Constraints from previous
work shown in Figure~\ref{fig:fig13} suggest that there may exist some level of
diffuse light at 0.6--1.6~\mum, at a level of $\sim$8--30 \nWsqmsr. (Note that
{\it all} diffuse light estimates plotted in color in Figure~\ref{fig:fig13}
have the full IGL already subtracted, and so truly represent the diffuse light
levels or limits reported by various groups). At this stage, it is not clear
whether this diffuse light is due to residual instrumental systematics that
have not been accounted for, a dim Zodiacal component (perhaps spherical or
spheroidal) seen from 1~AU that is not accounted for in the Zodiacal models, a
truly diffuse EBL component, or some combination of these possibilities. For
details on this topic, please see the discussions by, \eg\ 
\citet{Conselice2016}, \citet{Matsuura2017}, \citet{Sano2020},
\citet{Carleton2022}, \citet{Korngut2022}, \citet{Kramer2022}, 
\citet{Lauer2022}, \citet{OBrien2022}, and \citet{Windhorst2022}, and 
references therein.


\n \subsection{JWST sky-SB Estimates and Possible Limits to Diffuse Light }
\label{sec52} 

\sn Following Equation~(2) of \citet{Windhorst2022}, the sky-SB level {\it
between the detected objects} is a sum of Zodiacal Light, Diffuse Galactic
Light, and residual instrumental systematics including thermal and straylight
contributions. For JWST, that equation is: 
\begin{equation} 
{\rm SB}(\lambda, l^{\rm Ecl}, b^{\rm Ecl}, l^{\rm II}, b^{\rm II}, t, {\rm
SA}, T)={\rm Th}(\lambda, T)+ {\rm SL}(\lambda, l^{\rm Ecl}, b^{\rm Ecl},
t)+{\rm ZL}(\lambda, l^{\rm Ecl}, b^{\rm Ecl}, t, {\rm SA})+{\rm DGL}(\lambda,
l^{\rm II}, b^{\rm II})+{\rm dEBL}(\lambda)
\label{eq:eq3} 
\end{equation}
\n The left term in Equation~\ref{eq:eq3} is the total sky-SB that JWST
observes as a function of wavelength $\lambda$, Ecliptic coordinates ($l^{\rm
Ecl}$, $b^{\rm Ecl}$), Galactic coordinates ($l^{\rm II}$, $b^{\rm II}$), time
of the year ($t$ or Modified Julian Date MJD), solar elongation angle (SA), and
telescope and instrument temperatures, symbolized by $T$. The terms on the
right side include: thermal (Th) signal from blackbody photons in the
instruments and telescope that depends on wavelength and temperature;
straylight (SL) that depends on wavelength, pointing direction, and observing
date (Section~\ref{sec33}); Zodiacal light (ZL) as seen from L2 that depends on
wavelength, Ecliptic coordinates, and observing date via its effect on the SA 
and on the path through the Zodiacal dust cloud (especially JWST's position
above or below the Ecliptic plane; see Appendix~\ref{secAppC}); Diffuse
Galactic Light that depends on wavelength and Galactic coordinates; and any
diffuse EBL (dEBL) that is not already included in the discrete object catalogs
to AB\cle 28.5 mag in Section~\ref{sec4}, and therefore not yet masked out from
our NIRCam images. This last term includes the part of the discrete IGL
extrapolated for AB\cge 28.5 mag (\ie\ the eEBL), which is generally small and
subtracted below. 

At the start of JWST Cycle 1, we only have a limited number of JWST images and
so can only explore limits for the sky-SB values observed from L2 thus far by
PEARLS. When more JWST images become available, its full database can be
studied following the SKYSURF methods of \citet{Carleton2022}, who presented
the HST sky-SB between discrete objects in 34,000 WFC3/IR images, and set
constraints on diffuse light from the subset of HST 1.25--1.6~\mum\ images with
the darkest sky-SB values. At this stage, we will use our darkest available
JWST images to explore what constraints can be made currently and how these may
be improved in the future during JWST's lifetime.

We measured sky-SB in the NIRCam images following the SKYSURF procedures of
\citet{Windhorst2022} and \citet{Carleton2022}. In short, SKYSURF removes the 
light from all detected objects from the WFC3/IR images even in short HST
exposures (\texp$\simeq$500~s) to a total-object flux limit of $AB\la 26.5$~mag
at 1.25--1.6~\mum\ wavelengths. For JWST NIRCam, we used the same codes to
remove {\it the light from all detected} objects (Section~\ref{sec33} and
Appendix~\ref{secAppB3}) to $\rm AB\la 27.5$--28.5~mag at 0.9--4.5~\mum\ given
the detection limits in Table~1, at which flux levels \cge 97.5\% of the {\it
discrete IGL} is already detected in the JWST images, as the right panels of
Figures~\ref{fig:fig9}--\ref{fig:fig10} show. Details on the uncertainty in the
estimated sky-SB are given in Appendix~\ref{secAppB3}.

Tables~4--5 summarize the JWST NIRCam instrumental and astronomical background
levels as predicted from, or actually observed from L2 for each of the four
PEARLS targets observed as of\ 2022 July 31. Background components are assumed
to be uniform across the field-of-view but depend on wavelength. Tables~4--5
list the ETC predictions for the L2 Zodi, thermal and straylight, as well as
the straylight level of \citet[][using their Figure 5]{Rigby2022}. Details on
these predicted sky-SB component values and their uncertainties are given in
Appendix~\ref{secAppC}, and we summarize aspects relevant to current discussion
here. The sky-SB predictions for all these components and their uncertainties
(where relevant) are given in Tables~4--5, and include: 

\begin{itemize}

\item (1) The ETC-predicted JWST thermal radiation, which is more than
100$\times$ lower than the predicted total sky-SB even at 4.5~\mum\ (see
Figure~\ref{fig:fig12}) and is the dimmest component in Equation~\ref{eq:eq3}
for $\lambda$\cle 4\mum. 

\item (2) The L2 model prediction for the Zodiacal sky-SB for each target. This
was based on the position and orientation of JWST at the actual time of the
observation. These are based on the \citep{Kelsall1998} model, but for the
Zodiacal cloud geometry for L2 as seen by JWST at the time of the observation.
These predictions are uncertain by at least $\sim$9\%--4\% of the dimmest
Zodiacal sky-SB observed over the range 1.25--4.5\mum, respectively.

\item (3) The IPAC IRSA prediction for the DGL value at the Galactic
coordinates of the target. The DGL is generally a factor of 20--100$\times$
lower than the total predicted JWST sky-SB, and uncertain by up $\sim$0.3 
dex. 

\item (4) The actual straylight, which \citet{Rigby2022} noted is likely lower
than the pre-flight predictions. Indeed, using the full \citet{Rigby2022} SL
values would make the total sky-SB predictions from Equation~\ref{eq:eq3}
exceed the observed values {\ch in two of our PEARLS fields} in Tables~4--5. As
discussed above, the 0.9--3.5~\mum\ Zodiacal component is caused by Sunlight
scattered off the Zodiacal dust cloud components, and may have been
underestimated in some of the models. The 3.5--4.5\mum\ sky-SB is dominated by
the thermal contribution from the Zodiacal dust cloud components, while the
telescope+instrument thermal components are still negligible in these filters.
The thermal Zodiacal Light at $\lambda$\cge 3.5~\mum\ was the key component to
be modeled by \citet{Kelsall1998} for their COBE/DIRBE analysis. The minimum
sky-SB is predicted to occur around 3.5~\mum\ in wavelength
(Figures~\ref{fig:fig12}--\ref{fig:fig13}), so we will assume that: (a) the
thermal Zodiacal components at $\lambda$\cge 3.5~\mum\ are more accurately
predicted than the scattered Sunlight components at $\lambda$\cle 3.5~\mum; and
(b) the predicted thermal Zodiacal components should match the values observed
at 4.5~\mum\, without exceeding those observed in the minimum at 3.5\mum. Any
truly diffuse astrophysical source is expected to be much dimmer than this, so
we do not expect it to significantly affect our fitting. In order to not exceed
the observed PEARLS sky-SB values, we then find that the implied SL values are
generally {\ch $f$$\sim$50--100\%} of the \citep{Rigby2022} SL-values, with an
uncertainty in $f$ of at least $\sim$20\% in Equation~\ref{eq:eq4} below (see
also Appendix~\ref{secAppC}). With these adopted SL values in Tables~4--5, our
total predicted JWST sky-SB matches the observations in all 13 PEARLS filters in
Figure~\ref{fig:fig12} to within the uncertainties summarized above, using
Equations~\ref{eq:eq4}--\ref{eq:eq5} below. 

\item (5) We use the IGL integral for the seven fiducial filter wavelengths 
in Section~\ref{sec45} from Figures~\ref{fig:fig9}--\ref{fig:fig10}. For the
NIRCam wavelengths for which a full IGL integral is not yet available (F277W,
F410M, and the medium-band filters), we used the spline predictions at those
wavelengths from Figure~\ref{fig:fig11}. The IGL is assumed to be constant
across the sky, and therefore to be the same for each PEARLS target. The {\it
extrapolated discrete eEBL} (eEBL) of objects currently undetected in the JWST
images for AB\cge 28.5 mag is derived from
Figures~\ref{fig:fig9}--\ref{fig:fig10} (Section~\ref{sec46}). The
extrapolation for AB\cge 28.5 mag typically amounts to $\sim$2.5\% of the
total IGL\null. This is the {\it only} part of the IGL that still need to be
subtracted from the sky-SB, as our method already automatically removes
$\sim$97.5\% of the light from all brighter objects detected by NIRCam to
AB\cle 28.5 mag \citep[see][for details]{Carleton2022}. (Section~4.7 and
Equation~3 of \citet{Windhorst2022} also corrected this fraction for
SB-incompleteness, which can be $\sim$40\% at AB$\sim$28 mag, but the IGL
correction for known objects at AB\cge 28.5 mag remains very small).

\end{itemize}

Our best prediction of the observed JWST Zodiacal sky-SB is then:
\begin{equation} 
{\rm 
JWST(Pred) = ETC(Thermal)\ +\ f\times SL(Rigby2022)\ +\ Zodi(L2)\ +\ DGL\ +\ 
eEBL,} 
\label{eq:eq4} 
\end{equation}
\n Tables~4--5 give this sum as predicted from the above models in all filters
at the actual time of PEARLS observations. Finally, these tables list our upper
limits to any diffuse light (DL) as the difference between the observed JWST
sky-SB ``JWST(Obs)'' and the total prediction of Equation~\ref{eq:eq4} for L2:
\begin{equation}
{\rm
Diffuse\ Light\ limit\ \lesssim\ JWST(Obs)\ -\ JWST(Pred)\ \pm\ ErrorBudget
}
\label{eq:eq5}
\end{equation}
\n The results from Equation~\ref{eq:eq4}--\ref{eq:eq5} are listed on the
bottom lines in each tier of Tables~4--5. This includes the full error budget
of {\ch $\sim$6--8\%} for JWST(Obs) in the LW--SW modules, respectively, from
Appendix~\ref{secAppB3}, and the combined uncertainty of $\sim$10\% in 
JWST(Pred) from Appendix~\ref{secAppC}. The total uncertainty in the difference
of Equation~\ref{eq:eq5} is thus {\ch $\sim$12--13\%} of the total sky-SB in
the LW--SW modules, respectively, assuming that the Observed and Predicted
sky-SB values are independent. 

With our assumption that the total sky-SB model should fully predict the
observed sky-SB values in the four PEARLS filters at 3.5--4.5~\mum, the model
predictions generally also match the observed sky-SB in the seven PEARLS
filters at 0.9--3.5~\mum, including the medium-band filters in TNJ1338, 
within the combined uncertainties. Therefore, to within the error budget of the
current assessment, we have no firm detection of remaining DL by JWST NIRCam. 

Accordingly, all our diffuse light constraints are plotted as upper limits
using the combined uncertainties of Appendix~\ref{secAppB}--\ref{secAppC} in
Figure~\ref{fig:fig13}. Brown downward open triangles indicate upper limits
from our deepest filter-exposures in the JWIDF and the El Gordo non-cluster
field. Brown downward asterisk- and tripod-shape indicate the upper limits from
the shallower TNJ and VV191 exposures. We excluded in this process the
detectors that contained the overlapping nearby galaxy pair of
Figure~\ref{fig:fig5} and other large objects. Our PEARLS constraints in
Figure~\ref{fig:fig13} indicate upper limits to Diffuse Light captured by the
grey hashed area, and generally amount to 12--13\% of the total sky-SB observed
by NIRCam. Within the current uncertainties in the JWST NIRCam calibration {\it
and} in the total JWST sky-SB model, we cannot make firmer statements about the
diffuse light as seen by JWST. In particular, if the JWST SL were even lower
than we adopted here, firmer constraints on DL may be made. For this purpose,
future work will require a more accurate assessment of the NIRCam calibration
uncertainties, and more accurate models for the JWST straylight, the Zodiacal
Light as seen from L2, and for the DGL. 

The two JWIDF diffuse light points at 1.5 and 2.0~\mum\ are marginally above
the total model predictions in Figure~\ref{fig:fig12}d. Our F150W and F200W
NIRCam diffuse light limits in Figure~\ref{fig:fig13} are in line with the
1.1--1.6~\mum\ CIBER detections of \citet{Matsuura2017} and \citet{Sano2020}
(purple triangles in Figure~\ref{fig:fig13}), and with the SKYSURF upper limits
in the HST/WFC3 F140W and F160M filters of \citet[][]{Carleton2022}. These
papers, as well as \citet{Tsumura2018} and \citet{Korngut2022}, suggested that
some very dim spherical --- or nearly spherical --- Zodiacal component could be
missing from the \citet{Kelsall1998} model. \citet{Kelsall1998} also noted 
that a dim spherical Zodiacal component could have been missed in their model
of the COBE/DIRBE data. 

At the longer NIRCam wavelengths of 2.7--4.5~\mum, the PEARLS DL limits in
Figure~\ref{fig:fig13} are lower in value, reaching as low as 8--12 \nWsqmsr,
and consistently so between our four PEARLS fields within the current error
budget. This is because the total sky-SB is significantly darker
(Figure~\ref{fig:fig12}), and the total error budget of the LW modules
(Appendix~\ref{secAppB}) and the uncertainties in the sky-SB models are correspondingly
smaller at 2.7--4.5~\mum\ (Appendix~\ref{secAppC}).\deleted{That is, future
analysis of the sky-SB in many more NIRCam images across the sky may be able to
set firmer constraints on the amount of diffuse light that may be present at
2.7--4.5~\mum.} 


\n \section{Discussion } \label{sec6}

\mn It is remarkable how even the first JWST images of our PEARLS fields ---
with relatively short NIRCam exposures in a total of 13 filters --- give us a
fresh look on the distant Universe. The fact that JWST achieved its
diffraction limit at wavelengths well below 2.0~\mum\ is a tremendous
achievement for the JWST Project and of great value to the community. The JWST
NIRCam PSF is so sharp and stable that star--galaxy classification is
straightforward with existing methods, even in short exposures. The same is
true for making object catalogs and deriving galaxy counts. At 0.9, 1.1, 1.5,
2.0, 3.5, and 4.5~\mum\ wavelengths, comparison to existing ground-based,
WISE, and Spitzer galaxy counts is also straightforward. 

Our JWST galaxy counts of Section~\ref{sec45} agree well with previous work,
but go \cge 2~mag deeper even in our short NIRCam exposures. Combining two
fields that are separated widely in the sky decreases the Cosmic Variance
component of the uncertainty in the counts, which can be \cle 9\%, or more at
brighter levels \citep[\eg][see Section~\ref{sec2}]{Driver2010}. The combined
error in the counts from ZPs (\cle 4\%), transforming to the VISTA/IRAC filters
system (\cle 3--6\%), and CV (\cle 9\%) is \cle 10--12\%
(Appendix~\ref{secAppC}), which is our uncertainty in the IGL. The
galaxy counts at 0.9, 1.1, 1.5, 2.0, 3.5, and 4.5~\mum\ show some interesting
trends. The energy-normalized differential galaxy counts reach a maximum in the
range $\sim$19.3--20.3 AB mag. Objects in this range produce most of the IGL
per magnitude bin. The actual flux level in AB-mag {\it and the width of the
peak} are both functions of wavelength. This reflects the luminosity function
and redshift distribution of the galaxy population that dominates each of these
wavelengths. 

The galaxy population slowly changes from later-type galaxies at lower
redshifts dominating in the blue (including the HST-unique wavelengths) to
earlier-type galaxies at higher redshifts that dominate at 3.56 and
4.44~\micron. Yet, not all of these galaxies are ellipticals, as discussed
below. As the beautiful first NIRCam images already attest, JWST images will
thus see a greater dominance of, and emphasis on, earlier-type galaxies, which
will stand out the most in JWST images \citep[\eg][]{Ferreira2022}. This is in
contrast to the ``Faint Blue Galaxy'' population of actively star-forming
galaxies that have dominated HST's UV--optimally images for the past decades
\citep[\eg][]{Abraham1996b, Driver1995, Windhorst2011}. The morphology of
nearby galaxies can be strongly wavelength dependent \citep[\eg][and references
therein]{TaylorMager2007, Mager2018, Windhorst2002}, especially for the
earlier-type galaxies. We should therefore expect that JWST will put our
studies of ``old galaxies'' in a new light, and provide the first glimpse of the
first galaxies. 

Our 3.5--4.5~\mum\ IGL values in Figure~\ref{fig:fig13} are somewhat below the
\citet{Driver2016b} points, but not significantly so given the current error
budget. As a consequence, to provide a best fit to the total PEARLS IGL in
Figure~\ref{fig:fig13}, we needed to reduce the spheroid contribution in the
\citet{Andrews2018} model to 95\% of its value, and increase their disk 
component by 30\% to match the 3.5--4.5~\mum\ PEARLS points, while decreasing 
the unobscured AGN sky-SB to 75\% of \citet{Andrews2018} model in order to not
over predict the UV-optical IGL values of \citet{Driver2016b} and
\citet{Koushan2021}. To within the current uncertainties, these conclusions
are not unique, but they may point to the need for a more significant fraction 
of red spiral galaxies at 3.4--4.5~\mum, as other recent JWST work has 
suggested \citep[\eg][]{Ferreira2022, Fudamoto2022}. With the new NIRCam images
now at hand, future IGL models may need to include a larger fraction of
(dusty) spirals. 

\begin{figure*}[!hptb]
\vspace*{-1.00cm}
\n\cl{
\includegraphics[width=0.720\txw,angle=-0]{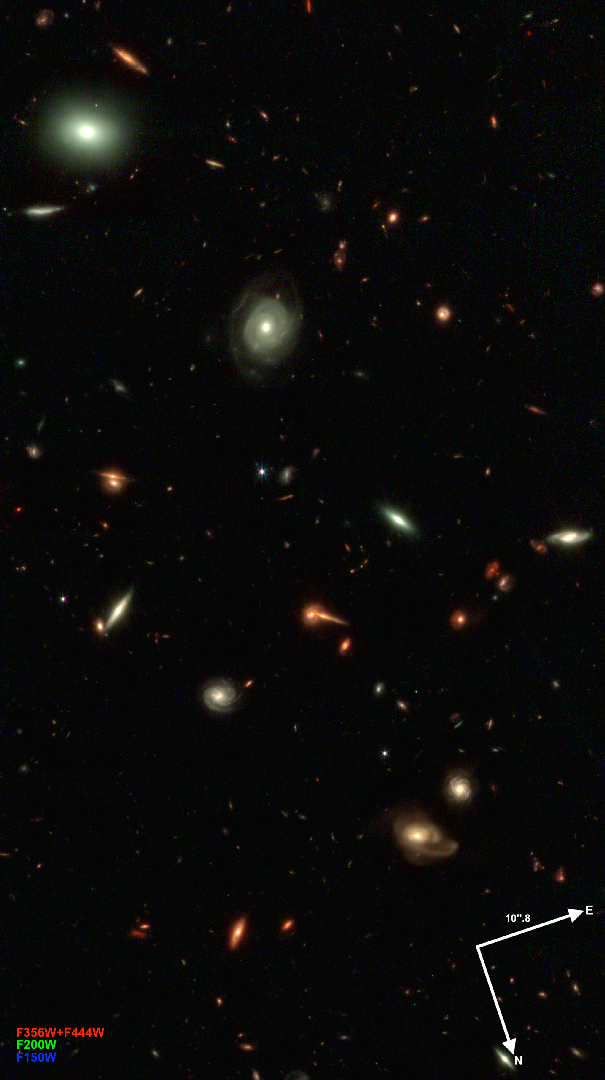}
}

\vspace*{+0.20cm}
\n \caption{
PEARLS NIRCam image of the IRAC Dark Field (JWIDF) Epoch-1 at the north Ecliptic
pole. Filter F150W is rendered as blue, F200W as green, and F356W+F444W as red
using a log scaling \citep[\eg][]{Lupton2004, Coe2015}. This 2040$\times$3644
pixel section covers 61\farcs2$\times$109\farcs3, and image orientation is
shown by the labeled arrows. Areas with remaining wisps and snowball imprints 
were masked before making object catalogs and counts. (Please magnify all PDF 
images to see details). 
}
\label{fig:fig2}
\end{figure*}

\begin{figure*}[!hptb]
\sn\cl{
\includegraphics[width=1.000\txw]{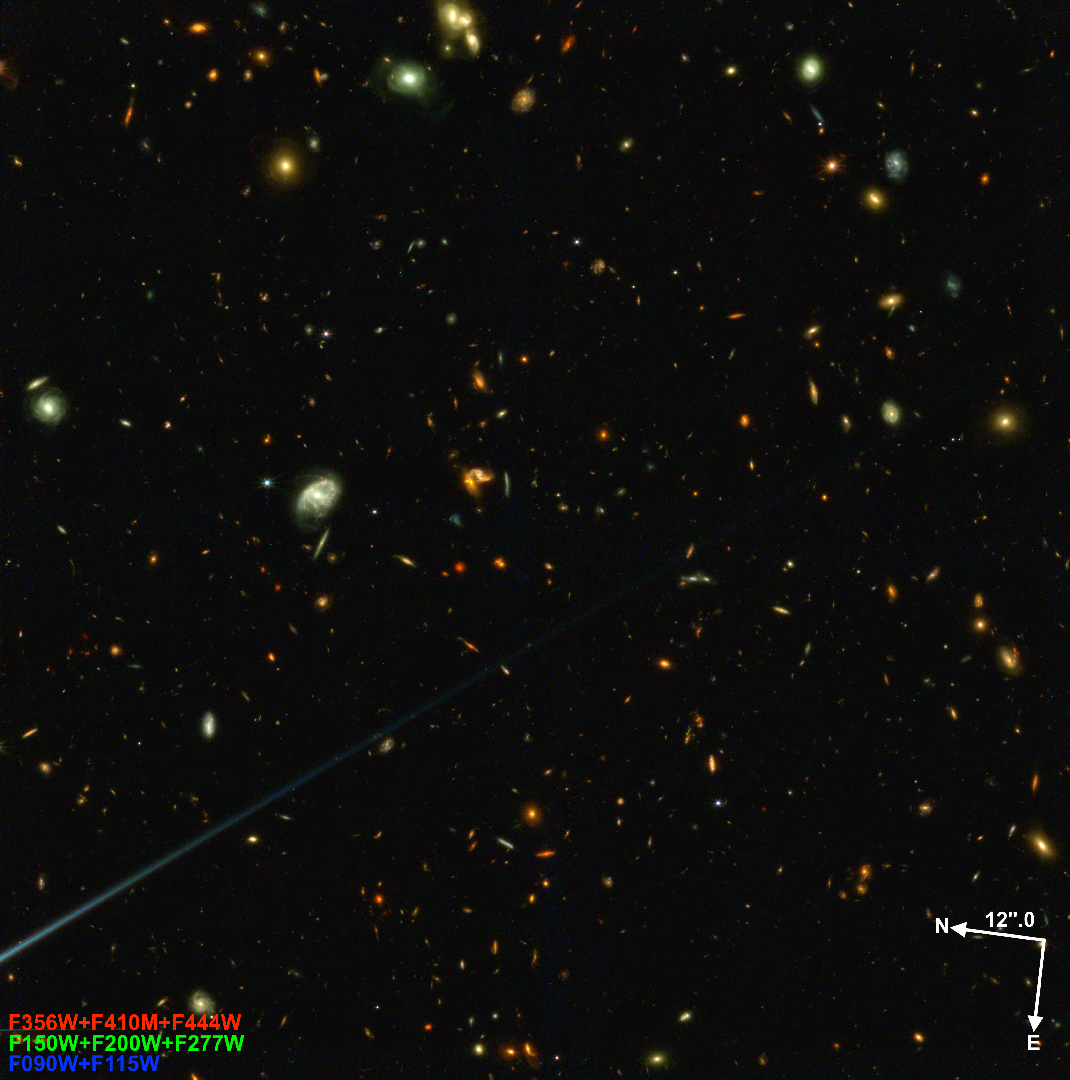}
}

\vspace*{+0.20cm}
\n \caption{
PEARLS NIRCam image of the El Gordo module 3\arcm\ away from the cluster.
Filters F090W+F1115W are rendered as blue, F150W+F200W+F277W as green, and
F356W+F410M+F444W as red. This El Gordo 4466$\times$4424 pixel section covers
134\farcs0$\times$132\farcs7, and image orientation is shown by the labeled
arrows. Areas with remaining detector border effects and bright-star
diffraction spikes (\eg\ the blue spike from a bright star just outside the
lower-left FOV) were masked before making object catalogs and counts. 
}
\label{fig:fig3}
\end{figure*}

\begin{figure*}[!hptb]
\sn\cl{
\includegraphics[width=1.000\txw]{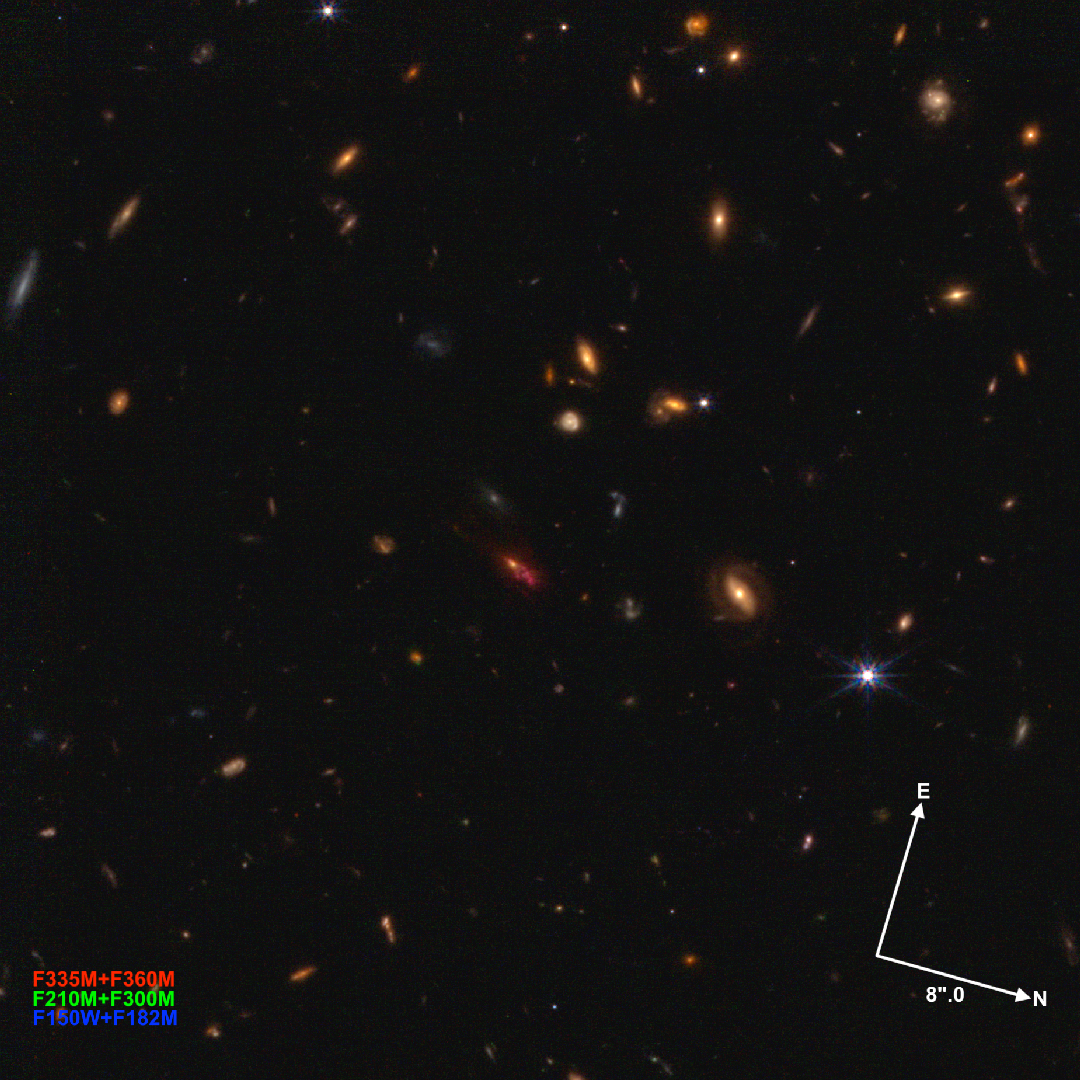}
}

\vspace*{+0.20cm}
\n \caption{
NIRCam image of the $z=4.1$ TNJ1338 protocluster. Filters F150W+F182M are
rendered as blue, F210M+F300M as green, and F335M+F360M as red. This 
1850$\times$1850 pixel section covers 55\farcs5$\times$55\farcs5, and the image
orientation is shown by the labeled arrows. The radio galaxy is the irregular
orange object in the center.
}
\label{fig:fig4}
\end{figure*}

\begin{figure*}[!hptb]
\n\cl{
\includegraphics[width=1.000\txw,angle=+0]{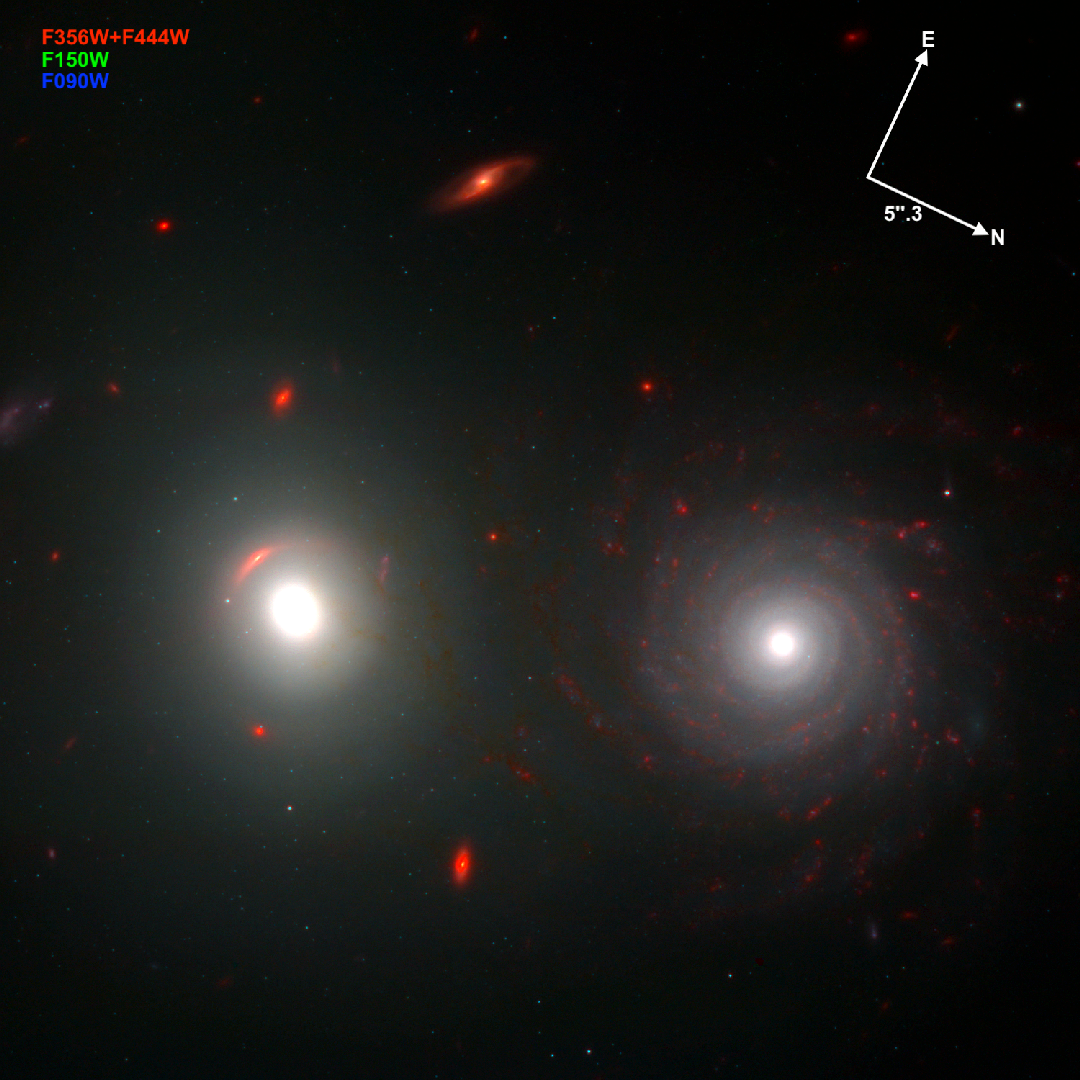}
}

\vspace*{+0.00cm}
\n \caption{
PEARLS NIRCam image of the VV~191 system. Filter F150W is rendered as blue,
F200W as green, and F356W+F444W as red. The elliptical galaxy VV~191a on the 
left backlights the spiral VV~191b on the right. Separation between the nuclei
is 20\farcs4, and the image orientation is shown by the labeled arrows. The
orange arclet south of VV~191a is gravitationally lensed by the bright
elliptical \citep[for details, see][]{Keel2022}. Note the background objects of
similar angular size but different color. This 1500$\times$1500 pixel section 
covers 45\farcs0$\times$45\farcs.
}
\label{fig:fig5}
\end{figure*}

Our 3.5--4.5~\mum\ PEARLS IGL values are \cle 40--50\% below the direct EBL
constraints in Figure~\ref{fig:fig13} from MAGIC \citep[\eg][]{Dwek2013,
Ahnen2015, Ahnen2016}, which are estimated from how intervening EBL photons
distort the $\gamma$-ray spectra of blazars over a range of redshifts. More
recent $\gamma$-ray blazar results are converging closer towards the total IGL
values \citep[\eg][]{FermiLAT2018}. Various sources of diffuse light may cause a
discrepancy between the IGL and the $\gamma$-ray constraints, as discussed by,
\eg\ \citet{Driver2016b}, \citet{Windhorst2018}, \citet{Carleton2022}, and
\citet{Windhorst2022}, and references therein. A possible source of Diffuse
Light are tidal tails of long-lived stars pulled out in galaxy interactions
over the entire redshift range where galaxy assembly happens. For instance,
\citet{Ashcraft2018a} and \citet{Ashcraft2022} analyzed ultradeep (32 hr)
ground-based LBT U-band and r-band images at various stacked seeing-FWHM
values, and find r-band tidal tails in galaxy pairs up to z\cle 0.5--0.9. They
suggest that $\sim$10--20\% of the galaxy light from brighter galaxies
(AB$\sim$20--23 mag, which cause most of the IGL in
Figures~\ref{fig:fig9}--\ref{fig:fig11}) may be at large radii to SB-limits of
AB\cle 31--32 \magarc. It is unlikely then that tidal tails between galaxies
produce well over 20\% of the IGL. Remarkably, though, JWST indeed sees tidal
tails between galaxy pairs and groups in the CEERS images of
\citet[][]{Finkelstein2022}, some of which can be also seen in our JWIDF image
of Figure~\ref{fig:fig2} here. If tidal tails consisting of older stars pulled
out during galaxy interactions are common place, future JWST imaging should
find many more such examples, and be able to better quantify the amount of DL
present in dim tidal tails of faint galaxies. At the median redshift of these
galaxies, NIRCam 0.9--4.5~\mum\ images are ideal for such a study. 

Our 3.5--4.5~\mum\ PEARLS IGL values are a factor of $\sim$2--3 below our
current PEARLS DL constraints, as shown in Figure~\ref{fig:fig13}. At our
reddest wavelengths of 2.7--4.5~\mum, our PEARLS diffuse light limits are about
$\sim$8--12 \nWsqmsr, \ie\ about the same level as the diffuse light level 
suggested by \citet{Lauer2022} at $\sim$51 AU, who found a signal of 8$\pm$2
\nWsqmsr\ at 0.6~\mum. If such diffuse light were caused by tidal tails or
other stellar populations during the history of cosmic star-formation, one may
expect it to have a similar wavelength dependence as the IGL, or be redder,
\ie\ the diffuse light level seen by \citet{Lauer2022} would amount to
$\sim$5--7 \nWsqmsr\ at 3.5 \mum. This is {\ch just} below our current diffuse
light limits, but higher than the PEARLS IGL values in Figure~\ref{fig:fig11}
\& \ref{fig:fig13}. When the JWST calibrations improve over time, and models
for its total sky-SB predictions are improved, future work should be able to
better assess how much truly diffuse light can be present in the infrared, and
what its nature may be. 


\n \section{Summary and Conclusions} \label{sec7} 

\mn In this paper, we present an overview and describe the rationale, methods,
and first results from the JWST GTO project ``PEARLS.'' The following are our
main highlights and results: 

\mn \bul (1) The first PEARLS NIRCam observations are those of the overlapping
galaxy pair VV~191, the radio-selected protocluster at $z=4.1$ around TNJ
1338$-$1942, the massive galaxy cluster known as El Gordo, and the IRAC Dark
Field. 

\mn \bul (2) Star--galaxy classification, object-catalog construction, and
galaxy counting are straightforward in the four fields observed so far
(excluding the areas affected by VV~191, TNJ 1338-1942, and the El Gordo cluster
itself). 

\mn \bul (3) The JWST galaxy counts at 0.9, 1.2, 1.5, 2.0, 3.5, and 4.5~\mum\
wavelengths are consistent with previous ground-based, HST and Spitzer/WISE 
galaxy counts to within \cle 10--20\%, given the combined error budget from ZPs,
filter flux-scale transformations, and Cosmic Variance. Our PEARLS galaxy
counts extend the previous work by \cge 2 mag to AB\cle 28.5--29 mag at
3.5--4.5~\mum\ wavelengths.

\mn \bul (4) The normalized differential galaxy counts, to first order and when
normalized by the converging count slope of 0.4 dex/mag, reach a maximum around
AB$\simeq$20 mag at wavelengths of 0.9--4.5~\mum. This peak corresponds to the
objects that produce most of the IGL. The PEARLS IGL converges to values within
\cle 10\% accuracy at 0.9--4.5~\mum. 

\mn \bul (5) Both the AB-magnitude at which most IGL is produced and the 
width over which the middle 50\% of the IGL is produced depend on wavelength.
This reflects the luminosity function and redshift distribution of the galaxy
populations that dominate each wavelengths.

\mn \bul (6) Our early JWST images, after removing discrete objects brighter
than $\rm AB\simeq29$~mag, yield 0.9--4.5~\mum\ diffuse light limits in good 
agreement with model predictions from Zodiacal light, JWST thermal- and 
straylight, and Diffuse Galactic Light. After removing available model
predictions for these components, and the small extrapolated contribution for
galaxies fainter than 28.5~mag (eEBL), our images provide upper limits to the
amount of diffuse light that may be present. Our best DL limits are in line
with previous work at 1--2~\mum\, and are lower in value at 2.7--4.5~\mum\
wavelengths, because of the much lower total sky-SB and the correspondingly
smaller uncertainties in both the NIRCam sky-SB data and the models at 2.7--4.5
\mum. The search for diffuse light as part of the cosmic infrared background
will become more accurate as JWST gathers many more images across the sky during
its lifetime, and when its calibration and models of the L2 Zodiacal light and
JWST's straylight levels improve. 

\mn \bul (7) During Cycle 1, PEARLS will provide NIRCam images, and some
NIRISS grism or NIRSpec spectra, for another 12 targets, which will be done
through 22 more pointings or epochs, as summarized in Table~2. v1 data products
on the NEP TDF and other targets will become available as soon as we have them. 

\mn With the enormous new range in both flux and wavelength that the JWST
images provide, the community will now have the resources to expand and deepen
the study of the morphology, SED, star-formation rates, masses, dust content,
and extinction at redshifts extending to the epoch of First Light, as well as
better constrain how much diffuse light may be present in the infrared.




\bn {Acknowledgements:}
We dedicate this paper to Karin Valentine, who during her life as Media
Relations Manager at the ASU School of Earth \& Space Exploration was a true
champion of outreach for NASA missions and was so eager to see the first JWST
images. We thank the JWST Project at NASA GSFC and JWST Program at NASA HQ for
their many-decades long dedication to make the JWST mission a success. We
especially thank Tony Roman, the JWST scheduling group and Mission Operations
Center staff at STScI for their continued dedicated support to get the JWST
observations scheduled. We thank Scott Kenyon and John Mackenty for helpful 
discussions. {\ch We thank the referee and also Dr. Jane Rigby for very 
thoughtful suggestions that helped us improve the submitted manuscript.}
This work is based on observations made with the NASA/ESA/CSA James Webb Space
Telescope. The data were obtained from the Mikulski Archive for Space Telescopes
(MAST) at the Space Telescope Science Institute, which is operated by the
Association of Universities for Research in Astronomy, Inc., under NASA contract
NAS 5-03127 for JWST. These observations are associated with JWST programs 1176
and 2738.
RAW, SHC, and RAJ acknowledge support from NASA JWST Interdisciplinary
Scientist grants NAG5-12460, NNX14AN10G and 80NSSC18K0200 from GSFC. Work by
RGA was supported by NASA under award number 80GSFC21M0002. JFB was supported
by grant PHY-2012955 issued by the National Science Foundation. RAB gratefully
acknowledges support from the European Space Agency (ESA) Research Fellowship.
CC is supported by the National Natural Science Foundation of China, No.
11803044, 11933003, 12173045, (in part) by the Chinese Academy of Sciences
(CAS) through a grant to the CAS South America Center for Astronomy (CASSACA),
and science research grants from the China Manned Space Project with NO.
CMS-CSST-2021-A05. CJC acknowledges support from the European Research Council
(ERC) Advanced Investigator Grant EPOCHS (788113). LD acknowledges the research
grant support from the Alfred P. Sloan Foundation (award number FG-2021-16495).
KJD acknowledges funding from the European Union's Horizon 2020 research and
innovation programme under the Marie Sk\l{}odowska-Curie grant agreement No.
892117 (HIZRAD). LF acknowledges funding from the Coordena\c cao de Aperfei\c
coamento de Pessoal de N\'\i vel Superior in Brazil (CAPES). LF acknowledges
support by Grant No. 2020750 from the United States-Israel Binational Science
Foundation (BSF) and Grant No. 2109066 from the United States National Science
Foundation (NSF). BLF thanks the Berkeley Center for Theoretical Physics for
their hospitality during the writing of this paper. MH acknowledges the
support from the Korea Astronomy and Space Science Institute grant funded by
the Korean government (MSIT; Nr. 2022183005) MI acknowledges support from the
National Research Foundation of Korea through grants 2020R1A2C3011091 and
2021M3f7A1084525. PK is supported by NSF grant AST-1908823. RLL is supported by
the National Science Foundation Graduate Research Fellowship under Grant No.
DGE-1610403. WPM acknowledges that support for this work was provided by the
National Aeronautics and Space Administration through Chandra Award Numbers
GO8-19119X, GO9-20123X, GO0-21126X and GO1-22134X issued by the Chandra X-ray
Center, which is operated by the Smithsonian Astrophysical Observatory for and
on behalf of the National Aeronautics Space Administration under contract
NAS8-03060. GM is supported by the Collaborative Research Fund under Grant No.
C6017-20G which issued by the Research Grants Council of Hong Kong S.A.R. MAM
acknowledges the support of a National Research Council of Canada Plaskett
Fellowship, and the Australian Research Council Centre of Excellence for All
Sky Astrophysics in 3 Dimensions (ASTRO 3D), through project number CE17010001.
AKM acknowledges support by the Ministry of Science \& Technology, Israel. IRS
acknowledges support from STFC (ST/T000244/1). LW was supported by grant AST
1817099 issued by the NSF and grant 80NSSC20K0538 issued by NASA. CNAW
acknowledges funding from the JWST/NIRCam contract NASS-0215 to the University
of Arizona. JSBW was supported by the Australian Research Council Centre of
Excellence for All Sky Astrophysics in 3 Dimensions (ASTRO 3D), through project
\#CE170100013. EZ acknowledges funding from the Swedish National Space Agency.
AZ acknowledges support by the Ministry of Science \& Technology, Israel, and
by Grant No. 2020750 from the United States-Israel Binational Science
Foundation (BSF) and Grant No. 2109066 from the United States National Science
Foundation (NSF). 
We also acknowledge the indigenous peoples of Arizona, including the Akimel
O'odham (Pima) and Pee Posh (Maricopa) Indian Communities, whose care and
keeping of the land has enabled us to be at ASU's Tempe campus in the Salt
River Valley, where much of our work was conducted. 

\sn \software{ Astropy: \url{http://www.astropy.org} \citep{Robitaille2013,
Astropy2018};\ IDL Astronomy Library: \url{https://idlastro.gsfc.nasa.gov}
\citep{Landsman1993};\ Photutils:
\url{https://photutils.readthedocs.io/en/stable/} \citep{Bradley20};\
\ProFound: \url{https://github.com/asgr/ProFound} \citep{Robotham2017, 
Robotham2018};\ \ProFit: \url{https://github.com/ICRAR/ProFit}
\citep{Robotham2018};\ \SExtractor: SourceExtractor:
\url{https://www.astromatic.net/software/sextractor/} or
\url{https://sextractor.readthedocs.io/en/latest/} \citep{Bertin1996}.}
\facilities{ Hubble and James Webb Space Telescope Mikulski Archive
\url{https://archive.stsci.edu}.}\ 
{\ch Our specific GTO PEARLS observations were retrieved from MAST at STScI,
and once they become public can be accessed via the following data sets:\ 
VV191:      \dataset[10.17909/dn6q-d483]{https://doi.org/10.17909/dn6q-d483};\ 
IDF-epoch1: \dataset[10.17909/z0aj-zm13]{https://doi.org/10.17909/z0aj-zm13};\ 
El-Gordo:   \dataset[10.17909/mw1y-gb32]{https://doi.org/10.17909/mw1y-gb32};\ 
TNJ1338:    \dataset[10.17909/7gez-0t67]{https://doi.org/10.17909/7gez-0t67};\ 
TDF-spoke1: \dataset[10.17909/d4w2-dz48]{https://doi.org/10.17909/d4w2-dz48}.
}



\vspace*{+0.200cm}
{\footnotesize 
\hspace*{-1.20cm}
\begin{verbatim}
Table 1. PEARLS Targets with NIRCam Images Taken as of 2022 July: Depth from ETC, SourceExtractor and Galaxy Counts
=============================================================================================================================
Instr.+Filters  R.A.  (J2000)  Decl.      Obs. Date  Visit   Area   SCeff Net ------------- Net t_exp (sec) -----------------
Target           h  m  s.sss   o  '   "   YYYY-MM-DD  Nr    (')x(') (%)   Hrs ------- 5-sigma point-source AB-limit ---------
-----------------------------------------------------------------------------------------------------------------------------

NIRCam Broad-band:                                                            F090W F115W F150W F200W F277W F356W F410M F444W
-----------------------------------------------------------------------------------------------------------------------------
VV191-Backlit  13 48 22.0990 +25 40 40.01 2022-07-02 341.1 2.15x4.30 32.0 0.52 0934  ----  0934  ----  ----  0934  ----  0934
PSF-FWHM (")                                                                  0.066  ---- 0.068  ----  ---- 0.164  ---- 0.163
ETC 5sig AB-lim                                                               27.62  ---- 28.01  ----  ---- 28.00  ---- 27.59
Cat 5sig AB-lim                                                               27.88  ---- 28.24  ----  ---- 29.01  ---- 28.81
Counts 80% compl                                                              27.3   ---- 27.6   ----  ---- 28.5   ---- 28.3 
DelABlim(80%-ETC)                                                             -0.3   ---- -0.4   ----  ---- +0.5   ---- +0.7 

IRAC-Dark-ep1  17 40 08.5352 +68 58 27.00 2022-07-08 121.1 2.15x4.30 61.2 1.76 ----  ----  3157  3157  ----  3157  ----  3157
PSF-FWHM (")                                                                   ----  ---- 0.063 0.075  ---- 0.166  ---- 0.164
ETC 5sig AB-lim                                                                ----  ---- 28.96 29.13  ---- 28.81  ---- 28.41
Cat 5sig AB-lim                                                                ----  ---- 28.75 28.93  ---- 29.67  ---- 29.43
Counts 80% compl                                                               ----  ---- 28.1  28.2   ---- 29.0   ---- 29.0 
DelABlim(80%-ETC)                                                              ----  ---- -0.9  -0.9   ---- +0.2   ---- +0.6 

El-Gordo       01 02 55.4000 -49 15 38.00 2022-07-29 241.1 2.15x4.30 57.3 2.50 2491  2491  1890  2104  2104  1890  2491  2491
PSF-FWHM (")                                                                  0.062 0.057 0.062 0.074 0.119 0.171 0.153 0.160
ETC 5sig AB-lim                                                               28.43 28.61 28.60 28.88 28.58 28.55 28.03 28.32
Cat 5sig AB-lim                                                               28.57 28.69 28.57 28.87 29.43 29.55 29.06 29.30
Counts 80% compl                                                              27.9  27.9  27.7  28.1  28.8  28.9  28.1  28.9 
DelABlim(80%-ETC)                                                             -0.5  -0.7  -0.9  -0.8  +0.2  +0.4  +0.1  +0.6 

NIRCam Medium-band:                                                           F115W F150W F182M F210M F300M F335M F360M F444W
----------------------------------------------------------------------------------------------------------------------------- 
TNJ1338-1942   13 38 26.1000 -19 42 28.00 2022-07-01 361.1 2.15x4.30 37.9 0.86 ----  1031  1031  1031  1031  1031  1031  ----  
PSF-FWHM (")                                                                   ---- 0.064 0.071 0.079 0.125 0.169 0.160  ----  
ETC 5sig AB-lim                                                                ---- 27.86 27.51 27.30 27.15 27.25 27.28  ----  
Cat 5sig AB-lim                                                                ---- 27.7  27.4  27.2  28.35 28.25 28.16  ----  
Counts 80% compl                                                               ---- 27.1  26.6  26.4  27.8  27.4  27.3   ----  
DelABlim(80%-ETC)                                                              ---- -0.8  -0.9  -0.9  +0.7  +0.2  +0.0   ----

-----------------------------------------------------------------------------------------------------------------------------
\end{verbatim} 
} 
\label{tab:tab1}
Notes: For each object, line 1 lists the J2000 (RA, Decl.) tangent point to
which the images were drizzled, the observing date, the APT visit number, the
area covered, the net exposure time per filter and the net total hours per
visit, as well as the visit's spacecraft efficiency. Line 2 lists for each
filter the stellar PSF-FWHM in arcsec as measured from unsaturated stars in the
drizzled images. Line 3 lists the 5-sigma point source sensitivity in AB-mag
predicted by the {\it pre-launch} ETC for the net integration time on the first
line of each target. NIRCam ETC Parameters used were aperture radii r=0\arcspt
08 for SW and r=0\farcs16 for LW, and sky annuli r=0\farcs3--0\farcs99 for SW
and r=0\farcs6--1\farcs98 for LW. Line 4 lists the 5-sigma detection limit
derived from the AB-level in Figure~4--6 where the median \SExtractor\ 
catalog flux error is 0.20 mag. Line 5 indicates the AB-level in Figure~4--8
where the galaxy counts are $\sim$80\% complete compared to a power law
extrapolation. Line 6 indicates the difference between the 80\% galaxy count
and predicted ETC 5-sigma point source completeness limits in AB-mag. All
PEARLS NIRCam images have a zeropoint of 28.0865 to convert the flux (in
MJy/sr) in each drizzled 0\farcs0300 pixel to AB-mag. 


\ve 


\vspace*{-1.20cm}
{\footnotesize 
\hspace*{-1.20cm}
\begin{verbatim}
Table 2. PEARLS Targets, Area Covered, Exposure Times, and Depth per Image or Grism: NIRCam, NIRISS, NIRSpec data to be taken
=============================================================================================================================
Instr.+Filters  R.A.  (J2000)  Decl.      Obs. Date  Visit   Area   SCeff Net ------------- Net t_exp (sec) -----------------
Target           h  m  s.sss   o  '   "   YYYY-MM-DD  Nr    (')x(') (%)   Hrs ------- 5-sigma point-source AB-limit ---------
-----------------------------------------------------------------------------------------------------------------------------
NIRCam Broad-band:                                                            F090W F115W F150W F200W F277W F356W F410M F444W
-----------------------------------------------------------------------------------------------------------------------------
IRAC-Dark-ep2   17 40 08.535 +68 58 27.00 2023-01-05 121.2 2.15x4.30 53.9 1.76 ----  ----  2512  2512  ----  2512  ----  2512
                                                                               ----  ---- 28.75 28.93  ---- 28.64  ---- 28.23
IRAC-Dark-ep3   17 40 08.535 +68 58 27.00 2023-07-01 121.3 2.15x4.30 53.9 1.76 ----  ----  2835  2835  ----  2835  ----  2835
                                                                               ----  ---- 28.87 29.04  ---- 28.74  ---- 28.34
NEP-TDF-ep1     17 22 47.896 +65 49 21.54 2023-05-21 111.1 2.15x6.36 64.9 3.49 2920  2920  3350  3350  3350  3350  2920  2920
                                                                              28.61 28.77 28.91 29.09 28.81 28.82 28.07 28.35
NEP-TDF-ep2     17 22 47.896 +65 49 21.54 2022-08-26 112.1 2.15x6.36 64.9 3.49 2920  2920  3350  3350  3350  3350  2920  2920
                                                                              28.64 28.80 28.94 29.11 28.84 28.84 28.09 28.37
NEP-TDF-ep3     17 22 47.896 +65 49 21.54 2022-11-22 113.1 2.15x6.36 64.9 3.49 2920  2920  3350  3350  3350  3350  2920  2920
                                                                              28.62 28.78 28.92 29.10 28.82 28.82 28.05 28.31
NEP-TDF-ep4     17 22 47.896 +65 49 21.54 2023-02-18 114.1 2.15x6.36 64.9 3.49 2920  2920  3350  3350  3350  3350  2920  2920
                                                                              28.60 28.76 28.91 29.09 28.81 28.81 28.04 28.30
WFC3-ERS-Field  03 32 42.397 -27 42 07.93 2023-07-29 131.1 2.15x4.30 63.3 3.48 3779  3779  2491  2491  2491  2491  3779  3779
                                                                              28.63 28.81 28.74 28.92 28.58 28.57 28.07 28.29
MACS0416-24-ep1 04 16 08.900 -24 04 28.70 2022-09-26 211.1 2.15x4.30 64.4 3.72 3779  3779  2920  2920  2920  2920  3779  3779
                                                                              28.70 28.87 28.91 29.10 28.74 28.77 28.24 28.50
MACS0416-24-ep2 04 16 08.900 -24 04 28.70 2022-12-10 212.1 2.15x4.30 64.4 3.72 3779  3779  2920  2920  2920  2920  3779  3779
                                                                              28.72 28.89 28.93 29.13 28.78 28.82 28.29 28.57
MACS0416-24-ep3 04 16 08.900 -24 04 28.70 2023-09-26 213.1 2.15x4.30 63.9 3.61 3779  3350  2920  2920  2920  2920  3350  3779
                                                                              28.70 28.80 28.91 29.10 28.74 28.77 28.17 28.50
Abell2744       00 14 21.200 -30 23 50.10 2023-07-29 221.1 2.15x4.30 62.1 3.25 3350  3350  2491  2491  2491  2491  3350  3350
                                                                              28.53 28.71 28.72 28.92 28.60 28.66 28.17 28.45
MACS1149+22     11 49 36.400 +22 23 59.00 2024-01-23 231.1 2.15x4.30 66.4 3.25 3350  3350  2491  2491  2491  2491  3350  3350
                                                                              28.47 28.66 28.68 28.88 28.56 28.62 28.13 28.40
PLCK-G165.7+67  11 27 15.000 +42 28 31.00 2023-03-25 251.1 2.15x4.30 57.3 2.50 2491  2491  1890  2104  2104  1890  2491  2491
                                                                              28.38 28.56 28.57 28.85 28.55 28.54 28.03 28.34
Clio            08 42 20.893 +01 38 32.66 2023-03-17 261.1 2.15x4.30 52.9 1.74 2491  ----  1890  1890  1890  1890  ----  2491
                                                                              28.30  ---- 28.48 28.67 28.36 28.43  ---- 28.27
RXC-J1212+27    12 12 19.250 +27 33 08.70 2023-01-01 271.1 2.15x4.30 52.9 1.74 2491  ----  1890  1890  1890  1890  ----  2491
                                                                              28.32  ---- 28.51 28.70 28.41 28.43  ---- 28.14
PLCK-G191.24+62 10 44 42.600 +33 50 53.40 2023-04-03 281.1 2.15x4.30 57.3 2.50 2491  2491  1890  2104  2104  1890  2491  2491
                                                                              28.32 28.50 28.51 28.79 28.49 28.49 27.99 28.29

NIRISS Grism:                                                                  ---- G150C G150R F200W  ----  ----  ----  ----
-----------------------------------------------------------------------------------------------------------------------------
NEP-TDF-ep1     17 22 47.896 +65 49 21.54 2023-05-21 111.2 2.22x4.90 58.7 3.49 ----  2835  2835  6456  ----  ----  ----  ----
                                                                               ---- 25.86 25.86 29.53  ----  ----  ----  ----
NEP-TDF-ep2     17 22 47.896 +65 49 21.54 2022-08-23 112.2 2.22x4.90 58.7 3.49 ----  2835  2835  6456  ----  ----  ----  ----
                                                                               ---- 25.86 25.86 29.53  ----  ----  ----  ----
NEP-TDF-ep3     17 22 47.896 +65 49 21.54 2022-11-22 113.2 2.22x4.90 58.7 3.49 ----  2835  2835  6456  ----  ----  ----  ----
                                                                               ---- 25.86 25.86 29.53  ----  ----  ----  ----
NEP-TDF-ep4     17 22 47.896 +65 49 21.54 2023-02-18 114.2 2.22x4.90 58.7 3.49 ----  2835  2835  6456  ----  ----  ----  ----
                                                                               ---- 25.86 25.86 29.53  ----  ----  ----  ----

NIRSpec Prism:                                                                            PRISM                              
-----------------------------------------------------------------------------------------------------------------------------
NDWFS1425+3254  14 25 16.408 +32 54 09.58 2023-04-27 311.1 0.10x0.10 39.9 1.14             4202                              
                                                                                          26.25 
SDSSJ0005-0006  00 05 52.340 -00 06 56.86 2023-07-10 321.1 0.10x0.10 39.9 1.14             4202                              
                                                                                          26.05 
-----------------------------------------------------------------------------------------------------------------------------
Total PEARLS                                            165.66 (')^2 59.5 68.9                                               
-----------------------------------------------------------------------------------------------------------------------------
\end{verbatim}
} 
\label{tab:tab2}
\vspace*{-0.20cm}
Notes: As for Table~1. Obs.Date is the earliest observation date in the Long
Range Plan (LRP) windows on the STSCI website. For the two NIRISS grisms G150C
and G150R we list the 1-sigma continuum sensitivity for unbinned spectral
pixels. For the IFU PRISM observations the NIRSpec ETC suggests a 5$\sigma$
sensitivity at 2~\mum\ for unresolved emission lines with a line flux of
$\sim$1.2$\times$10$^{-17}$ erg/cm$^2$/s, and a 2$\sigma$ sensitivity at
2~\mum\ for a continuum source of (9--10)$\times$10$^{-21}$ erg/cm$^2$/s (or
AB$\simeq$26.05--26.25 continuum mag at 1.5~\mum). The totals on the bottom
line indicate the total area, spacecraft efficiency, and net observing hours
for the entire PEARLS GTO program 1176+2738. 



\vspace*{+0.50cm}
{\footnotesize 
\hspace*{-1.20cm}
\begin{verbatim}
Table 3. Parameters of 0.9-4.5 micron galaxy counts and IGL including JWST PEARLS data
=================================================================================================================
lambda_c  AB_25%  AB_50%   AB_75%  IGL_FWHM  AB_peak  IGL_peak      Tot_IGL_int   IGL_tot AB22slope Err    Filter
micron    AB-mag  AB-mag   AB-mag  AB-mag    AB-mag  W/Hz/m2/d2/mag W/Hz/m2/deg2  nW/m/sr dex/mag   mag    VISTA
# (1)      (2)      (3)      (4)     (5)       (6)      (7)            (8)          (9)     (10)    (11)   (12)
-----------------------------------------------------------------------------------------------------------------
0.883     18.01    20.23    22.46    4.45     20.27     5.847e-28     4.688e-27    10.45    0.206   0.06    z    
        +/-0.03  +/-0.02  +/-0.01  +/-0.02  +/-0.01   +/-0.1%       +/-0.4%      +/-0.04 +/-0.001                
1.020     17.89    20.08    22.37    4.48     19.99     7.404e-28     5.870e-27    11.33    0.231   0.07    y    
        +/-0.03  +/-0.02  +/-0.02  +/-0.02  +/-0.01   +/-0.1%       +/-0.5%      +/-0.06 +/-0.002               
1.250     17.75    19.92    22.21    4.46     19.78     8.895e-28     7.117e-27    11.21    0.225   0.07    j    
        +/-0.03  +/-0.02  +/-0.01  +/-0.02  +/-0.01   +/-0.1%       +/-0.3%      +/-0.04 +/-0.002                
1.650     17.59    19.63    21.85    4.26     19.40     1.233e-27     9.206e-27    10.98    0.234   0.06    h    
        +/-0.02  +/-0.02  +/-0.01  +/-0.02  +/-0.01   +/-0.1%       +/-0.4%      +/-0.04 +/-0.002                
2.150     17.70    19.50    21.50    3.80     19.22     1.590e-27     1.063e-26    9.735    0.226   0.05    k    
        +/-0.02  +/-0.01  +/-0.01  +/-0.01  +/-0.01   +/-0.1%       +/-0.3%      +/-0.10 +/-0.002                
3.540     18.95    20.40    22.15    3.20     19.90     1.464e-27     8.242e-27    4.583    0.205   0.04    IRAC1
        +/-0.01  +/-0.01  +/-0.01  +/-0.01  +/-0.01   +/-0.1%       +/-0.2%      +/-0.09 +/-0.002                
4.490     19.31    20.79    22.43    3.12     20.18     1.200e-27     6.964e-27    3.053    0.189   0.04    IRAC2
        +/-0.01  +/-0.01  +/-0.01  +/-0.01  +/-0.01   +/-0.1%       +/-0.2%      +/-0.05 +/-0.002                   
-----------------------------------------------------------------------------------------------------------------
\end{verbatim}
} 
\label{tab:tab3}
Notes: See Section~\ref{sec46} for definition of these parameters. The formal
fitting errors are listed below each parameter value. Because of the vast
statistics and dynamic range in the combined galaxy counts from AB$\simeq$10--29
mag, the fitting errors are much smaller than the NIRCam ZP- and AB-flux scale
transformation uncertainties of Appendix~\ref{secAppB2}, which are listed in
column 11. These combined errors are therefore used for the IGL parameters of
Figure~\ref{fig:fig11}.


\ve 


\vspace*{-1.00cm}
{\footnotesize 
\hspace*{-1.20cm}
\begin{verbatim}
Table 4a. PEARLS sky-SB: ETC Predictions, JWST Observations, DGL, eEBL, Kelsall 1998 Model, and Diffuse Light limits
=====================================================================================================================
Field/sky-SB  Obs.Date  F090W       F115W       F150W       F200W       F277W       F356W       F410M       F444W
---------------------------------------------------------------------------------------------------------------------
lambda_c (micron):      0.8985      1.1434      1.4873      1.9680      2.7279      3.5287      4.0723      4.571
MJy/(nW/m^2):           3337        2622        2016        1523        1099        849.6       736.2       655.9
---------------------------------------------------------------------------------------------------------------------

VV191-Backlit 2022-07-02      
---------------------------------------------------------------------------------------------------------------------
Rigby22-straylight      0.052496    0.080266    0.073986    0.061728    0.058747    0.038679    0.037862    0.035899
ETC-straylight          0.045678    0.048888    0.043277    0.032939    0.026557    0.019997    0.025215    0.037771
ETC-zodi                0.280842    0.232496    0.199449    0.156536    0.101723    0.094726    0.152429    0.311742
ETC-thermal             0.000000    0.000000    0.000000    0.000000    0.000003    0.000234    0.001226    0.004315 

L2-Zodi-Pred            0.194767    0.164668    0.137195    0.103823    0.068512    0.067173    0.131417    0.297195 
                       (0.008508)  (0.006757)  (0.005367)  (0.004135)  (0.003078)  (0.002592)  (0.004548)  (0.006025)    
1.0*Straylight(R22)     0.052496    0.080266    0.073985    0.061723    0.058747    0.038679    0.037862    0.041315
                       (0.010499)  (0.016053)  (0.014797)  (0.012346)  (0.011752)  (0.007736)  (0.007572)  (0.007180)
DGL-Predict             0.000361    0.000661    0.000809    0.000809    0.001068    0.001509    0.001826    0.002140 
                       (0.000181)  (0.000331)  (0.000405)  (0.000405)  (0.000535)  (0.000755)  (0.000913)  (0.001004)
PEARLS-IGL              0.003198    0.004292    0.005560    0.006817    0.006979    0.005455    0.004803    0.004636
PEARLS-IGL(>28.5)       0.000078    0.000107    0.000139    0.000170    0.000175    0.000136    0.000120    0.000114
---------------------------------------------------------------------------------------------------------------------
Total-Predict-skySB     0.24770     0.24570     0.21213     0.16653     0.12851     0.10773     0.17245     0.34508  
                       (0.01351)   (0.01742)   (0.01575)   (0.01302)   (0.01215)   (0.00819)   (0.00888)   (0.01106)         
---------------------------------------------------------------------------------------------------------------------
PEARLS_observ-skySB     0.2007      ----        0.2273      ----        ----        0.1124      ----        0.3193
                       (0.0252)     ----       (0.0174)     ----        ----       (0.0059)     ----       (0.0132)
---------------------------------------------------------------------------------------------------------------------
Obs-Pred (MJy/sr)      <0           ----        0.0152      ----        ----        0.0047      ----        <0       
Obs-Pred (nW/m^2/sr)   <0           ----        31          ----        ----        4           ----        <0       
DL-upper-limit (nW)    <95          ----       <47          ----        ----       <9           ----       <11
=====================================================================================================================
\end{verbatim}
} 
\label{tab:tab4}
Notes: The top two lines give the effective wavelength of each NIRCam filter,
and the factors needed to convert units of MJy/sr to \nWsqmsr. Obs.Date is the
actual observing date of the PEARLS target. For each target, line 1 gives the
\citet{Rigby2022} straylight in MJy/sr for each filter. Lines 2--4 give
quantities predicted by the JWST-ETC: straylight; Zodiacal Light; and thermal
radiation from a telescope model. Line 5+6 give the Zodiacal foreground
predicted for L2 at the time of the observation from the Spitzer IPAC model and
its uncertainty. All (model) uncertainties are between parentheses, with
details on the error budgets in Appendix~\ref{secAppB}--\ref{secAppC}. Line 7+8
give the adopted straylight level and its multiplier, $f$, following
Section~\ref{sec52}. Line 9+10 give the Diffuse Galactic Light level predicted
by the IPAC IRSA model. Line 11+12 give the IGL levels from Section~\ref{sec46}
and Table~3, as well as the eEBL, \ie\ the IGL fraction that comes from beyond
our typical NIRCam detection limits (AB\cge 28.5 mag). Line 13+14 give the
total predicted NIRcam sky-SB in each, following Equation~\ref{eq:eq5}. Line
15+16 give the observed NIRCam sky-SB in each JWST image measured between the
detected objects \citep[Section~\ref{sec52} and][]{Windhorst2022}. Line 17+18
give the difference between the Observed--Predicted sky-SB in MJy/sr and
\nWsqmsr, respectively. Line 18 gives our 1$\sigma$ upper limit to diffuse
light in \nWsqmsr\ for each observed PEARLS filter, accounting for the full
error budgets in both the Observed and Predicted sky-SB values in
Appendix~\ref{secAppB}--\ref{secAppC}. 


\ve 


\vspace*{-1.00cm}
{\footnotesize 
\hspace*{-1.20cm}
\begin{verbatim}
Table 4b. PEARLS sky-SB: ETC Predictions, JWST Observations, DGL, eEBL, Kelsall 1998 Model, and Diffuse Light limits
=====================================================================================================================
Field/sky-SB  Obs.Date  F090W       F115W       F150W       F200W       F277W       F356W       F410M       F444W
---------------------------------------------------------------------------------------------------------------------
lambda_c (micron):      0.8985      1.1434      1.4873      1.9680      2.7279      3.5287      4.0723      4.571
MJy/(nW/m^2):           3337        2622        2016        1523        1099        849.6       736.2       655.9
---------------------------------------------------------------------------------------------------------------------

JWIDF-epoch1  2022-07-08
---------------------------------------------------------------------------------------------------------------------
Rigby22-straylight      0.052496    0.080266    0.073986    0.061728    0.058747    0.038679    0.037862    0.035899
ETC-straylight          0.045758    0.059683    0.055036    0.043173    0.038475    0.026917    0.030043    0.035379
ETC-zodi                0.175812    0.146354    0.127120    0.102019    0.069139    0.069044    0.114380    0.231812
ETC-thermal             0.000000    0.000000    0.000000    0.000000    0.000003    0.000234    0.001226    0.004315

L2-Zodi-Pred            0.090777    0.077508    0.065341    0.050712    0.033910    0.041297    0.095256    0.220162 
                       (0.008508)  (0.006757)  (0.005367)  (0.004125)  (0.003085)  (0.002592)  (0.004548)  (0.006025)
1.0*Straylight(R22)     0.052496    0.080266    0.073986    0.061727    0.058747    0.038679    0.037862    0.041315 
                       (0.010499)  (0.016053)  (0.014797)  (0.012345)  (0.011754)  (0.007736)  (0.007572)  (0.007180)
DGL-Predict             0.001096    0.002004    0.002454    0.002454    0.003246    0.004577    0.005537    0.006490 
                       (0.000548)  (0.001002)  (0.001227)  (0.001227)  (0.001623)  (0.002289)  (0.002769)  (0.003045)
PEARLS-IGL              0.003198    0.004292    0.005560    0.006817    0.006979    0.005455    0.004803    0.004636
PEARLS-IGL(>28.5)       0.000078    0.000107    0.000139    0.000170    0.000175    0.000136    0.000120    0.000114
---------------------------------------------------------------------------------------------------------------------
Total-Predict-skySB     0.14445     0.15989     0.14192     0.11506     0.09608     0.08492     0.14000     0.27240  
                       (0.01352)   (0.01745)   (0.01579)   (0.01307)   (0.01225)   (0.00847)   (0.00926)   (0.01147)
---------------------------------------------------------------------------------------------------------------------
PEARLS_observ-skySB     ----        ----        0.1672      0.1350      ----        0.0857      ----        0.2433    
                        ----        ----       (0.0107)    (0.0087)     ----       (0.0039)     ----       (0.0101)
---------------------------------------------------------------------------------------------------------------------
Obs-Pred (MJy/sr)       ----        ----        0.0253      0.0199      ----        0.0008      ----        <0       
Obs-Pred (nW/m^2/sr)    ----        ----        51          30          ----        1           ----        <0       
DL-upper-limit (nW)     ----        ----       (38)        (24)         ----       <8           ----       <10

El-Gordo:    2022-07-29   
---------------------------------------------------------------------------------------------------------------------
ETC-straylight          0.039684    0.043976    0.039169    0.029558    0.024191    0.017063    0.019863    0.026836
ETC-zodi                0.233085    0.193308    0.164389    0.126978    0.081585    0.071304    0.106510    0.211112
ETC-thermal             0.000000    0.000000    0.000000    0.000000    0.000003    0.000234    0.001226    0.004315 

L2-Zodi-Pred            0.159645    0.133880    0.110478    0.082512    0.053512    0.047257    0.084041    0.188139 
                       (0.008508)  (0.006757)  (0.005367)  (0.004126)  (0.003080)  (0.002592)  (0.004548)  (0.006025)
0.8*Straylight(R22)     0.041997    0.064213    0.059188    0.049387    0.047033    0.030943    0.030290    0.033052 
                       (0.010499)  (0.016053)  (0.014797)  (0.012346)  (0.011755)  (0.007736)  (0.007572)  (0.007180)
DGL-Predict             0.000248    0.000454    0.000556    0.000556    0.000734    0.001037    0.001255    0.0014710 
                       (0.000124)  (0.000227)  (0.000278)  (0.000278)  (0.000369)  (0.000519)  (0.000628)  (0.000691)
PEARLS-IGL              0.003198    0.004292    0.005560    0.006817    0.006979    0.005455    0.004803    0.004636
PEARLS-IGL(>28.5)       0.000078    0.000107    0.000139    0.000170    0.000175    0.000136    0.000120    0.000114
---------------------------------------------------------------------------------------------------------------------
Total-Predict-skySB     0.20197     0.19865     0.17036     0.13263     0.10146     0.07961     0.11693     0.22709  
                       (0.01351)   (0.01742)   (0.01574)   (0.01302)   (0.01216)   (0.00818)   (0.00886)   (0.01103)  
---------------------------------------------------------------------------------------------------------------------
PEARLS_observ-skySB     0.1671      0.1822      0.1784      0.1387      0.0840      0.0798      0.1243      0.2163   
                       (0.0125)    (0.0110)    (0.0108)    (0.0116)    (0.0038)    (0.0046)    (0.0066)    (0.0090)
---------------------------------------------------------------------------------------------------------------------
Obs-Pred (MJy/sr)      <0          <0           0.0080      0.0061     <0           0.0002      0.0074     <0        
Obs-Pred (nW/m^2/sr)   <0          <0           16          9          <0           0           5          <0        
DL-upper-limit (nW)    <61         <54         <38         <27         <14         <8          <8          <9       
=====================================================================================================================
\end{verbatim}
} 
Notes: These are given below Table~4a.


\ve 


\vspace*{-0.50cm}
{\footnotesize 
\hspace*{-1.20cm}
\begin{verbatim}
Table 5. PEARLS sky-SB: ETC Predictions, JWST Observations, DGL, eEBL, Kelsall 1998 Model, and Diffuse Light limits
======================================================================================================================
Field/sky-SB  Obs.Date  F090W       F115W       F150W       F182M       F210M       F300M       F335M       F360M 
----------------------------------------------------------------------------------------------------------------------
lambda_c (micron):      0.8985      1.1434      1.4873      1.8389      2.0908      2.9818      3.3538      3.6148    
MJy/(nW/m^2)            3337        2622        2016        1630        1434        1005        893.9       829.3
----------------------------------------------------------------------------------------------------------------------

TNJ1338-1942  2022-07-01  
----------------------------------------------------------------------------------------------------------------------
Rigby22-straylight      0.052496    0.080266    0.073986    0.063820    0.061010    0.040995    0.019612    0.038679
ETC-straylight          0.067593    0.075166    0.067042    0.057724    0.052513    0.034902    0.031816    0.033530 
ETC-zodi                0.445802    0.369774    0.317401    0.270570    0.231493    0.142155    0.118576    0.144350 
ETC-thermal             0.000000    0.000000    0.000000    0.000000    0.000000    0.000015    0.000087    0.000321 

L2-Zodi-Pred            0.438665    0.369418    0.303688    0.247075    0.213238    0.120340    0.107623    0.105611
                       (0.008508)  (0.006757)  (0.005367)  (0.004445)  (0.003920)  (0.002833)  (0.002551)  (0.002847)
0.5*Straylight(R22)     0.026248    0.040133    0.036993    0.031910    0.030505    0.020496    0.020354    0.019340
                       (0.010499)  (0.016053)  (0.014797)  (0.012764)  (0.012202)  (0.008192)  (0.007922)  (0.007736)
DGL-Predict             0.001249    0.002285    0.002798    0.002798    0.002798    0.004154    0.007570    0.005385
                       (0.000624)  (0.001142)  (0.001399)  (0.001399)  (0.001399)  (0.002086)  (0.003785)  (0.002693)
PEARLS-IGL              0.003198    0.004293    0.005560    0.006554    0.007014    0.006589    0.005822    0.005326 
PEARLS-IGL(>28.5)       0.000078    0.000107    0.000139    0.000164    0.000175    0.000165    0.000146    0.000133 
---------------------------------------------------------------------------------------------------------------------
Total-Predict-skySB     0.46624     0.41194     0.34362     0.28195     0.24672     0.14517     0.13578     0.13079
                       (0.01353)   (0.01746)   (0.01580)   (0.01359)   (0.01289)   (0.00892)   (0.00936)   (0.00867)
---------------------------------------------------------------------------------------------------------------------
PEARLS_observ-skySB     ----        ----        0.3106      0.2750      0.2301      0.1287      0.1361      0.1380    
                        ----        ----       (0.0186)    (0.0289)    (0.0263)    (0.0074)    (0.0064)    (0.0086)
---------------------------------------------------------------------------------------------------------------------
Obs-Pred (MJy/sr)       -----       -----      <0          <0          <0          <0           0.0003      0.0072 
Obs-Pred (nW/m^2/sr)    <0          <0         <0          <0          <0          <0           0           6       
DL-upper-limit (nW)     -----       -----      <49         <52         <42         <12         <10         <10
=====================================================================================================================
\end{verbatim}
} 
\label{tab:tab5}
Notes: These are given below Table~4a.


\ve 


\vspace*{-0.00cm}
\hspace*{-0.00cm}
\n




\vspace*{-0.00cm}
\hspace*{-0.00cm}
\n




\vspace*{-1.00cm}
\hspace*{-0.00cm}
\n




\vspace*{-1.00cm}
\hspace*{-0.00cm}
\n




\vspace*{-2.50cm}
\hspace*{-0.00cm}
\n


\ve 


\vspace*{-3.000cm}
\hspace*{-0.00cm}
\n\begin{figure*}[!hptb]
\sn\cl{\includegraphics[width=1.000\txw]{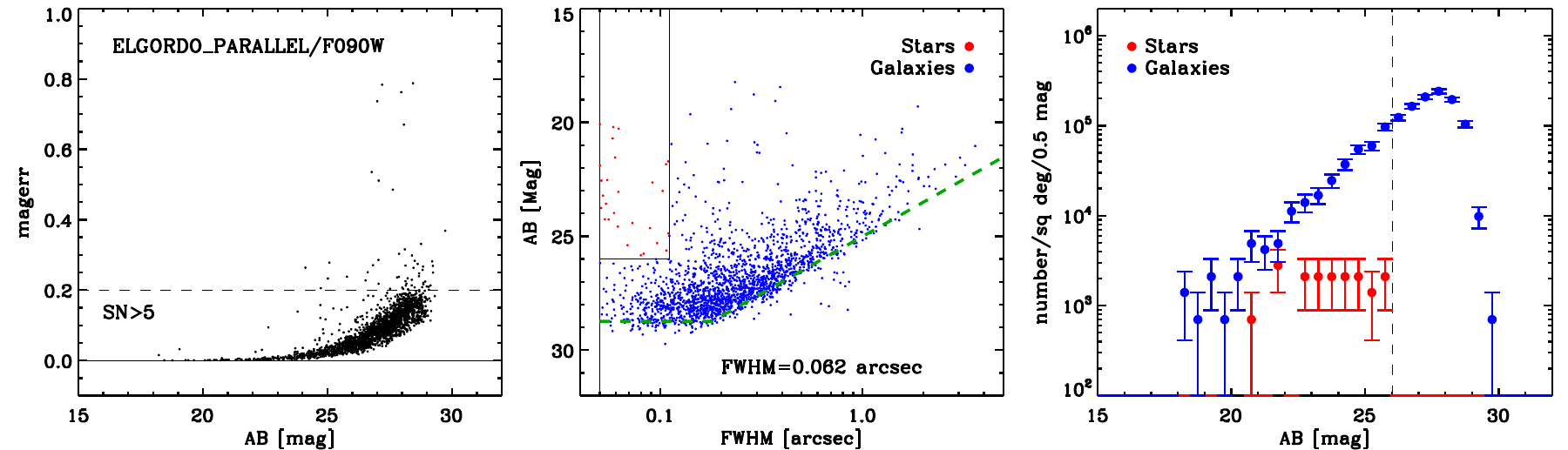}}

\sn\cl{\includegraphics[width=1.000\txw]{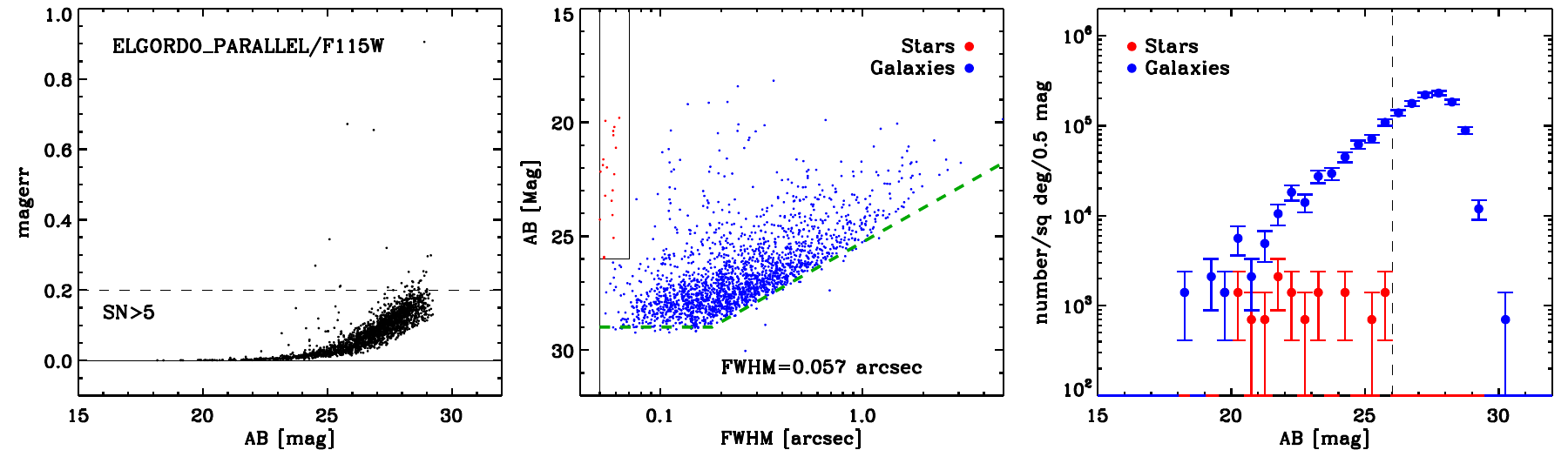}}

\sn\cl{\includegraphics[width=1.000\txw]{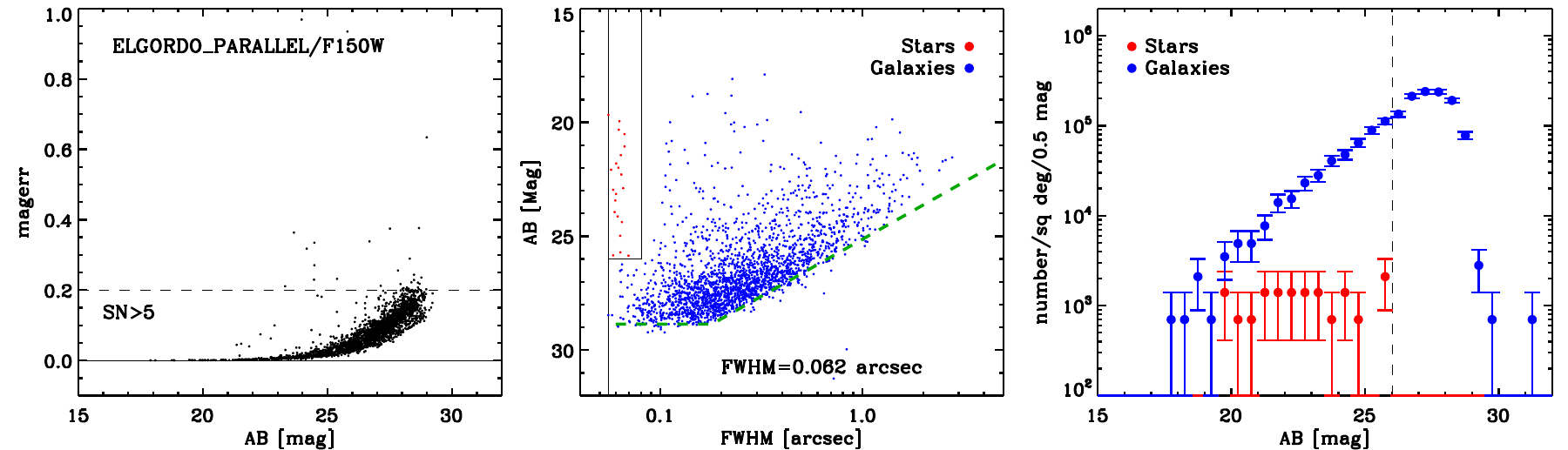}}

\sn\cl{\includegraphics[width=1.000\txw]{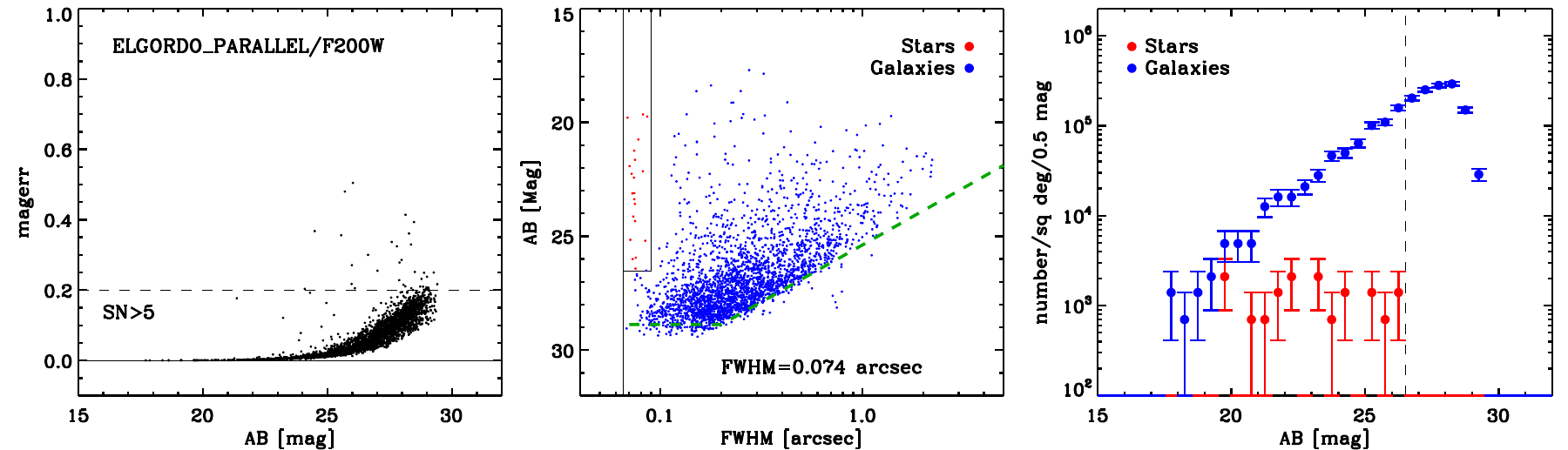}}

\vspace*{+0.10cm}
\n \caption{
Object detection, classification, and counts in the El Gordo non-cluster module
in the F090W, F115W, F150W, and F200W filters. 
{\bf (a) [Left]}:\ \SExtractor\ AB-magnitude error bars vs.\ MAG\_AUTO AB-mag
resulting from the adopted \SExtractor-parameters (Section~\ref{sec31}).
Horizontal dashed lines show the adopted 5$\sigma$ point-source detection 
limits (Table~1). 
{\bf (b) [Middle]}:\ Star--galaxy classification diagram based on \SExtractor\ 
MAG\_AUTO AB-magnitudes vs.\ image FWHM\null. Left solid vertical lines
indicate the NIRCam diffraction limit for each image with its current sampling.
Blue points represent galaxies. The box to the right of the vertical line
identifies objects classified as stars (red points). Objects with FWHM $<$
FWHM(PSF) have been flagged and removed from this plot as spurious detections
or border imperfections. The green dashed lines indicate for each image the
effective point source (horizontal) and SB (slanted) detection limits.
{\bf (c) [Right]}:\ Resulting star counts (red) and galaxy counts (blue).
{\ch The vertical dashed line is the limit to which stellar objects are defined.}
}
\label{fig:fig6}
\end{figure*}


\ve 


\vspace*{-1.50cm}
\hspace*{-0.00cm}
\n\begin{figure*}[!hptb]
\sn\cl{\includegraphics[width=1.000\txw]{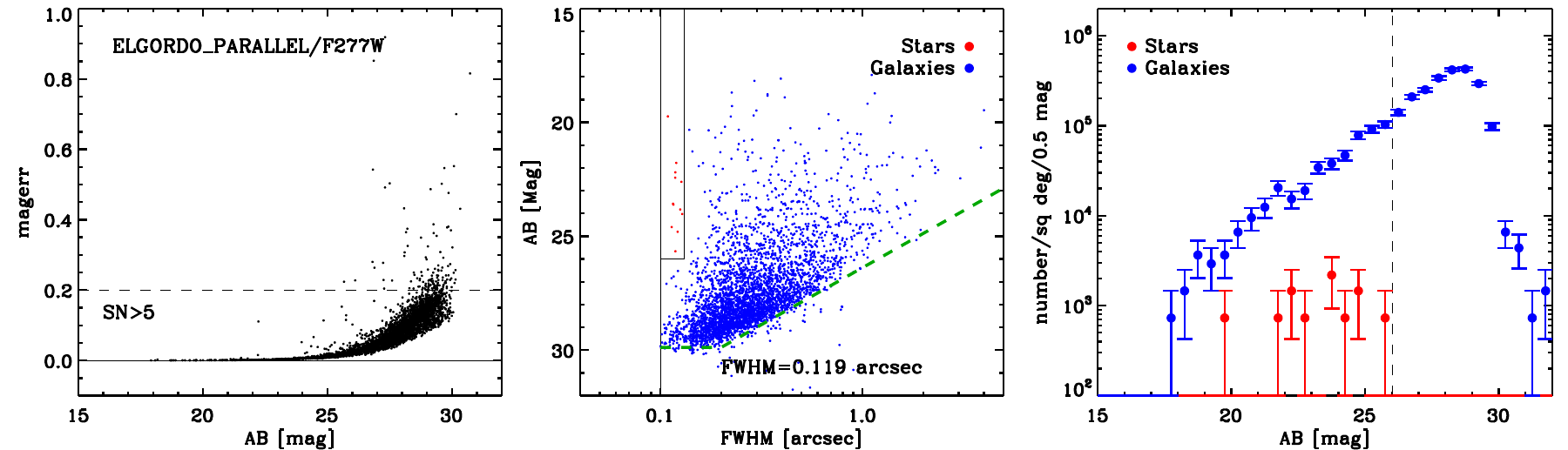}}

\sn\cl{\includegraphics[width=1.000\txw]{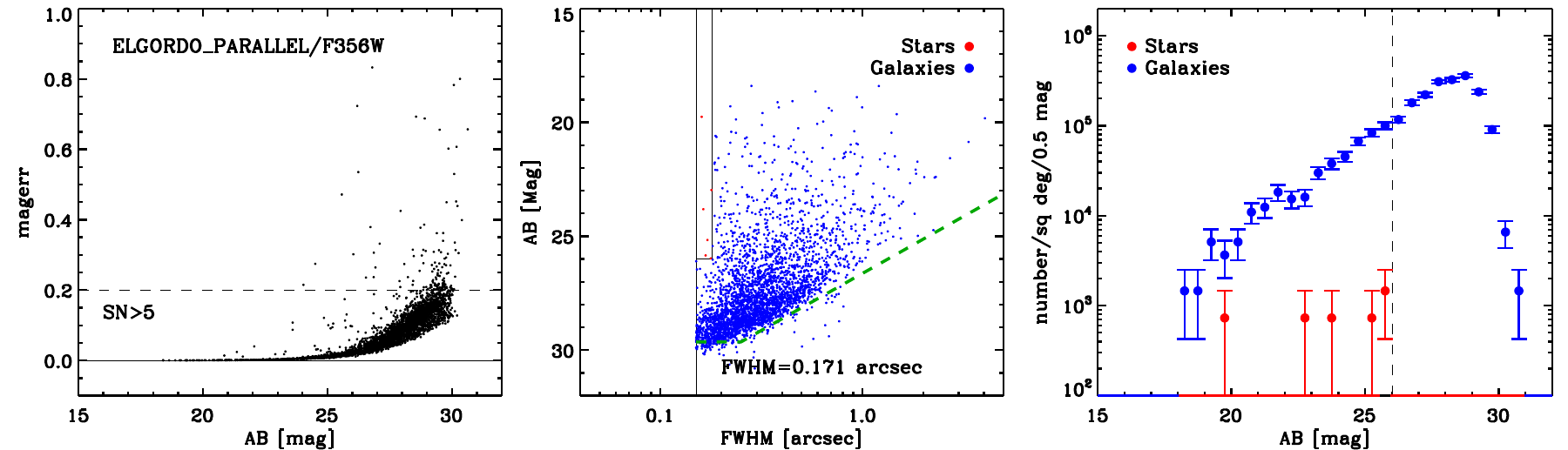}}

\sn\cl{\includegraphics[width=1.000\txw]{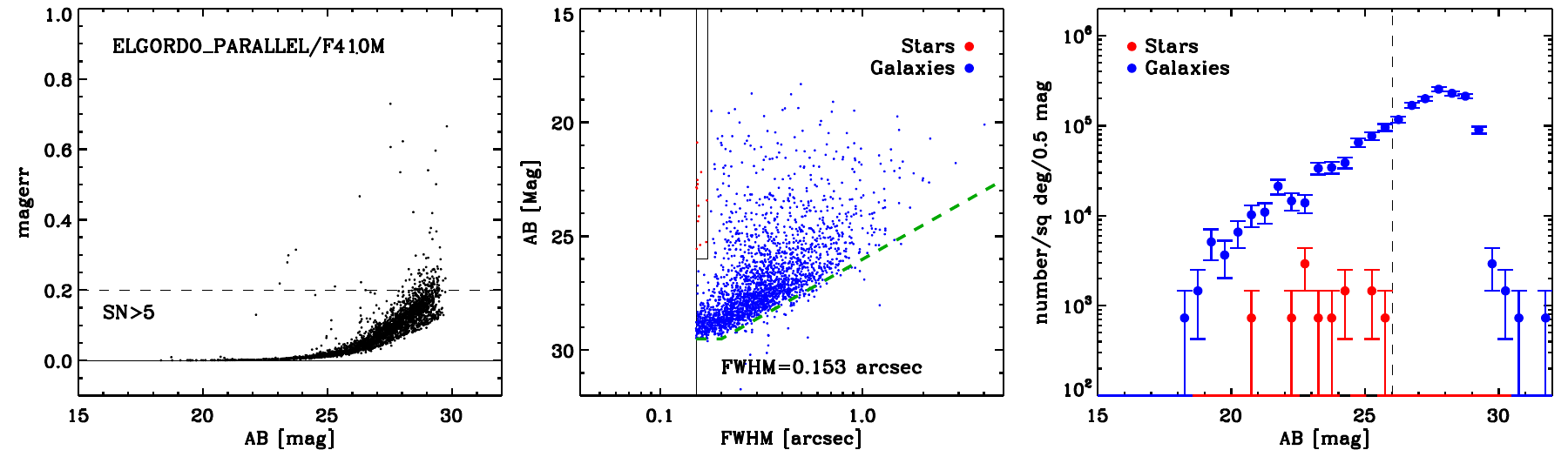}}

\sn\cl{\includegraphics[width=1.000\txw]{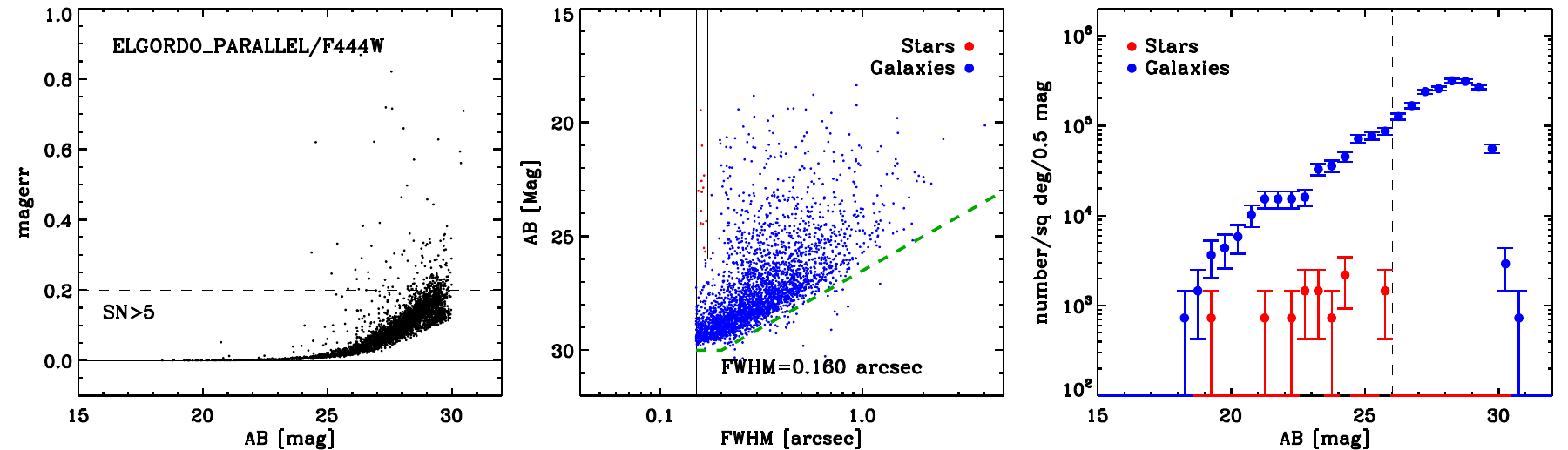}}

\vspace*{+0.10cm}
\n \caption{
Same as Figure~\ref{fig:fig6} for the object detection, classification, and
counts in the El Gordo non-cluster module in the F277W, F356W, F410M, and 
F444W filters. (Please magnify these PDF plots to see all data points.)
}
\label{fig:fig7}
\end{figure*}

\ve 


\vspace*{-1.50cm}
\hspace*{-0.00cm}
\n\begin{figure*}[!hptb]
\sn\cl{\includegraphics[width=1.000\txw]{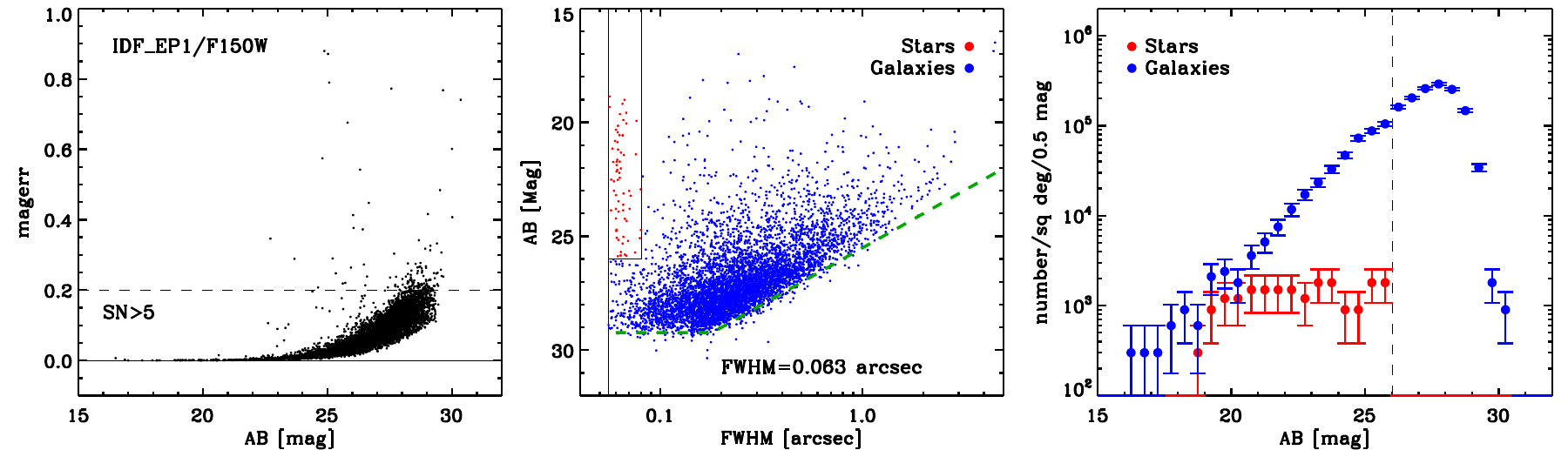}}

\sn\cl{\includegraphics[width=1.000\txw]{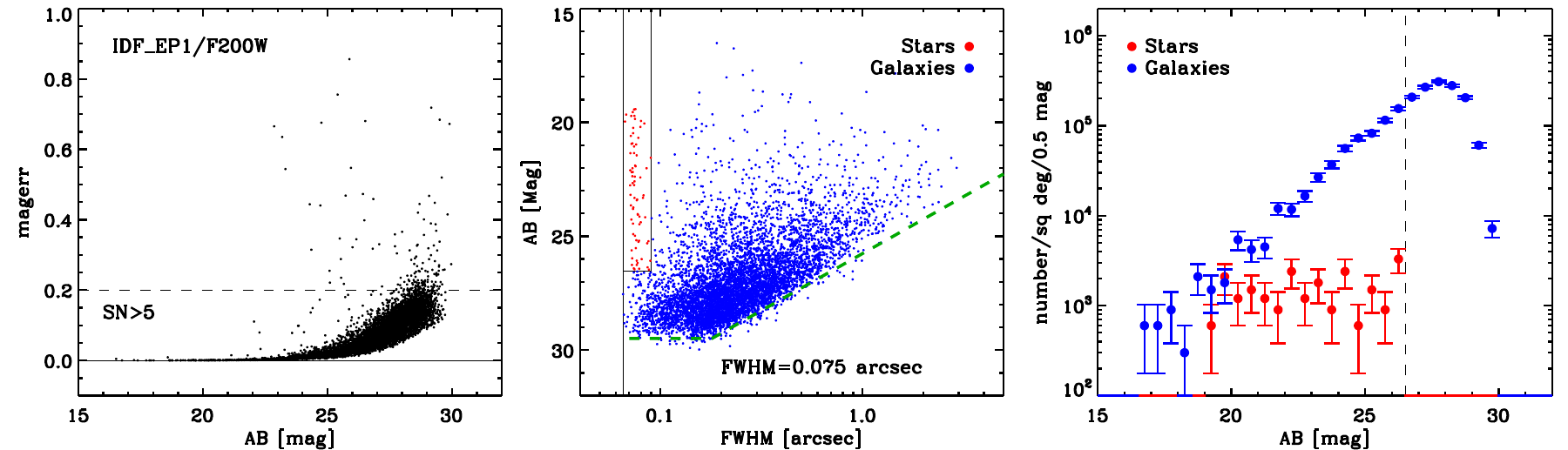}}

\sn\cl{\includegraphics[width=1.000\txw]{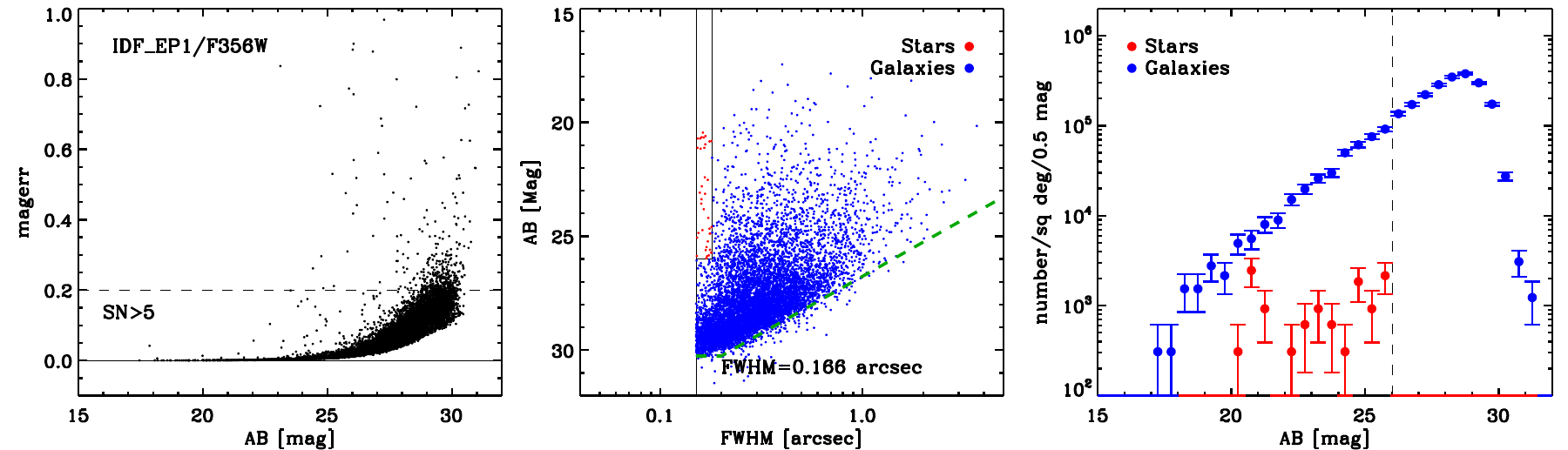}}

\sn\cl{\includegraphics[width=1.000\txw]{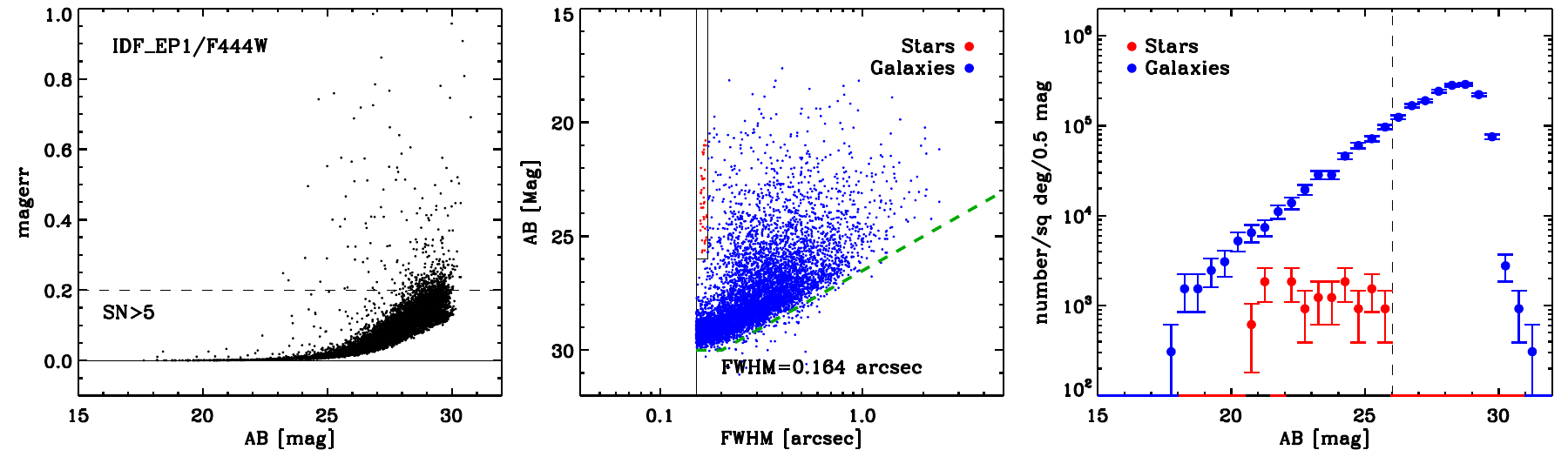}}

\vspace*{+0.10cm}
\n \caption{
Same as Figure~\ref{fig:fig6} for the object detection, classification, and
counts for all detectors covering the JWIDF in the F150W, F200W, F356W, and
F444W filters. (Please magnify these figures to see all data points.)
}
\label{fig:fig8}
\end{figure*}



\vspace*{-1.00cm}
\hspace*{-0.00cm}
\n\begin{figure*}[!hptb]
\vspace*{-1.00cm}
\n\cl{
\includegraphics[width=1.100\txw]{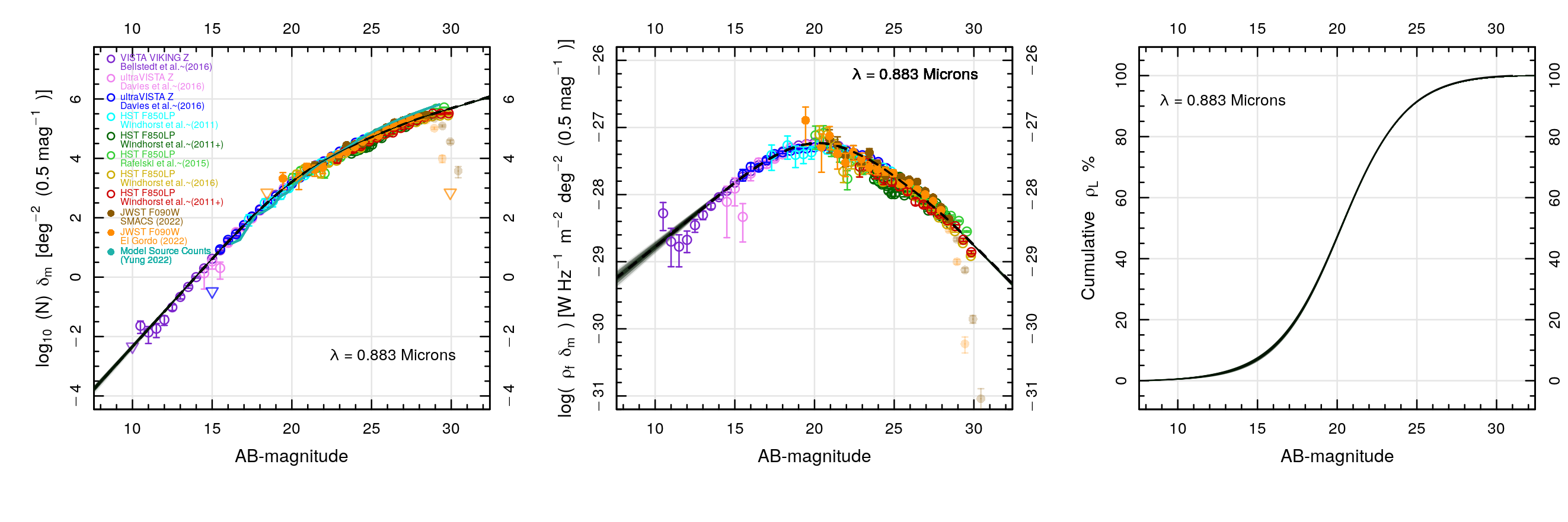}
}

\vspace*{-0.50cm}
\n\cl{
\includegraphics[width=1.100\txw]{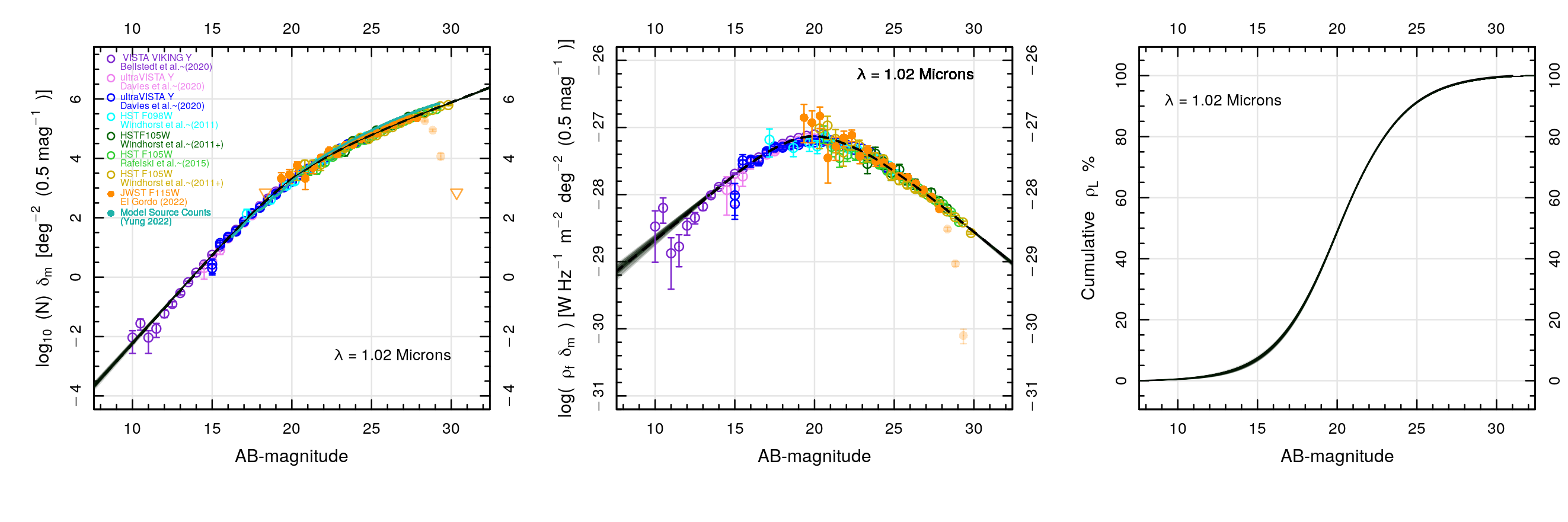}
}

\vspace*{-0.50cm}
\n\cl{
\includegraphics[width=1.100\txw]{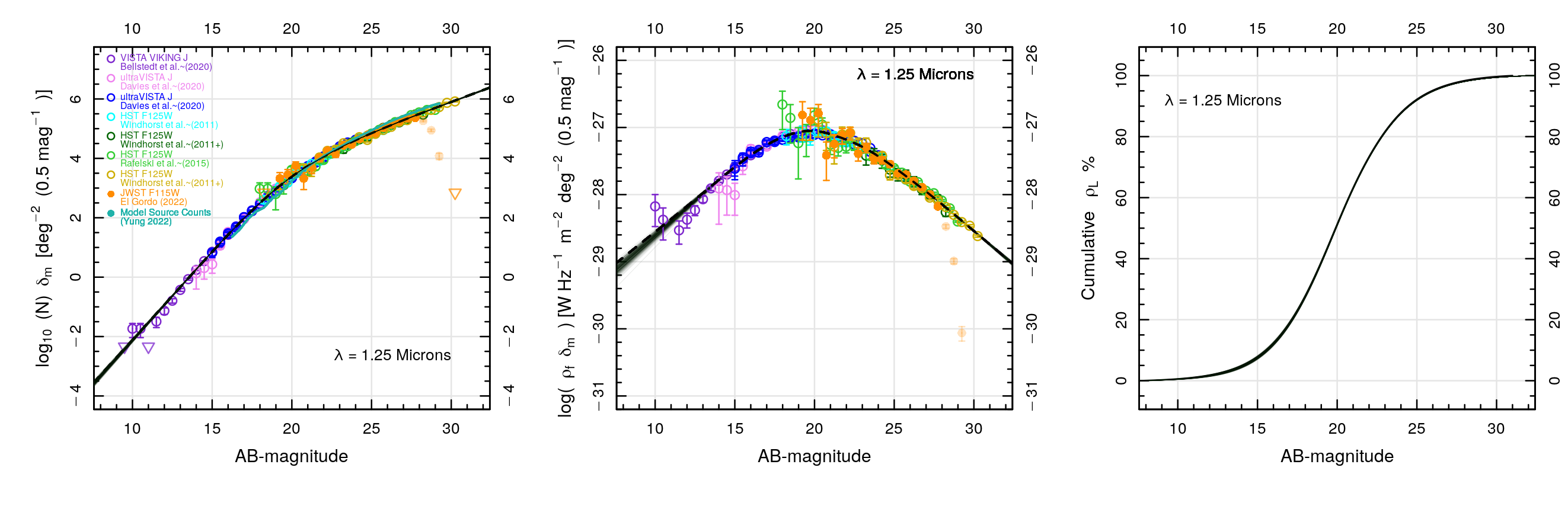}
}

\vspace*{-0.50cm}
\n \caption{
NIRCam galaxy counts in the JWIDF and the El Gordo non-cluster module (orange
and brown filled circles, respectively). Each row of three panels shows one
wavelength as indicated in the panels. 
{\bf (a) [Left panels]:}\ Differential galaxy counts in 0.5-mag bins. Open
diamonds show a combination of previous ground-based, HST, and Spitzer/WISE
counts \citep{Driver2016b, Koushan2021} with different surveys shown in
different colors as indicated in the legends. The green lines represent the
hierarchical-model predictions for the 0.9--4.5~\mum\ galaxy counts of
\citet{Yung2022}. When broad-band filters in different instruments are similar
but not identical, small corrections for effective wavelength differences may
be needed \citep[][see Appendix~\ref{secAppB2} here]{Koushan2021,
Robotham2020}. 
{\bf (b) [Middle panels]:}\ Energy counts after dividing the left panels by a
0.40~dex/mag slope. Units used are described in Section~\ref{sec4} and
\citet{Koushan2021}. Triangles without error bars indicate bins having only a
single object. The brightest bins of the PEARLS counts at 18\cle AB\cle 20 mag
show Cosmic Variance (Section~\ref{sec2} \& \ref{sec45}), but do not weigh into
the IGL fits, which at these flux levels are dominated by the faint end of the
brighter surveys. PEARLS counts beyond the respective 80\% completeness limits
(as derived in Table~1 from a best-fit power-law extrapolation) are plotted as
lightly shaded points, and are not included in the spline extrapolations to
estimate the total IGL.
{\bf (c) [Right panels]:}\ Integral of the middle panels normalized to 100\% of
the IGL energy received. 
}
\label{fig:fig9}
\end{figure*}




\vspace*{-2.50cm}
\hspace*{-0.00cm}
\n\begin{figure*}[!hptb]
\vspace*{-0.30cm}
\n\cl{
\includegraphics[width=1.100\txw]{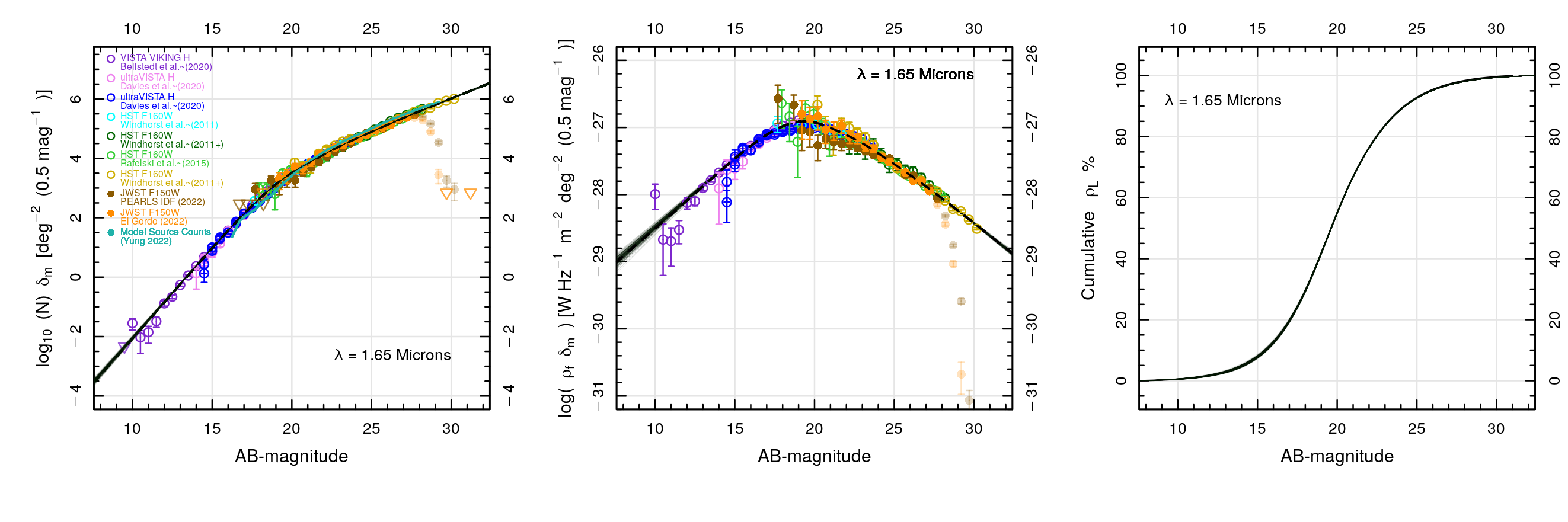}
}

\vspace*{-0.50cm}
\n\cl{
\includegraphics[width=1.100\txw]{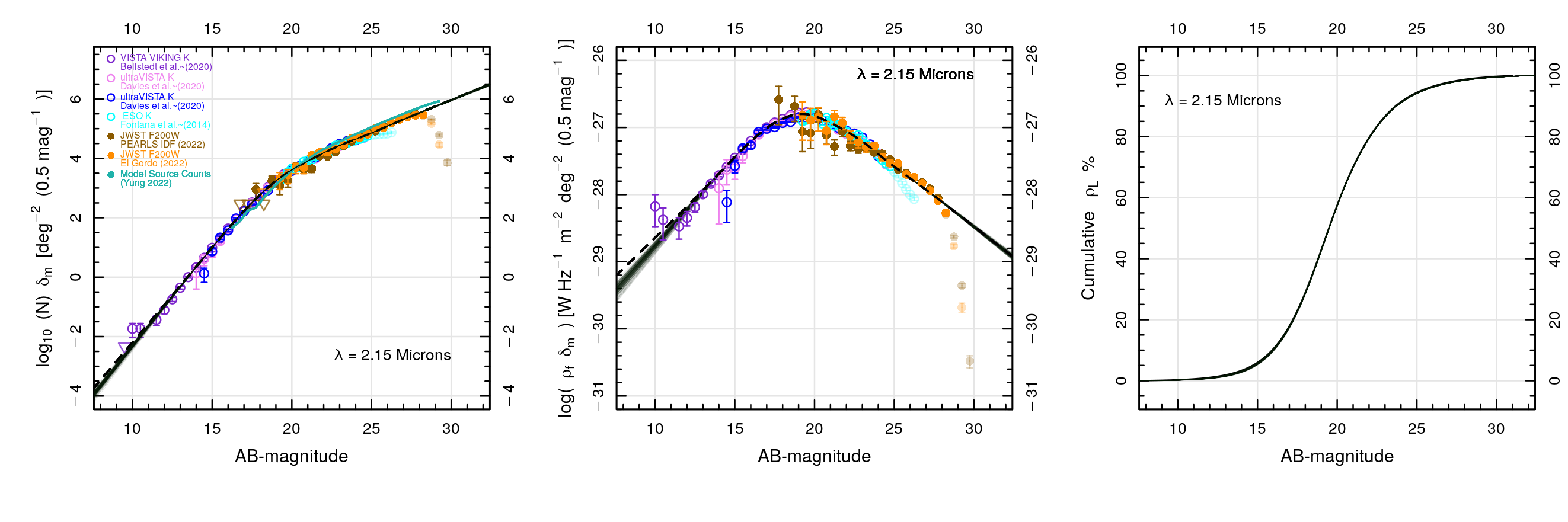}
}

\vspace*{-0.50cm}
\n\cl{
\includegraphics[width=1.100\txw]{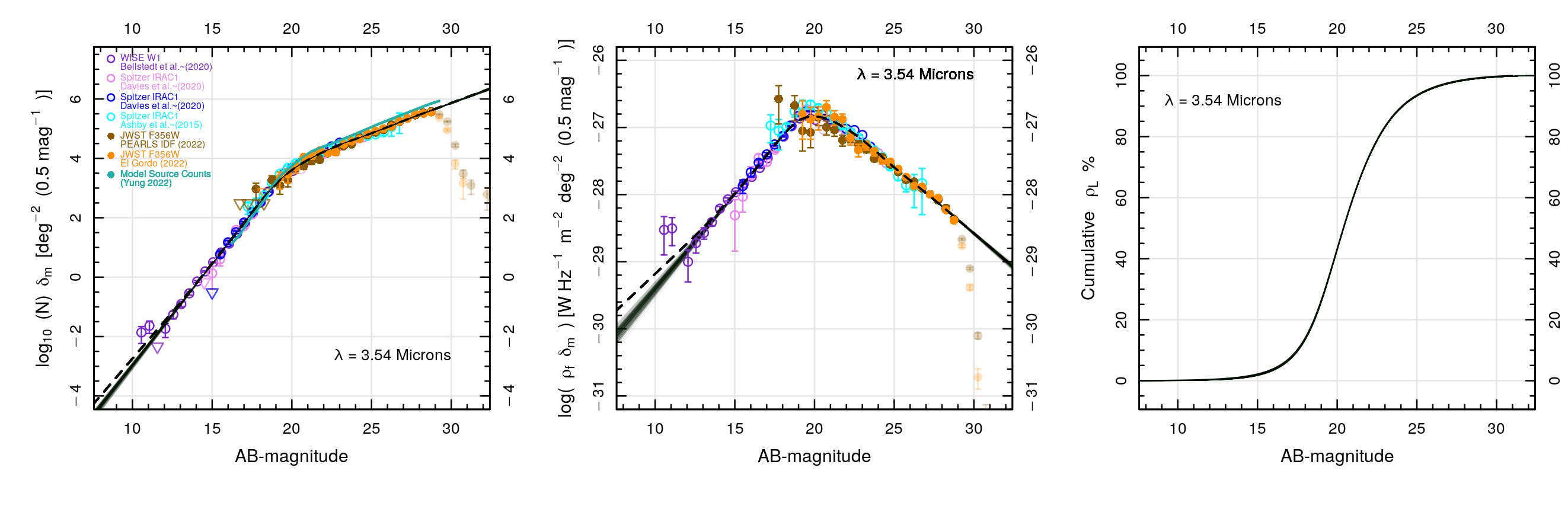}
}

\vspace*{-0.50cm}
\n\cl{
\includegraphics[width=1.100\txw]{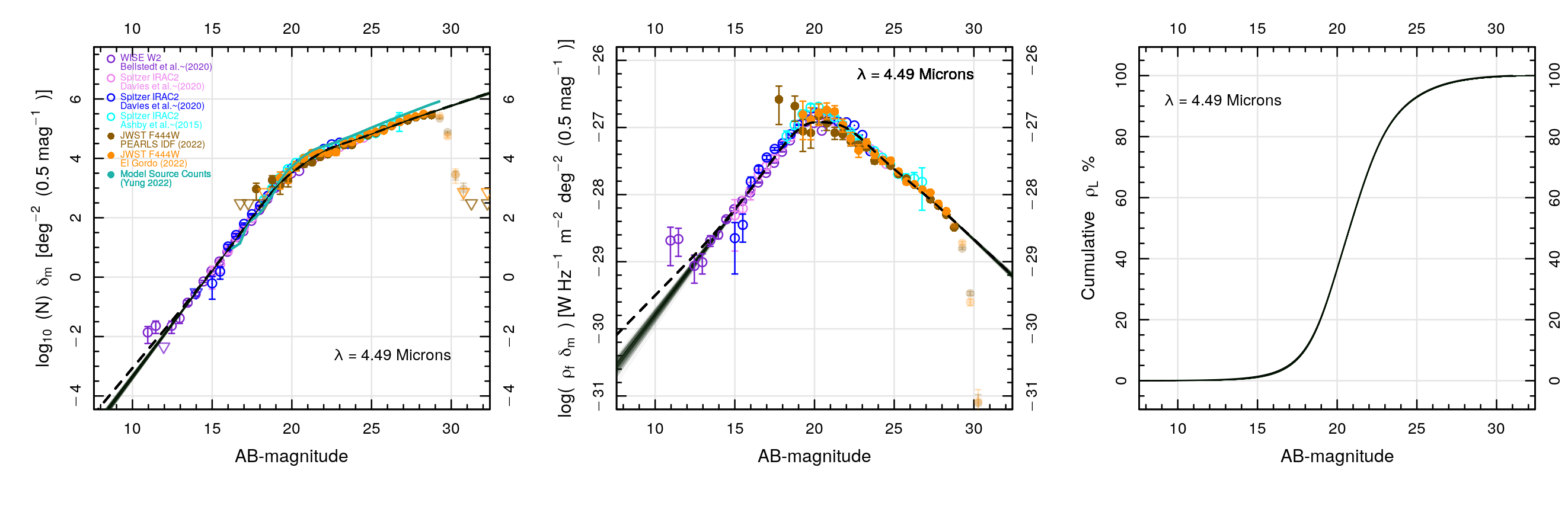}
}

\vspace*{-0.70cm}
\n \caption{
Same as Figure~\ref{fig:fig9} for the ground-based+HST 1.6 \& 2.2~\mum\ and
Spitzer 3.5 \& 4.5~\mum\ counts with JWST NIRCam counts in F150W, F200W, 
F356W and F444W as brown/olive filled circles in the JWIDF and El Gordo 
non-cluster module. 
}
\label{fig:fig10}
\end{figure*}



\vspace*{-0.00cm}
\hspace*{-0.00cm}
\n\begin{figure*}[!hptb]
\n\cl{
\includegraphics[width=0.980\txw]{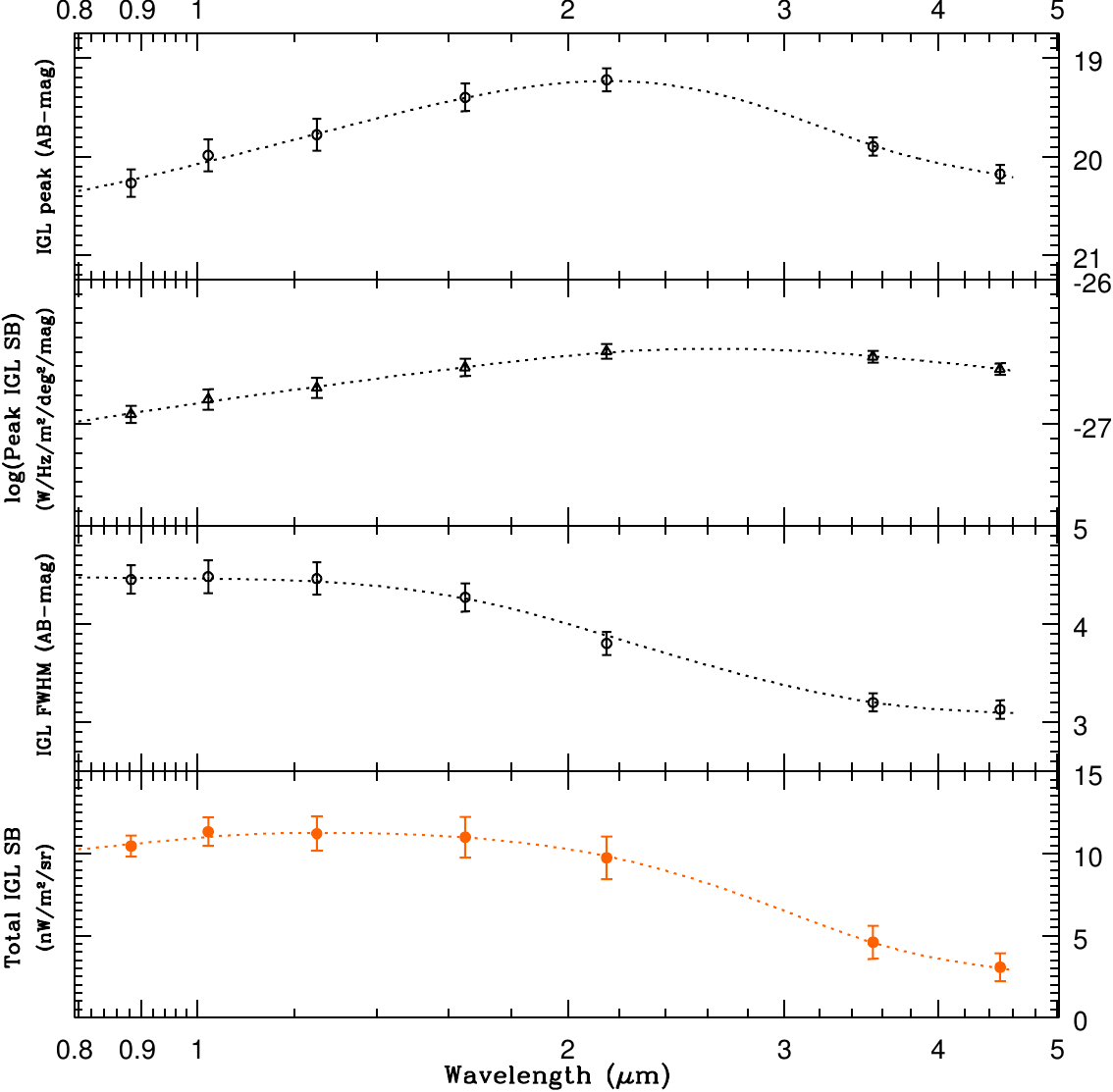}
}

\vspace*{+0.10cm}
\n \caption{
Parameters of the 0.9--4.5~\mum\ galaxy counts and integrated galaxy light
(IGL) as derived from Figures~\ref{fig:fig9}--\ref{fig:fig10}. Top panel shows
the AB-magnitude level at which the normalized differential counts peak (see
middle panels in Figures~\ref{fig:fig9}--\ref{fig:fig10}). This is where most
of the discrete IGL is generated at each wavelength. Second panel shows the peak
SB-value of the IGL derived from the middle panels of
Figures~\ref{fig:fig9}--\ref{fig:fig10}. Third panel shows the width around
the peak magnitude or interquartile (\ie\ 25\%--75\%) range where 50\% of the
discrete IGL is generated. We refer to this range as the ``IGL FWHM\null.'' The
bottom panel shows the total IGL values (in units of \nWsqmsr) of the
converging integrals in the right panels of 
Figures~\ref{fig:fig9}--\ref{fig:fig10}. Error bars were determined by combining
the NIRCam ZP uncertainties of Appendix~\ref{secAppB1} with the uncertainties
in transforming the NIRCam flux scale to the fiducial flux scale of the
VISTA/IRAC filters in Appendix~\ref{secAppB2} (see 
Sections~\ref{sec45}--\ref{sec46}).
}
\label{fig:fig11}
\end{figure*}




\vspace*{-1.00cm}
\hspace*{-0.00cm}
\n\begin{figure*}[!hptb]
\n\cl{
\includegraphics[width=1.050\txw]{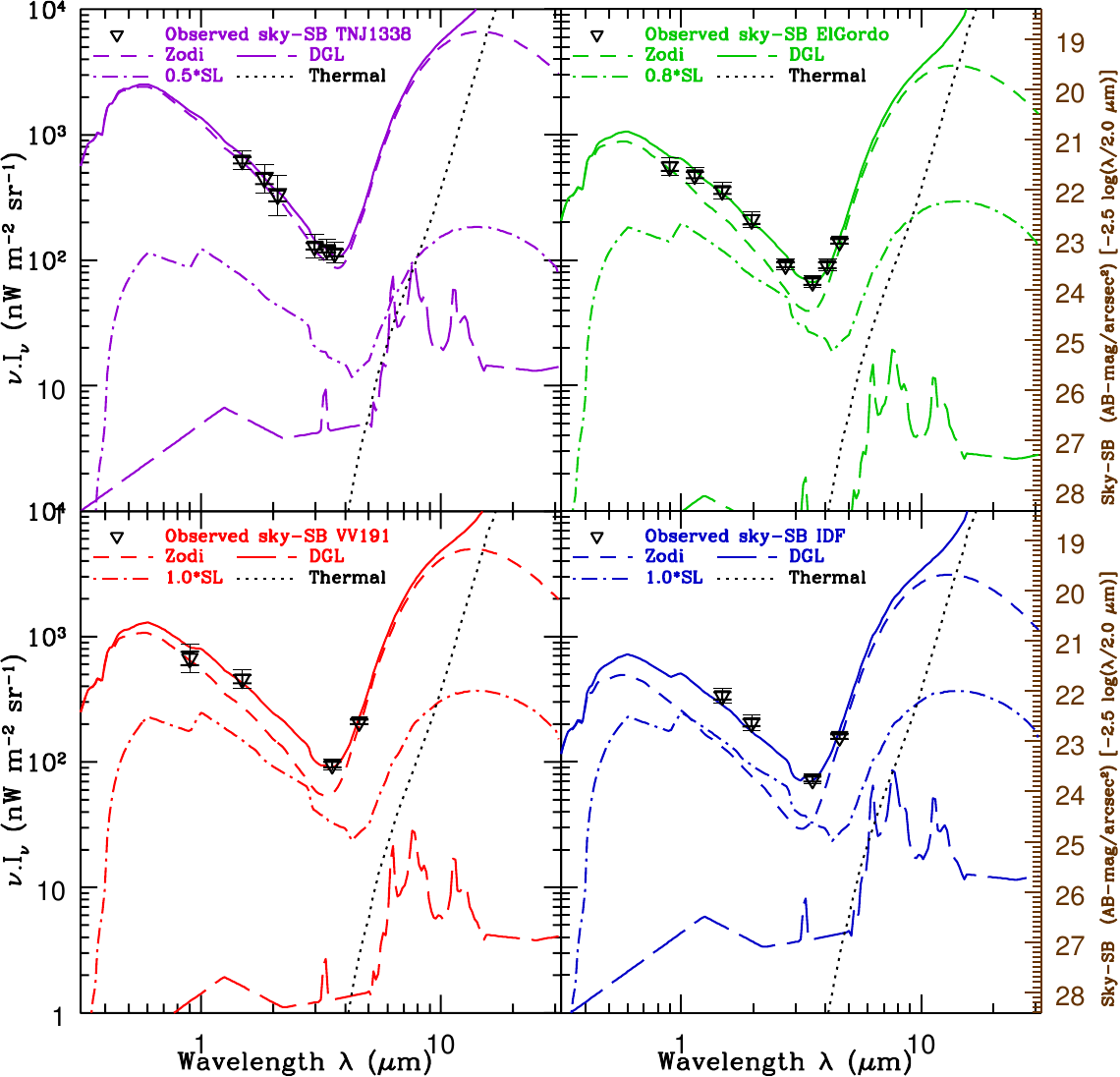}
}

\vspace*{+0.10cm}
\n \caption{
Comparison of the JWST NIRCam sky-SB values observed in our four PEARLS fields 
to models. Black triangles indicate the PEARLS sky-SB measurements of
Section~\ref{sec5}. The smaller (inner) error bars reflect the variation of the
object-free sky-SB measurements within the NIRCam detectors, while the larger
error bars reflect the median value of all sky-rms values across the images in
each filter. These error bars include the ZP and other sky-SB uncertainties of
Appendix~\ref{secAppB3}. Models are plotted for: TNJ1338 (upper left in
purple), El Gordo (upper right in green), VV191 (lower left in red), and the
JWIDF (lower right in blue). Short-dashed lines indicate the Zodiacal sky-SB
from L2 as predicted by the Spitzer model (Appendix~\ref{secAppC}). Zodiacal
light is the highest amplitude component for all targets. Dot-dashed lines
indicate the JWST straylight (SL) model from Figure~4 of \citet{Rigby2022};
long-dashed lines indicate the DGL levels predicted by the IRSA model
(Appendix~\ref{secAppC}); black dotted lines indicate the JWST thermal
contributions to the sky-SB predicted by the ETC; solid colored lines are the
sum of all four components. The SL level in each field was scaled down by
factors {\ch $f$$\simeq$0.5--1.0} compared to the \citet{Rigby2022} SL
amplitude to obtain a best fit of the sum of the four model components to our
3.5--4.5~\mum\ observations, where the sky-SB is lowest. Details are given in
Section~\ref{sec5} and Appendix~\ref{secAppC}. 
}
\label{fig:fig12}
\end{figure*}


\vspace*{-0.50cm}
\hspace*{-0.00cm}
\n\begin{figure*}[!hptb]
\n\cl{
\includegraphics[width=1.00\txw]{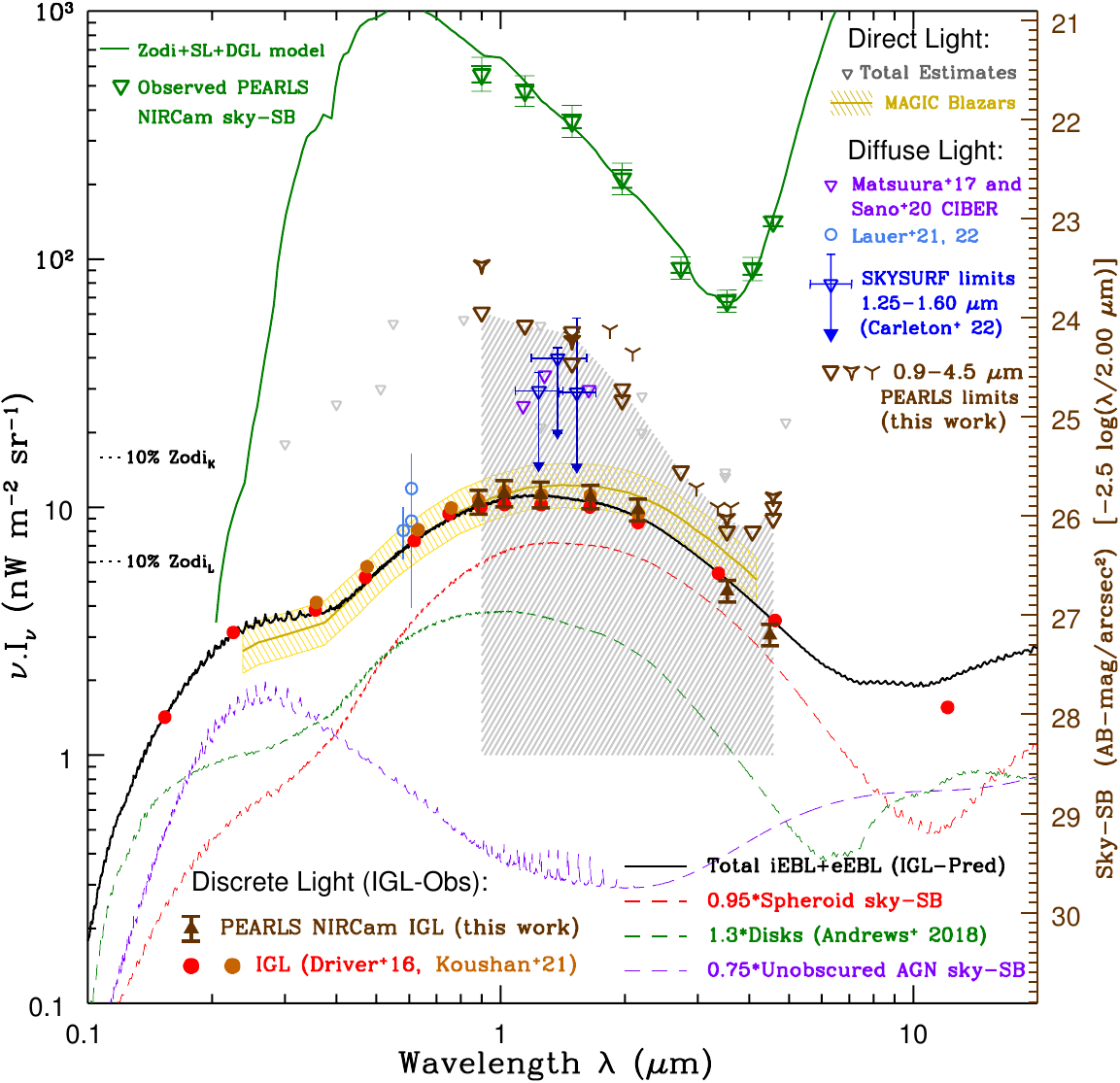}
}

\vspace*{+0.10cm}
\n \caption{
Summary of astrophysical foreground and background energy relevant to PEARLS.
Dark brown filled {\it upward-pointing} triangles with error bars indicate the
PEARLS NIRCam 0.9--4.5~\mum\ IGL measurements of Section~\ref{sec4} (Discrete
Light), and are lower limits to the total EBL. Dark brown {\it
downward-pointing} triangles with the grey error wedge indicate our current
JWIDF and El Gordo non-cluster 0.9--4.5~\mum\ {\it upper limits} to Diffuse
Light, with all known components subtracted (see error budget in
Section~\ref{sec5} and Appendix~\ref{secAppC}), and are in line with previous
limits to Diffuse Light. (Brown starred and tripod symbols indicate the less
accurate VV191 and TNJ limits). Green triangles show our 0.9--4.5~\mum\ PEARLS
NIRCam sky-SB observations in the El Gordo non-cluster module compared to the
models of Section~\ref{sec5} and Figure~\ref{fig:fig12} (green solid line). The
left scale indicates the total energy $\nu$.\Inu\ in \nWsqmsr, and the right
scale shows the corresponding sky-SB in AB-\magarc\ at 2.00~\mum\ (which can be
scaled to other wavelengths as indicated). Filled circles show previous IGL
counts of \citet{Driver2016b} (red) and \citet{Koushan2021} (orange). Solid
and dashed colored lines show the (component and total) discrete EBL models
from \citet{Andrews2018}. The orange line and hashed area show $\gamma$-ray
Blazar EBL constraints from the MAGIC TeV experiments \citep[see \eg][for a
summary]{Dwek2013}, and light grey triangles indicate {\it total EBL estimates}
that require accurate modeling of DGL and ZL and still include the IGL (Direct
Light). Purple triangles show the \citet{Matsuura2017} and \citet{Sano2020}
CIBER estimates of Diffuse Light in excess of the \citet{Kelsall1998} model
prediction. Dark blue upper limits are the SKYSURF 1.25--1.6~\mum\ Diffuse
Light limits from 34,000 WFC3/IR images of of \citet{Carleton2022} and
\citet{Windhorst2022}. Light blue circles with error bars at 0.61~\mum\ are the
\citet{Lauer2021, Lauer2022} Diffuse Light estimates with New Horizons at
43--51~AU. All Diffuse Light estimates plotted in color have the IGL
(=iEBL+eEBL) already subtracted. See \citet{Carleton2022} for a discussion of
possible causes of any remaining Diffuse Light.
}
\label{fig:fig13}
\end{figure*}

\bn \ 

\bn \ 

\bn \

\ve 

\bibliographystyle{aasjournal}


\bibliography{references_pearls_overview}{}




\vspace*{-1.00cm}
\appendix

\vspace*{-0.50cm}
\n \section{NIRCam Pipeline Processing Details: $1/f$ Corrections }\
\label{secAppA} 

\begin{figure*}[!hptb]
\n\cl{
\includegraphics[width=0.400\txw]{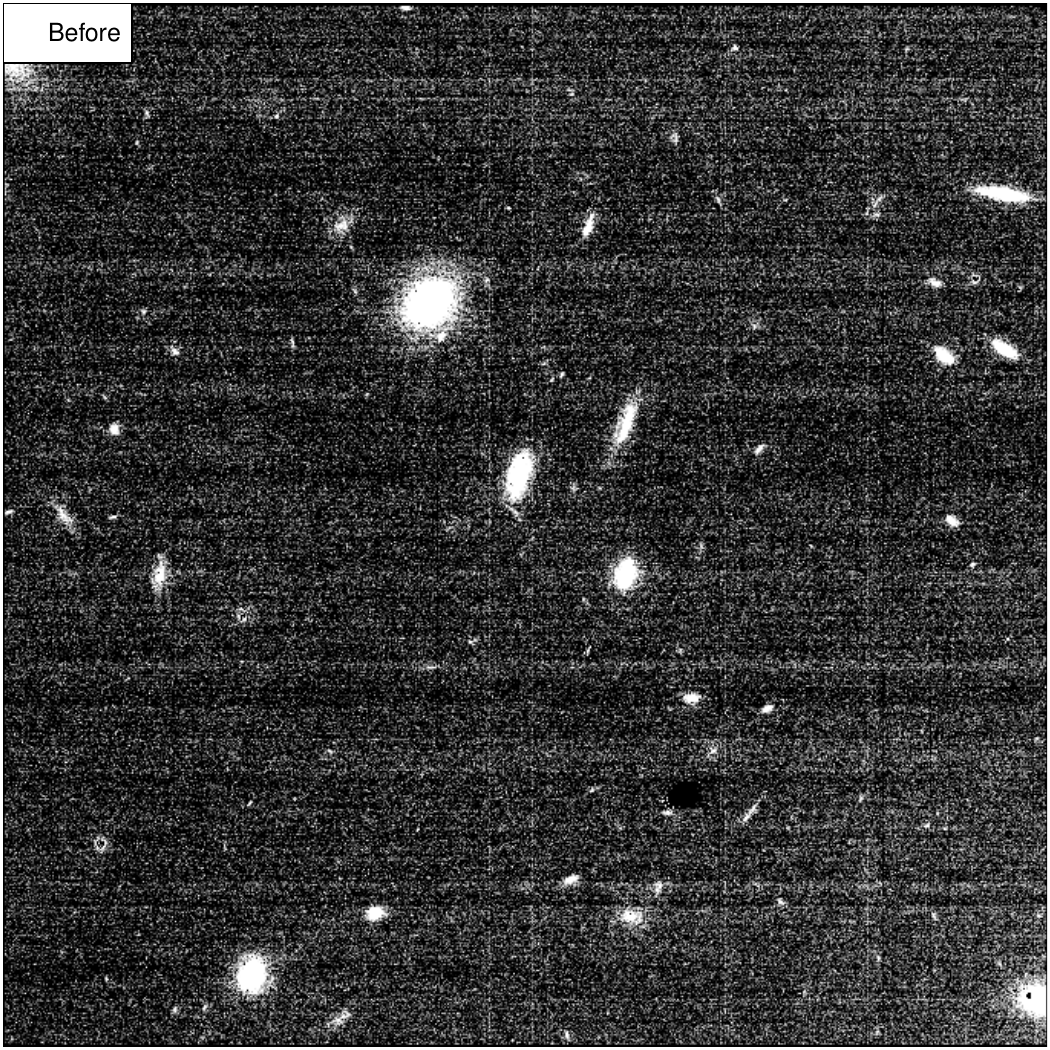}\
\includegraphics[width=0.400\txw]{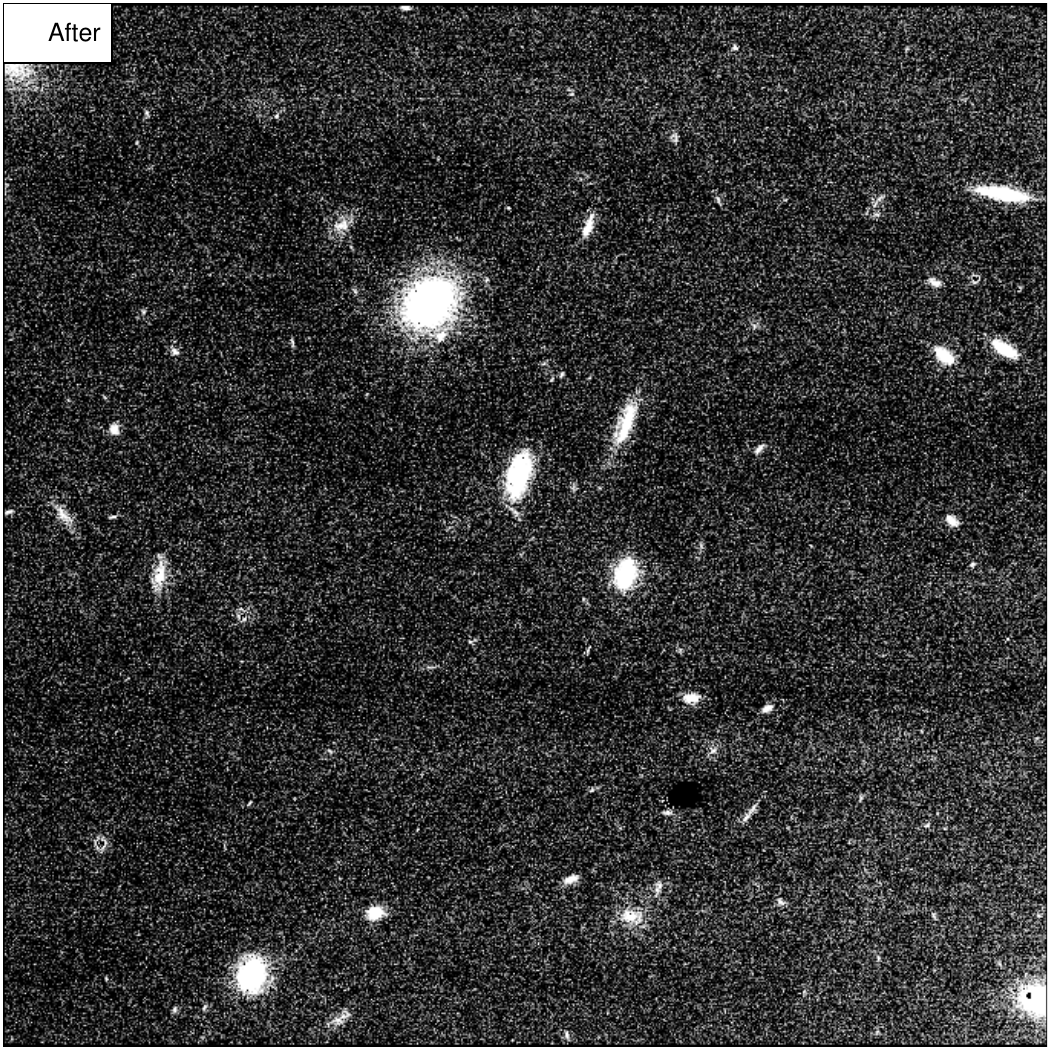}
}
\sn\cl{
\includegraphics[width=0.400\txw]{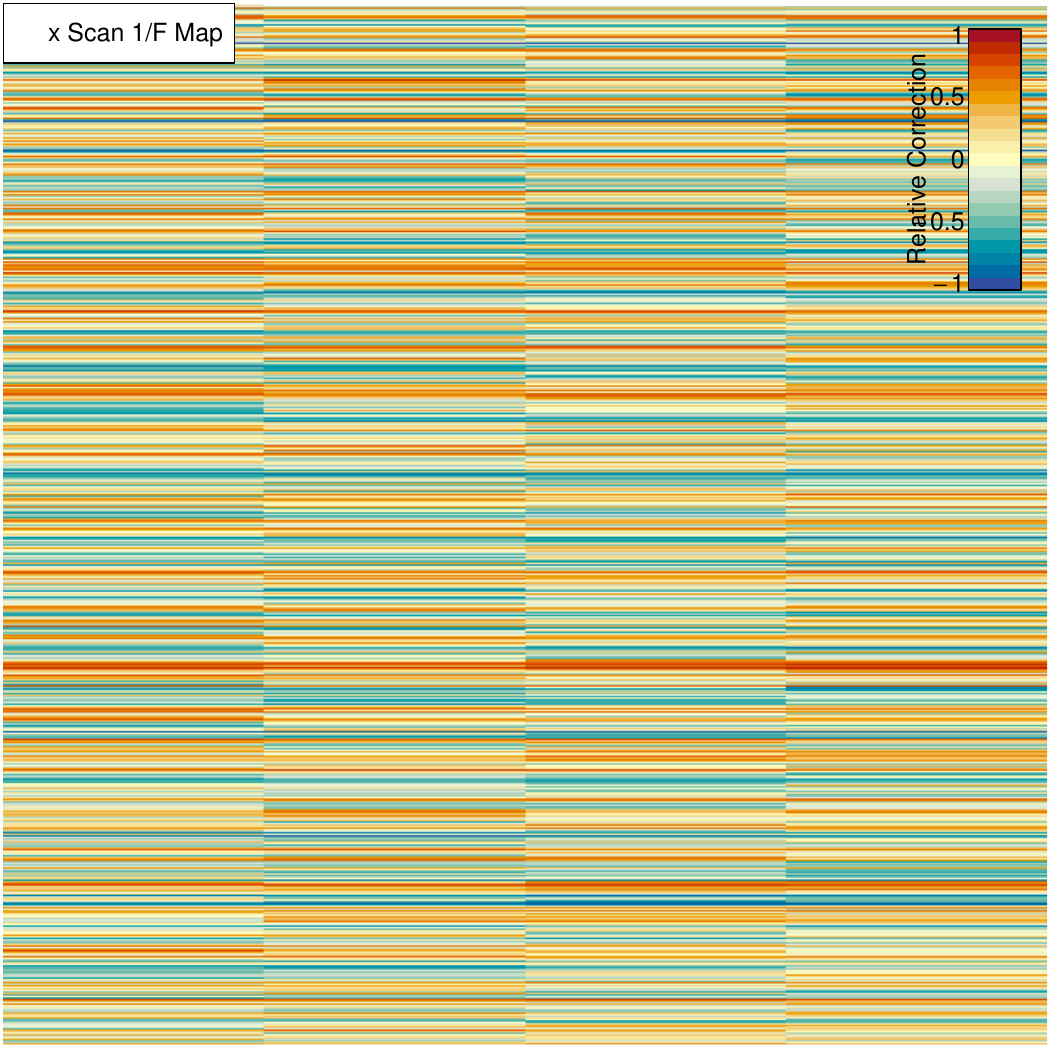}\
\includegraphics[width=0.400\txw]{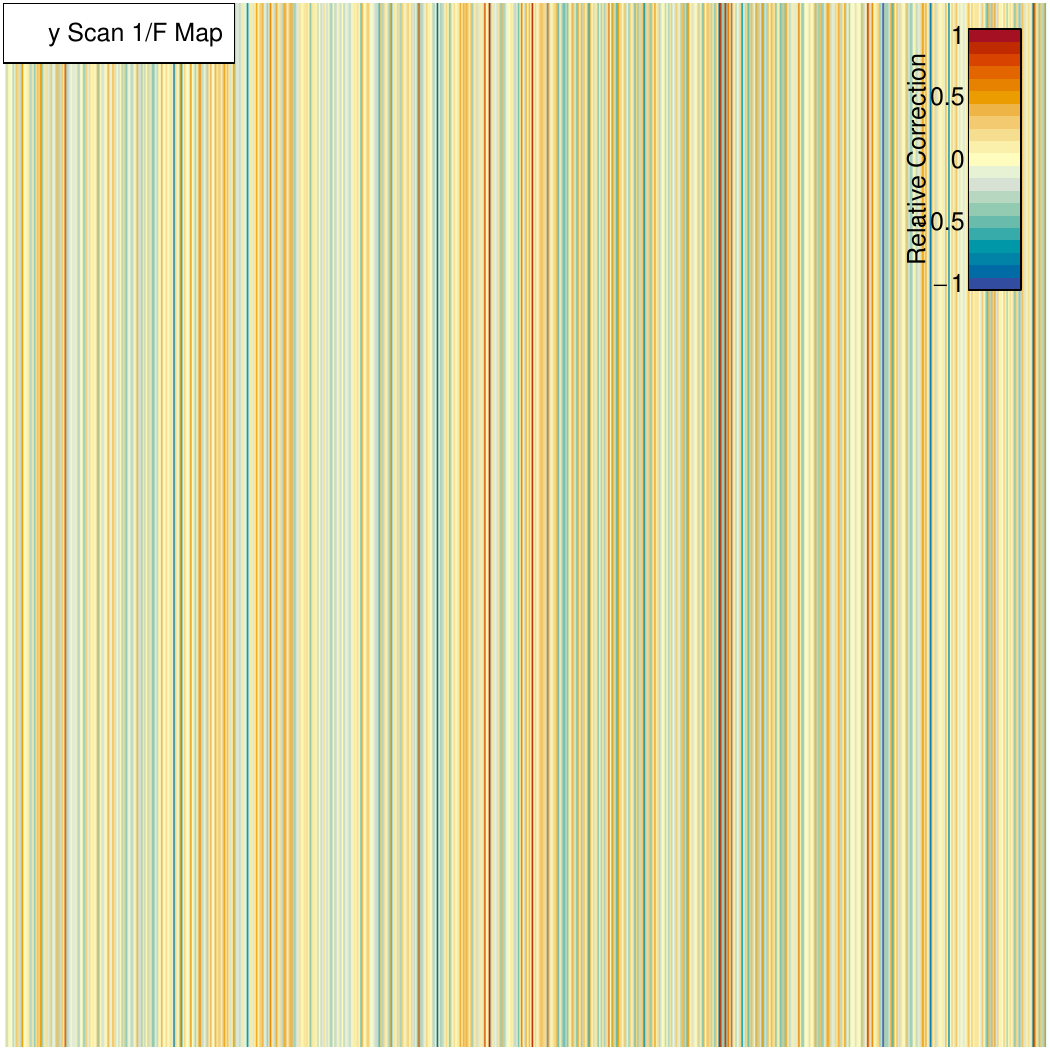}
}
\n\cl{\includegraphics[clip=true,trim=0 0 0 48,width=0.400\txw]{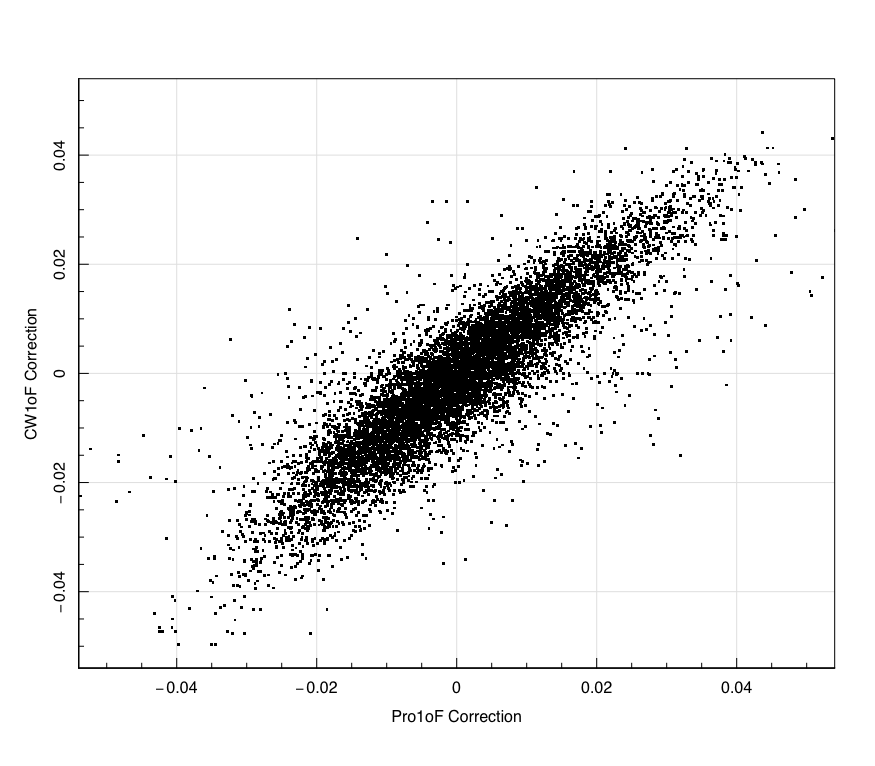}
}

\vspace*{-0.10cm}
\n \caption{
{\bf [(a) Top Left]:}\ Part of the JWIDF image before 1/f removal;
{\bf [(b) Top Right]:}\ Same part of the JWIDF image after 1/f removal with the
\ProFound-based 1/f subtraction code. (Please magnify these figures to see 
the details);
{\bf [(c) Middle Left]:}\ Row-wise 1/f pattern subtracted in the $x$-direction.
The inset color bars show the relative level of the corrections applied. The 
imprint of the four 5512$\times$2048 pixel sections read by the ASIC is
apparent;
{\bf [(d) Middle Right]:}\ Column-wise pattern subtracted in the $y$-direction;
{\bf [(e) Bottom]:}: Comparison of the Willott 1/f-removal algorithm compared to
the \ProFound-based 1/f subtraction. The two algorithms leave similar noise
imprints on the resulting sky-SB at the level of \cle 0.02--0.04~MJy/sr, where
1.0~MJy/sr typically corresponds to a 5$\sigma$ source of $\rm AB\sim28$~mag.
}
\label{fig:fig14}
\end{figure*}

We investigated several schemes to remove the NIRCam $1/f$ noise effects caused
by readout artifacts. Given the characteristics of the readout process and the
visual discontinuities evident in the images, the optimal approach proved to be
the following:
\begin{itemize}

\item All pixels with data quality flag DQ = 0 are masked.

\item The brightest 10\% of pixels are masked to filter out the real objects. 
Any number for masking between 10\% and 30\% works well in practice for a
diversity of images (including, \eg\ large galaxies in the field of view and
extremely empty frames).

\item Each $2040 \times 2040$ calibrated frame is divided into four $510 \times
2040$ sections (\ie\ shorter runs in the image $x$-dimension).

\item In scan blocks of $512 \times 1$, the $q=0.4$ quintile value is
calculated ignoring all masked pixels above. This creates a vector of length
2040 for each of the four sections analyzed.

\item For each scan block vector of length 2040 the running median is computed
with a window size of 101 pixels. This smooth distribution reflects large-scale
structure we wish to preserve and is removed from the vectors. Window sizes
between 51 and 201 pixels work well for the full diversity of images available.

\item Each vector is expanded along the $x$-dimension to create $510 \times
2040$ sections. The four sections are then combined to create a single $2040
\times 2040$ $1/f$ noise image. (See example in Figure~\ref{fig:fig14}c.)

\item The final $x$-direction noise map is removed from the original image.

\end{itemize}

A similar procedure is then carried out on the $y$-dimension except that the
entire $y$-column is analyzed. (See example in Figure~\ref{fig:fig14}d.) This
noise pattern is also removed, creating our final $1/f$-corrected frame. The
above process is all run by the function profoundSkyScan that is part of the
\ProFound\ package \citep{Robotham2017, Robotham2018}. For a particularly
difficult frame, Figure~\ref{fig:fig14}a shows an image before $1/f$ removal and
Figure~\ref{fig:fig14}b after $1/f$ removal. 

Figure~\ref{fig:fig14}e shows a comparison of the Willott $1/f$ removal
algorithm compared to the \ProFound-based $1/f$ subtraction. The average effect
of the $1/f$ correction is typically well below the pixel rms value. The rms
of this particular image is 0.06 in units of MJy/sr. The scatter away from the
$y=x$ line in Figure~\ref{fig:fig14}e shows that the two algorithms leave a
residual noise imprint on the resulting sky-SB that differs at the level of
$\sim$0.005 MJy/sr, so that systematics resulting from the $1/f$ correction
algorithms are \cle 10\% of the pixel rms level. Because 1.0 MJy/sr typically
corresponds to a compact $\sim$5$\sigma$ source of AB$\sim$28~mag
(Equation~\ref{eq:eq2} and Table~1), the $1/f$-removal algorithms create \cle
1\% flux changes for the faintest sources in the field and substantially less
for brighter sources. Also, the large scale structure of sources is preserved,
and no discontinuities are present between scan regions, at least those not
caused by the $1/f$ correction itself. 

The one difference between the Willott and \ProFound-based $1/f$ removal
algorithms is that the Willott algorithm does not remove large-scale gradients,
while the \ProFound-based algorithm has the option to remove such gradients,
although these are then preserved as a separate extension in the output {\sc
fits} files. As discussed in Section~4 of \citet{Windhorst2022}, such
large-scale gradients can be due to the real astrophysical scene (\eg\ cluster
ICL) and/or due to imperfections in the subtracted dark frames or color terms
in the applied flat-field corrections. Large-scale gradients are seldom more
than a few percent of the sky-SB across the image. Such gradients are in 
general not an issue for our purpose of constructing reliable, complete object
catalogs for faint and small objects. We preserve the information on
large-scale gradients when measuring the lowest estimated sky-SB (LES)
following the methods of \citet{Windhorst2022} that we apply for our JWST
sky-SB study of Section~\ref{sec5}. 


\vspace*{-0.80cm}
\hspace*{-0.00cm}
\n

\ve 

\vspace*{-1.50cm}
\n \section{NIRCam Pipeline Processing Details: ZP corrections and Error 
Budget} \label{secAppB} 

\n {\ch Appendix~\ref{secAppB1} compares the results from our recent v1
calibration with the early Pipeline reductions.} Appendix~\ref{secAppB2},
discusses if further corrections are needed to the NIRCam AB-magnitude scale in
order to compare our PEARLS 0.9-4.5~\mum\ galaxy counts to work done the last
few decades at the fiducial wavelengths of the VISTA+IRAC surveys. 
Appendix~\ref{secAppB3} summarizes the modeling of NIRCam sky-SB components and
their uncertainties.

{\ch \sn \subsection{Comparison of the Early Pipeline Calibrations and
jwst\_0995.pmap\_filters}\ \label{secAppB1} }

JWST photometric calibration (``zeropoint'' or ZP) has to be established in
flight and will evolve during the mission. Because standard-star observations
were not available in time, the earliest JWST observations had to be calibrated
with pre-flight ZPs. The MAST pipeline for NIRCam began using in-flight ZPs on
2022 July~27. At NIRCam wavelengths $\le$2.0~\mum, the on-orbit throughput was
near pre-launch expectations \citep{Rigby2022}, but it was up to 10\% smaller
for some filters. At wavelengths $>$2.0~\mum, the on-orbit throughput was about
15--30\% higher than the pre-launch expectations \citep[][their
Figure~8]{Rigby2022}. \citet{Rigby2022} also wrote that the throughput
stability is no worse than 4\% and is likely much better.

{\ch With the new ZPs derived from on-orbit standard star observations in all
NIRCam detectors and its main filters \citep{Boyer2022}, we reprocessed all our
PEARLS images with jwst\_0995.pmap\_filters (Section~\ref{sec3}). For the
record, the results from our original processing with jwst\_0916.pmap\_filters
through jwst\_0952.pmap\_filters are still available in the first submission of
this paper on \url{https://arxiv.org/abs/2209.04119}, which also had a table of
additional ZP corrections that this earlier processing required. {\it The
application of jwst\_0995.pmap\_filters has made these additional ZP
corrections obsolete}. It is nevertheless useful to give a brief comparison of
the main differences between jwst\_0995.pmap\_filters (v1) and our earlier
PEARLS images (v0.5):

\mn \bul (1) The latest v1 calibration more accurately corrects for ZP
variations between each of the 10 NIRCam detectors (typically by \cle 10--20\%
per detector). The improvement propagates into our images and science results
although in a rather subtle way, because our previous processing already had
averaged over 2--8 detectors and 2--4 fields. In general the change reduces
the dispersions in the sky-SB values but does not reduce their medians very
much, as described below. 

\sn \bul (2) The JWIDF has IRAC observations that allow direct comparison in
two NIRCam filters. These IRAC observations were accumulated every two weeks 
over 15 years and so are very deep \citep[\eg][]{Yan2018}. For unsaturated,
isolated, and matched objects in the magnitude range 18\cle AB\cle 20 mag in
both sets of images, Source Extractor MAG\_AUTO gives average total flux 
differences of F356W--IRAC1$\simeq$--0.026 mag and F444W--IRAC2$\simeq$--0.029 
mag. Small differences between the JWST and Spitzer fluxes can be caused by a
combination of filter differences, aperture corrections, and source confusion
due to the vastly different PSFs of the two telescopes. The NIRCam PSF has a
2.36$^2$$\simeq$5.56$\times$ larger area in F444W compared to F090W (Table~1 and
Figures~\ref{fig:fig6}--\ref{fig:fig8}). F444W also has a $\sim$140$\times$
smaller PSF area than the Spitzer IRAC2 filter. Despite these PSF differences,
the 2022 October NIRCam F356W and F444W zeropoints are consistent (within
2.6--2.9\%) with the deepest Spitzer images available. 

\sn \bul (3) The new calibration also tightened the dispersion between the
resulting galaxy counts when compared to the brighter galaxy counts in the
ground-based+HST+Spitzer filters and improved our estimates of the integrated
galaxy light (Section~\ref{sec4}), although these changes are well within the 
uncertainties quoted in Table~3 and hardly visible in 
Figures~\ref{fig:fig9}--\ref{fig:fig11}. 

\sn \bul (4) From the object-free sky-SB measurements in Section~\ref{sec5}, 
we confirm that the ZPs of the NIRCam SW detectors have become $\sim$10--20\%
more sensitive. That is, the ratio of August to October 2022 object-free sky-SB
values is typically 1.09--1.22 for the NIRCam SW filters, while this ratio is
typically 0.88--0.97 for the LW filters, which have thus become somewhat less
sensitive compared to the earlier values. Such ZP changes can affect the colors
of discrete objects, which we defer to future papers. 

\sn \bul (5) The new calibration has improved the rms variation between the
sky-SB measurements compared to our earlier calibrations. Specifically, the
relative errors on the object-free sky-SB values have significantly improved
with the new calibrations, with the October to August 2022 ratios of these
relative errors typically being a factor 0.63--0.83 for the SW filters and
sometimes less than 0.5 for the LW filters. Stated differently, the new ZP
calibrations and flat-field corrections have {\it reduced} the dispersion in
the sky-SB estimates significantly, especially for the LW filters. As a
consequence, our limits on diffuse light in Section~\ref{sec5} and
Figure~\ref{fig:fig13} have also improved, especially in the LW filters. 

\sn \bul (6) The more accurate detector-to-detector ZPs also improved the
overall quality of our sky-SB model fits in Figure~\ref{fig:fig12}, as
described in Section~\ref{sec52} \& Appendix~\ref{secAppC}. In particular, the
range in scale factors $f$ by which we needed to multiply our adopted
\citet{Rigby2022} straylight in Equation~\ref{eq:eq4} has increased from
$f=0.3$, 0.6, 0.8, 0.9 in our v0.5 reduction to $f\simeq0.5$, 0.8, 1.0, 1.0 in
v1, \ie\ providing a tighter range in the predicted SL values for our four
PEARLS fields. The SL spectrum, which we assumed to be constant other than
this factor $f$, in fact somewhat depends on JWST's pointing direction (J. Rigby
\etal\ 2022, private communication and in preparation). We refer to this work
for an in-depth discussion of the JWST straylight. For the very red thermal
Zodiacal SED \cite[see Figure 5 of][]{Rigby2022}, the effective wavelength 
\citep[Equation~A15 of][]{Bessell2012} of the F444W filter shifts from
$\sim$4.4~\mum\ for objects with relatively flat NIR-spectra like stars and 
galaxies to $\sim$4.57~\mum\ for the total sky-SB, which we accounted for in 
Figures~\ref{fig:fig12}--\ref{fig:fig13}.

In conclusion, we independently confirm the \citet{Rigby2022} and
\citet{Boyer2022} flux scale and adopt their suggested 4\% uncertainty in the
current JWST NIRCam flux scale in our error analysis below. While
model-dependent (Appendix~\ref{secAppC}), our sky-SB analysis does give a
nearly PSF-independent check on the zeropoints, to the extent that the sky-SB
is estimated in areas largely devoid of bright-object PSF-wings. The results
from the new calibrations have also been propagated into the error budgets for
our sky-SB values in Appendix~\ref{secAppB3}. The total uncertainties then are
\cle 8\% for the SW filters and \cle 6\% for the LW filters
(Section~\ref{secAppB3}). }

\sn \subsection{Comparison of NIRCam AB-mag Counts to VISTA/IRAC Counts}\
\label{secAppB2} 

\n In order to compare our PEARLS NIRCam 0.9--4.5~\mum\ object counts in 
Figures~\ref{fig:fig9}--\ref{fig:fig10} to previous work from ground-based
telescopes and Spitzer/IRAC, the NIRCam flux densities may need to be
transformed to the same flux scale as used for the filters in those previous
surveys. The NIRCam filter wavelengths are similar, but not identical, to those
used previously. The most extensive survey and number counts for
$\lambda<3$~\micron\ is that of \citet[][and references therein]{Koushan2021},
who combined many data sets. For $\lambda>3$~\micron, we use the compilation of
\citet{Driver2016a, Driver2016b}, who included number counts for AB\cle 18 mag
from WISE \citep{Jarrett2017} and for 18\cle AB\cle 26 mag from IRAC
\citep{Ashby2009, Ashby2015}. 

Table~6 lists the effective wavelengths $\lambda_c$ of the relevant VISTA/IRAC
and WISE filters in which these previous counts were done, as displayed in
Figures~\ref{fig:fig9}--\ref{fig:fig10}. We therefore use the effective 
wavelengths of the VISTA/IRAC filters in Table~6 as {\it fiducial} value to 
compare our NIRCam object counts to. The flux scale of the WISE W1 and W2 
filters was already transformed to the flux scale of these fiducial VISTA/IRAC 
filters. We used the ICRAR filter transform tool \citep{Robotham2020}
\footnote{\url{http://transformcalc.icrar.org} and 
\url{https://github.com/asgr/ProSpect}} to calculate the corrections needed to
bring the NIRCam AB-magnitude scale onto that of the VISTA/IRAC filters that
are closest in wavelength. The tool uses the filter and telescope transmission
curves folded with the detector QE curves for a large number of facilities to
perform numerical integration over a range of SEDs and redshifts, and produces
AB-flux scale offsets, $\Delta\rm AB$, and their uncertainties, 
$\sigma_{\Delta\rm AB}$, which we represent as:
\begin{equation} 
{\rm 
NIRCam\ \ ABmag = VISTA/IRAC\ ABmag\ +\ \Delta AB\ \pm \sigma_{\Delta AB}
} 
\label{eq:eq6} 
\end{equation}
\n The transformation requires an assumption for the redshift distribution of
the galaxy population at $\rm AB\simeq20$--28~mag, for which we used a median
redshift of \zmed$\simeq$1--2 \citep[\eg][]{Skelton2014, Inami2017}. The
resulting values for $\Delta\rm AB$ and $\sigma_{\Delta\rm AB}$ are given in
Table~6.\footnote{The NIRCam F115W filter is compared to VISTA $J$, because the
F115W filter is closer to $J$-band than the $Y$-band or VISTA 1.022~\micron,
which would have $\Delta\rm AB=+0.091$~mag.} For most filters, the uncertainty
in the transformation is 3--6\%, \ie\ similar to or larger than the actual flux
scale correction needed, which is --0.04 to +0.03 mag. This is mainly because
of the wide redshift range sampled. Therefore, no AB-mag scale corrections were
applied to our PEARLS NIRCam number counts to compare them to the VISTA/IRAC
counts. But we do add this uncertainty to our error budget. 

Table~6 also lists the Cosmic Variance uncertainty for our two PEARLS fields
used in the deep NIRCam galaxy counts thus far (Section~\ref{sec2} \&
\ref{sec45}). Assuming both ZP uncertainties and the CV uncertainty are
independent, we show the combined total error on the bottom line of Table~6.
These are our IGL errors used in Figures~\ref{fig:fig12}--\ref{fig:fig13}.
These errors are likely conservative, since other deep fields from HST, VLT and
Spitzer fold into the galaxy counts of Figures~\ref{fig:fig9}--\ref{fig:fig10}
thereby reducing CV, except at the faint end of the counts at wavelengths \cge
2.0\mum, where we only have NIRCam. 

\sn In summary, the uncertainty in the JWST NIRCam zeropoints is at least 4\%,
while the uncertainty of transforming the NIRCam AB-mag scale onto the fiducial
VISTA/IRAC filters that have been used for galaxy counts at brighter levels is
$\la$3--6\%. The combined uncertainty to compare counts that were done with
slightly different filters systems on different telescopes is thus
$\sim$3--7\%. Magnitude offsets of that size are hardly noticeable over the
very wide magnitude range plotted in Figures~\ref{fig:fig9}--\ref{fig:fig10}.
Future improvements in the NIRCam ZPs through further standard star monitoring
and more detailed comparison to the fluxes in the fiducial VISTA/IRAC filters
can provide a more accurate comparison, and observing more JWST fields will
decrease the uncertainty in the counts from Cosmic Variance. 


{
\vspace{+0.20cm}
\begin{verbatim}
Table 6. ZP & Transformation Uncertainties of JWST NIRCam to VISTA system for Galaxy Counts/IGL 
=================================================================================================
NIRCam filter:           F090W    F115W    F150W    F200W    F277W    F356W    F410M    F444W 
lambda_c (mum)           0.8985   1.1434   1.4873   1.9680   2.7279   3.5287   4.0723   4.3504
NIRCam ZP uncertainty    0.04     0.04     0.04     0.04    <0.04    <0.04    <0.04    <0.04     
-------------------------------------------------------------------------------------------------
VISTA/IRAC filter:      VISTA-Z  VISTA-J  VISTA-H  VISTA-K   -----    IRAC-1   -----    IRAC-2
lambda_c (mum)           0.883    1.254    1.648    2.154    -----    3.544    -----    4.487
-------------------------------------------------------------------------------------------------
Delta_AB                +0.026   -0.006   -0.044   -0.003    -----    0.001    -----   +0.018
Transform uncertainty    0.05     0.06     0.04     0.03     -----    0.01     -----    0.003
-------------------------------------------------------------------------------------------------
Total ZP uncertainty     0.06     0.07     0.06     0.05    <0.04    <0.04    <0.04    <0.04     
-------------------------------------------------------------------------------------------------
CV error                 0.09     0.09     0.09     0.09     0.09     0.09     0.09     0.09     
-------------------------------------------------------------------------------------------------
Total IGL error          0.11     0.12     0.11     0.10    <0.10    <0.10    <0.10    <0.10     
-------------------------------------------------------------------------------------------------
\end{verbatim}
} 
\label{tab:tab7}
\vspace{-0.30cm}
\n Note: The first tier of this Table lists the NIRCam ZP uncertainties from
Section~\ref{secAppB1} for each filter. The second tier lists the effective 
wavelengths of the VISTA/IRAC filters used as fiducial for the PEARLS NIRCam
galaxy counts in Section~\ref{sec45}. The third tier lists the correction
$\Delta\rm AB$ that would need to be added to the calibrated JWST AB magnitudes
to bring them onto the same AB-scale as used for the VISTA \citep{Koushan2021}
and IRAC \citep{Ashby2009, Ashby2015} galaxy counts using the ICRAR filter
transformation tool, together with its transformation error. The fourth tier
lists the combined NIRCam ZP uncertainties and the ICRAR transformation error.
The fifth tier lists the Cosmic Variance error expected for the 0.9--4.5~\mum\
galaxy counts in our two current PEARLS NIRCam fields (Section~\ref{sec45}).
The bottom tier lists the combined fractional error, assuming all contributions
are independent, and is used to assess the errors in our IGL parameters
(Section~\ref{sec46}). 

\sn \subsection{Uncertainties in the Observed NIRCam sky-SB Estimates }\
\label{secAppB3} 

\n For the uncertainties in our observed JWST NIRCam sky-SB values we need to
consider other error sources than those that apply to the flux-scale errors in
our galaxy counts in Appendix~\ref{secAppB2}. We wish to make an estimate of
the absolute sky-SB in our 13 NIRCam filters, and so the main sources of error
are different. For details of an assessment of this kind, we refer to
Section~4 and Table~5 of \citep{Windhorst2022}, where the sources of error in
the absolute sky-SB as measured by WFC3/IR were summarized for the F125W--F160W
filters. In short, their total errors in the estimated WFC3/IR sky-SB were
3--4\% in these filters, and dominated by the flat-field (\cle 2\%) and ZP
errors (\cle 1.5\%). This was through careful tracking of the WFC3/IR
performance and its calibration over 12 years in orbit. At this stage, such
errors for JWST are surely less well known, so we estimate our error budget by
giving conservative limits to each main component that affects our estimated
sky-SB values:

\mn {\bf (1) Algorithm to get Lowest Estimated Sky-SB (LES):}\ With the LES
algorithm of \citet{Windhorst2022} and \citep{OBrien2022}, we divided the
2048$\times$2048 pixel image from each individual NIRCam detector into
32$\times$32 boxes of 64$\times$64 pixels and used these to determine the
lowest estimated sky (LES) values following the percentile clip method of
\citep{OBrien2022}. From Monte Carlo simulations with realistic object
densities and CR distributions, they showed that the LES method gives reliable
estimates of the object-free sky-SB, to within 0.4\% of the simulated sky-SB,
even in the presence of 10\% gradients across the field. While the object
density in the NIRCam images of Section~\ref{sec45} is $\sim$3$\times$ higher
to $\rm AB\la 28.5$~mag compared to the HST WFC3/IR image density at its
detection limit of $\rm AB\la 26.5$~mag, the NIRCam SW and LW pixel size is
also $\sim$16--4$\times$ smaller in area compared to WFC3/IR, respectively, so
that at least similar amounts of empty sky are available to the current depth
in the NIRCam images to measure object-free LES sky-SB values. Hence we adopt
an uncertainty of \cle 0.4\% of the algorithm itself estimates the sky-SB in
the object free areas. When applying this algorithm, we find that it does
ignore areas with residual wisps and snowballs well (as it flags those as
potential objects with positive flux to be avoided in the sky-SB estimate). 

\mn {\bf (2) ZP Uncertainties:}\ The 4\% NIRCam ZP uncertainties of Table~6
also apply to the observed sky-SB values of Figure~\ref{fig:fig12}. Since we
plot all data points at their actual effective NIRCam wavelengths the Transform
uncertainties of Table~6 do not apply. Most of the JWST sky-SB comes from the
Zodiacal belt at distances \cle 3--5 AU \citep[\eg][]{Windhorst2022}, and the
IGL is $\sim$10--70$\times$ dimmer than than the ZL (Figure~\ref{fig:fig13}).
Hence, a \cle 9\% CV error in the IGL (Section~\ref{sec46}) is very small
compared to these other errors in the total sky-SB estimates. 

\mn {\bf (3) Flat-Field and Residual 1/f and Pedestal Uncertainties:}\ We
verified that the LES algorithm of Section~\ref{sec5} and
\citet{Windhorst2022} finds the cleanest regions to estimate the sky-SB in
each detector after the 1/f corrections of Appendix~\ref{secAppA}. With the
most recent reduction of context file jwst\_0995.pmap\_filters, the flat field
uncertainty has improved to $\sim$2\% compared to the 7--8\% uncertainty in our
earlier reductions with context file jwst\_0942.pmap\_filters (B. Sunnquist;
private communication). Since flat-field uncertainties can be a dominant
component in our error budget for absolute sky-SB estimates, we check for this
as following. We find that the LES sky-SB estimates have a \cle {\ch 2--4\%}
variation between the LW detectors in our 2.7--4.5~\mum\ filters, including our
TNJ1338 medium-band LW filters. However, these variations increase to \cle {\ch
4--7\%} for between the SW detectors in SW 0.9--2.0~\mum\ filters. This is
likely due to the larger number of detectors, some of which still have residual
offsets after the flat-fielding and pedestal removal procedure of
Section~\ref{sec31}. Hence, we adopt a {\ch 7\%} uncertainty in the flat-field
induced sky-SB estimate for all SW filters, and a {\ch 4\%} uncertainty for all
LW filters. 

\mn {\bf (4) Bias and Dark-Current Frame Subtraction Uncertainties:}\ In the
12 years on-orbit data analyzed for the WFC3/IR detectors, these errors were 
\cle 1\% following Section~4 and Table~5 of \citet{Windhorst2022}. The NIRCam
bias and dark current levels and their uncertainties listed in
Section~\ref{sec31} and its websites are also very low, typically \cle
1.4--2.1\% of the sky-SB at 3.5--2.0~\mum\ in Table~4,
respectively.\footnote{This estimate uses the detector gains
(Section~\ref{sec31}) and $PHOTMJSR$ and $PIXAR\_SR$ keywords in the FITS
headers.} Hence, uncertainties in the NIRCam dark current removal are much
smaller than the ZP and the flat-field plus residual pedestal uncertainties
above. 

\mn In summary, following the discussion of \citet{Windhorst2022}, we will
assume that the above errors in estimating the sky-SB are independent. This is
justified because the standard stars from which the NIRCam ZP are derived are
measured over an area much smaller than the above dominant flat-field/residual
pedestal errors. The resulting uncertainty in our combined error on the
absolute NIRCam sky-SB is thus {\ch $\sim$6\%} of the observed sky-SB for the LW
filters and {\ch $\sim$8\%} for the SW filters. We propagate these errors for
each filter and PEARLS field into the NIRcam sky-SB estimates of Tables~4--5 and
Figure~\ref{fig:fig13}. 


\sn \section{Thermal, Straylight, Zodiacal, and DGL Models to Interpret the 
JWST sky-SB}\ \label{secAppC} 

\mn In this section, we summarize the main components in the error budget when
modeling the sky-SB values observed by JWST NIRCam from L2. The ETC output file
\texttt{``backgrounds.fits''} contains an array in its second header which
contains the predicted ETC-straylight, ETC-thermal, and in-field ETC-Zodiacal
components, as well as the combined ETC-total foreground, respectively. Where
relevant, the uncertainties that we derived for these components below are
listed in Tables~4--5 between parentheses on the lines directly below the model
prediction for each component. 

\mn {\bf (1) Thermal Component:}\ The JWST ETC provides predictions for the 
thermal contribution from its own components at their various temperatures. 
JWST component temperatures are monitored continuously. They are typically
$\sim$42--45K for the OTE and 6--39 K for JWST's science instruments,
\footnote{\url{https://webb.nasa.gov/content/webbLaunch/whereIsWebb.html}} {\it
\ie\ considerably colder and more constant} than the varying ambient
temperatures of HST across its orbit \citep[see, \eg\ Appendix~\ref{secAppA}
of][]{Carleton2022}. As a consequence, the Thermal values for JWST NIRCam in
Tables~4--5 are predicted to be much lower than those for HST. The JWST thermal
radiation is in fact more than 100$\times$ lower than the predicted total
sky-SB even at 4.5~\mum, as can be seen in Figure~\ref{fig:fig12}. JWST thermal
sensors on the website above report typical NIRCam temperatures stable at 38.5
K to well within 1 K for many days after its initial cool-down period. We will
thus adopt the ETC thermal sky-SB predictions for NIRcam, and assume that we
may ignore its uncertainties in our total error budget as it is the smallest of
all components. Note that this situation is quite different for HST, where some
component temperatures remain at room temperature and can vary by $\pm$a few
K within an orbit, resulting in non-negligible thermal dark signal in the
WFC3/IR F160W filter \citep{Carleton2022}. As a consequence, JWST can make more
accurate sky-SB observations that are less sensitive to thermal signal than HST
and can do so at much longer wavelengths. 

\mn {\bf (2) Stray Light Model Prediction and its Uncertainty:}\ The JWST SL
model is created by ray-tracing the infrared sky from 2MASS and WISE onto 
JWST, and estimates the fraction of light that can make it onto the detector
\citep[\eg][]{Lightsey2016}. This depends on the dust deposition on JWST mirrors,
which after launch appeared to be much smaller than the requirements
\citep{Rigby2022}. This straylight is significantly out of focus, and to first
order generates an elevated sky-SB onto the NIRCam detectors with a predicted
overall spectrum. The uncertainty in the predicted SL amplitude is not well
known from first principles. During its development, the JWST Project designed
the telescope and sunshield with a requirement that the SL in general be \cle
40\% of the Zodiacal sky-SB at 2.0~\mum\ wavelength. Figure~5 of
\citet{Rigby2022} suggests that the JWST 1--5~\mum\ SL may be substantially
lower than this requirement. Hence, in Figure~\ref{fig:fig12} we adopt the
lower of the two \citet{Rigby2022} SL curves as our fiducial. 

Because the uncertainty in the SL prediction is not well known, we will assume
that at 3.5~\mum\ --- where the total JWST sky-SB in Equation~\ref{eq:eq4} is
lowest --- the JWST sky-SB prediction JWST(Pred) should not exceed but match
the observed sky-SB value JWST(Obs). We found that it was not possible to do
this by assuming the full \citet{Rigby2022} SL as fiducial --- in {\ch two}
panels of Figure~\ref{fig:fig12} the predicted JWST sky-SB would be much higher
than the observed NIRCam sky-SB values, if we assumed 100\% of the
\citet{Rigby2022} SL. We therefore allowed the fraction $f$ in
Equation~\ref{eq:eq4} to vary, while assuming $f$\cle 1.0. We then attempted
to find the fraction $f$ by which we need to multiply the lower
\citet{Rigby2022} SL-value to get a best fit to our observed 3.5--4.5~\mum\
sky-SB values in Tables~4--5. That is, we set $f$ to produce the best match to
the difference in JWST(Obs)--JWST(Pred) in Equation~\ref{eq:eq5} at
3.5--4.5~\mum\ using the sum in Equation~\ref{eq:eq4}. 

For the TNJ1338, El Gordo, VV191, and JWIDF fields we find that the 
3.5--4.5~\mum\ JWST(Pred) values in Equation~\ref{eq:eq4} best match the
observed sky-SB JWST(Obs) in Tables~4--5 when we use multipliers for the
\citet{Rigby2022} SL of {\ch $f$$\simeq$0.5, 0.8, 1.0 and 1.0}, respectively. We
estimate that the $f$$\times$SL values used in Tables~4--5 are uncertain by at
least 0.2$\times$the \citet{Rigby2022} SL itself. Based on this variation, we
adopt 0.2$\times$SL as the straylight error in our error budget in Tables~4--5.
The $f$$\times$SL values are generally a factor of 2--10$\times$ lower than the
total predicted JWST sky-SB in Figure~\ref{fig:fig12}. Hence, the assumption of
a straylight uncertainty of 20\% of the total \citet{Rigby2022} SL value
results in the JWST SL being the dominant uncertainty in predicting the JWST
sky-SB, as shown in Tables~4--5. 

\mn {\bf (3) L2 Zodiacal Light Model Prediction and its Uncertainty:}\ Zodiacal
light intensities for PEARLS' JWST observations were calculated using the 
Spitzer background model. That model was derived from the \citet{Kelsall1998}
model,\footnote{
\url{https://lambda.gsfc.nasa.gov/product/cobe/dirbe\_zodi\_sw.html}} which was
designed for the COBE/DIRBE observations from low Earth orbit (LEO)\null. The
Spitzer model updated the scattering component to increase the contrast between
Ecliptic plane and poles and generalized the model to a wider and continuous
range of wavelengths and to arbitrary locations in the Solar system, as needed 
for the slowly changing Spitzer position around the Sun compared to the Earth. 
This model includes the L2 location, which is $\sim$1,500,000~km from Earth.
Details of this model are given on the IRSA website.\footnote{
\url{https://irsa.ipac.caltech.edu/data/SPITZER/docs/dataanalysistools/tools/contributed/general/zodiacallight/}\ 
and\
\url{https://irsa.ipac.caltech.edu/data/SPITZER/docs/files/spitzer/background.pdf}}
The Spitzer model was run using the ephemeris of JWST's L2 orbit from the ESA
website\footnote{
\url{https://www.cosmos.esa.int/web/spice/operational-kernels-data}\ using the
kernel {\it jwst\_horizons\_20211225\_20240221\_v01.bsp}. This JWST ephemeris
may need to be corrected from time to time due to JWST station-keeping burns
that happen every few weeks.} and the actual times of our PEARLS observations
in Table~1. Figure~\ref{fig:fig12} shows the resulting Zodiacal Light
intensities as predicted for JWST's position in L2. 

The Zodiacal-light brightness depends on not only on distance from the Sun but
also on the density and temperature profiles of the interplanetary dust (IPD)
cloud and on the specific line of sight through the cloud. Solar elongation
angle in particular is a significant factor. L2's distance from the Sun is on
average $\sim$1\% larger than the Earth's, but the other details matter for
specific observations. Comparing the four PEARLS observations so far to what
would have been seen in LEO, the scattered sunlight component was $\sim$1\%
fainter for El Gordo and TNJ1338 but $\sim$1--2\% brighter for VV~191 and the
JWIDF\null. (A larger-than-average path length through the IPD cloud at JWST's
south-of-ecliptic orbital position at that time probably explains the latter.)
The thermal Zodiacal component at 3~\micron\ was 3--7\% dimmer at L2 compared
to LEO, but around $\sim$1.5 and 30\micron\ there was at most 2\% difference.
The larger difference at the shorter wavelengths occurs because the lower
temperatures at the larger solar distance of L2 have a stronger impact on the
Wien side of the thermal spectrum. Overall, the differences between L2 and LEO
are modest.

\citet[][their Table~7]{Kelsall1998} reported uncertainties in their ZL model
of 15~\nWsqmsr\ at 1.25~\mum, 6~\nWsqmsr\ at 2.2~\mum, 2.1~\nWsqmsr\ at 
3.5~\mum, and 5.9~\nWsqmsr\ at 4.9~\mum, respectively. These come from their
IPD-cloud modeling uncertainties. These uncertainties are also present in the
Spitzer model predictions and correspond to Zodiacal model uncertainties at our
NIRCam wavelengths in Table~1 of $\sim$11~\nWsqmsr\ at 1.49~\mum,
$\sim$7~\nWsqmsr\ at 2.0~\mum, $\sim$2.2~\nWsqmsr\ at 3.53~\mum, and
$\sim$4.1~\nWsqmsr\ at 4.57~\mum, respectively. At the darkest Zodiacal sky-SB
measured in the JWIDF of $\sim$131~\nWsqmsr\ at 1.49~\mum, $\sim$76.7~\nWsqmsr\
at 2.0~\mum, $\sim$34.3~\nWsqmsr\ at 3.53~\mum, and $\sim$107~\nWsqmsr\ at
4.57~\mum, these L2 Zodiacal modeling uncertainties are $\sim$8\%, $\sim$9\%,
$\sim$6\%, and $\sim$4\% at these four wavelengths, respectively. Blueward of
the bluest COBE/DIRBE 1.25\mum\ filter, the Zodiacal sky-SB values predicted
for the NIRCam F115W and F090W filters are less reliable and should be viewed
with caution. Our four PEARLS fields observed so far span a wide range of
Ecliptic latitudes, and therefore the Zodiacal sky-SB differs significantly
among the fields. These differences are much more than the adopted 4--9\%
uncertainties here, as shown in Figure~\ref{fig:fig12}. We fold these L2
Zodiacal model uncertainties into the total error budget to predict the NIRCam
sky-SB in Section~\ref{sec5}. 

\mn {\bf (5) Diffuse Galactic Light Model Prediction and its Uncertainty:}\ 
The DGL intensities for each PEARLS target came from the Spitzer IPAC IRSA
model prediction (where it is referred to as
``ISM''),\footnote{\url{https://irsa.ipac.caltech.edu/applications/BackgroundModel/}}
as discussed by \citet[][and references therein]{Carleton2022}. The work of
\citet{Sano2016}, \citet{Sano2017}, and \citet{Onishi2018} has suggested that
the DGL as derived from the IRSA model of \citet{Brandt2012} can be uncertain
by a factor of two. Hence, in our error budget will include an DGL uncertainty
of $\pm$0.3 dex in the predicted JWST(Pred) values used in
Equation~\ref{eq:eq4} in Tables~4--5. The DGL is generally a factor of
20--100$\times$ lower than the total predicted JWST sky-SB in
Figure~\ref{fig:fig12}, so that a factor of two DGL uncertainty is not the
dominant error in predicting the total JWST sky-SB. The IRSA model predicted
the highest DGL for TNJ1138 amongst all our PEARLS targets, in fact so high
that zero SL would be required for TNJ1138, which seems unrealistic. Assuming
that the IRSA DGL prediction is too high by 0.3 dex for this field alone, we
adopt this lower DGL value, in which case still only a {\ch 
$\sim$0.5}$\times$SL level is required in Figure~\ref{fig:fig12}a. In any case,
the Zodiacal level in the TNJ1138 field is the brightest of all PEARLS fields,
about 10$\times$higher than the nominal SL and $\sim$10--100$\times$ higher
than the DGL prediction. This illustrates the limitations of our current
assessment. Better SL and DGL models are needed to more accurately predict JWST
NIRCam's observed sky-SB levels in future work. 

\mn {\bf (6) Subtracted eEBL and its Uncertainty:}\ The last component in
Equation~\ref{eq:eq5} is the eEBL, \ie\ the fraction of the IGL extrapolated in
Section~\ref{sec4} that comes from discrete objects that remain undetected in
the NIRCam images for AB\cge 28.5 mag. This fraction is only 2.5\% of the total
IGL, which itself has an uncertainty of $\sim$10\% (section~\ref{secAppB3}). 
Hence, the eEBL uncertainty is a very small part of the total error budget,
and is not listed in Tables~4--5. 

\mn {\bf (7) Resulting Error Budget for the NIRCam sky-SB Model:}\ In
conclusion, in our error budget of the diffuse light, the uncertainty in the
JWST SL model prediction is the largest uncertainty. Under the assumption that
the uncertainties in modeling the L2 Zodiacal Light, the JWST Stray Light, and
the Diffuse Galactic Light are independent, we add them in quadrature in
Tables~4--5. The combined uncertainties in the modeled NIRCam sky-SB values are 
listed on the line below each component of the predicted JWST sky-SB 
(Total-Predict-skySB), and are typically $\sim$10\%. 

\mn With the summaries of Appendix~\ref{secAppB}--\ref{secAppC} we have all
the tools to make estimates of the object-free NIRCam sky-SB and compare these
to the currently available models. This is discussed in Section~\ref{sec5},
Tables~4--5, and Figure~\ref{fig:fig13}. 

\ve 

\end{document}